\documentclass[twocolumn]{aastex631}

\usepackage{comment} % added by FDE
\usepackage{xspace} % added by FDE
\let\oldAA\AA
\renewcommand{\AA}{\text{\oldAA}\xspace} % AA in math mode and with 

\newcommand{\txn}[1]{\textnormal{#1}}
\newcommand{\Mtot}{\hbox{$\txn{M}_\txn{tot}$}}
\newcommand{\Mstar}{\hbox{{$\txn{M}_{\ast}$}}}
\newcommand{\Msun}{\hbox{$\txn{M}_{\odot}$}}
\newcommand{\tauV}{\hbox{$\hat{\tau}_{\scriptscriptstyle V}$}}

\newcommand{\Zsun}{\hbox{$\txn{Z}_\odot$}}
\newcommand{\logUs}{\hbox{$\log \txn{U}_{\scriptscriptstyle S}$}}
\newcommand{\Zstar}{\hbox{$\txn{Z}_{\ast}$}}
\newcommand{\Zmetal}{\hbox{$\txn{Z}$}}
\newcommand{\Zism}{\hbox{$\txn{Z}_\textsc{ism}$}}

\newcommand{\age}{\hbox{$\txn{t}$}}
\newcommand{\SFR}{\hbox{$\txn{SFR}$}}
\newcommand{\fesc}{\hbox{$\txn{f}_{\mathrm{esc}}$}}
\newcommand{\yrInv}{\hbox{$\txn{yr}^{-1}$}}

\newcommand{\OII}{\hbox{[O\,{\sc ii]}}}

\accepted{September 30, 2024}

\begin{document}

\title{Searching for Emission Lines at $z>11$: The Role of Damped Lyman-$\alpha$ and Hints About the Escape of Ionizing Photons}

\author[0000-0003-4565-8239] {Kevin N.\ Hainline}
\affiliation{Steward Observatory, University of Arizona, 933 N. Cherry Ave, Tucson, AZ 85721, USA}

\author[0000-0003-2388-8172] {Francesco D'Eugenio}
\affiliation{Kavli Institute for Cosmology, University of Cambridge, Madingley Road, Cambridge CB3 0HA, UK}
\affiliation{Cavendish Laboratory, University of Cambridge, 19 JJ Thomson Avenue, Cambridge CB3 0HE, UK}

\author[0000-0002-6780-2441] {Peter Jakobsen}
\affiliation{Cosmic Dawn Center (DAWN), Copenhagen, Denmark}
\affiliation{Niels Bohr Institute, University of Copenhagen, Jagtvej 128, DK-2200, Copenhagen, Denmark}

\author[0000-0002-7636-0534]{Jacopo Chevallard}
\affiliation{Department of Physics, University of Oxford, Denys Wilkinson Building, Keble Road, Oxford OX1 3RH, UK}

\author[0000-0002-6719-380X] {Stefano Carniani}
\affiliation{Scuola Normale Superiore, Piazza dei Cavalieri 7, I-56126 Pisa, Italy}

\author[0000-0002-7595-121X] {Joris Witstok}
\affiliation{Kavli Institute for Cosmology, University of Cambridge, Madingley Road, Cambridge CB3 0HA, UK}
\affiliation{Cavendish Laboratory, University of Cambridge, 19 JJ Thomson Avenue, Cambridge CB3 0HE, UK}

\author[0000-0001-7673-2257] {Zhiyuan Ji}
\affiliation{Steward Observatory, University of Arizona, 933 N. Cherry Ave, Tucson, AZ 85721, USA}

\author[0000-0002-9551-0534] {Emma Curtis-Lake}
\affiliation{Centre for Astrophysics Research, Department of Physics, Astronomy and Mathematics, University of Hertfordshire, Hatfield AL10 9AB, UK}

\author[0000-0002-9280-7594] {Benjamin D.\ Johnson}
\affiliation{Center for Astrophysics $|$ Harvard \& Smithsonian, 60 Garden St., Cambridge MA 02138 USA}

\author[0000-0002-4271-0364] {Brant Robertson}
\affiliation{Department of Astronomy and Astrophysics, University of California, Santa Cruz, 1156 High Street, Santa Cruz CA 96054, USA}

\author[0000-0002-8224-4505] {Sandro Tacchella}
\affiliation{Kavli Institute for Cosmology, University of Cambridge, Madingley Road, Cambridge CB3 0HA, UK}
\affiliation{Cavendish Laboratory, University of Cambridge, 19 JJ Thomson Avenue, Cambridge CB3 0HE, UK}

\author[0000-0002-2678-2560] {Mirko Curti}
\affiliation{European Southern Observatory, Karl-Schwarzschild-Strasse 2, 85748 Garching, Germany}

\author[0000-0003-3458-2275] {Stephane Charlot}
\affiliation{Sorbonne Universit\'e, CNRS, UMR 7095, Institut d'Astrophysique de Paris, 98 bis bd Arago, 75014 Paris, France}

\author[0000-0003-4337-6211] {Jakob M.\ Helton}
\affiliation{Steward Observatory, University of Arizona, 933 N. Cherry Ave, Tucson, AZ 85721, USA}

\author[0000-0001-7997-1640]{Santiago Arribas }
\affiliation{Centro de Astrobiolog\'ia (CAB), CSIC–INTA, Cra. de Ajalvir Km.~4, 28850- Torrej\'on de Ardoz, Madrid, Spain}

\author[0000-0003-0883-2226] {Rachana Bhatawdekar}
\affiliation{European Space Agency (ESA), European Space Astronomy Centre (ESAC), Camino Bajo del Castillo s/n, 28692 Villanueva de la Cañada, Madrid, Spain; European Space Agency, ESA/ESTEC, Keplerlaan 1, 2201 AZ Noordwijk, NL}

\author[0000-0002-8651-9879] {Andrew J.\ Bunker}
\affiliation{Department of Physics, University of Oxford, Denys Wilkinson Building, Keble Road, Oxford OX1 3RH, UK}

\author[0000-0002-0450-7306] {Alex J.\ Cameron}
\affiliation{Department of Physics, University of Oxford, Denys Wilkinson Building, Keble Road, Oxford OX1 3RH, UK}

\author[0000-0003-1344-9475] {Eiichi Egami}
\affiliation{Steward Observatory, University of Arizona, 933 N. Cherry Ave, Tucson, AZ 85721, USA}

\author[0000-0002-2929-3121] {Daniel J.\ Eisenstein}
\affiliation{Center for Astrophysics $|$ Harvard \& Smithsonian, 60 Garden St., Cambridge MA 02138 USA}

\author[0000-0002-8543-761X] {Ryan Hausen}
\affiliation{Department of Physics and Astronomy, The Johns Hopkins University, 3400 N. Charles St. Baltimore, MD 21218}

\author[0000-0002-5320-2568]{Nimisha Kumari}
\affiliation{AURA for European Space Agency, Space Telescope Science Institute, 3700 San Martin Drive. Baltimore, MD, 21210}

\author[0000-0002-4985-3819] {Roberto Maiolino}
\affiliation{Kavli Institute for Cosmology, University of Cambridge, Madingley Road, Cambridge CB3 0HA, UK}
\affiliation{Cavendish Laboratory, University of Cambridge, 19 JJ Thomson Avenue, Cambridge CB3 0HE, UK}
\affiliation{Department of Physics and Astronomy, University College London, Gower Street, London WC1E 6BT, UK}

\author[0000-0003-4528-5639]{Pablo G. P\'erez-Gonz\'alez}
\affiliation{Centro de Astrobiolog\'ia (CAB), CSIC–INTA, Cra. de Ajalvir Km.~4, 28850- Torrej\'on de Ardoz, Madrid, Spain}

\author[0000-0002-7893-6170] {Marcia Rieke}
\affiliation{Steward Observatory, University of Arizona, 933 N. Cherry Ave, Tucson, AZ 85721, USA}

\author[0000-0001-5333-9970] {Aayush Saxena}
\affiliation{Department of Physics, University of Oxford, Denys Wilkinson Building, Keble Road, Oxford OX1 3RH, UK}
\affiliation{Department of Physics and Astronomy, University College London, Gower Street, London WC1E 6BT, UK}

\author{Jan Scholtz}
\affiliation{Kavli Institute for Cosmology, University of Cambridge, Madingley Road, Cambridge CB3 0HA, UK}
\affiliation{Cavendish Laboratory, University of Cambridge, 19 JJ Thomson Avenue, Cambridge CB3 0HE, UK}

\author[0000-0001-8034-7802]{Renske Smit}
\affiliation{Astrophysics Research Institute, Liverpool John Moores University, 146 Brownlow Hill, Liverpool L3 5RF, UK}

\author[0000-0002-4622-6617] {Fengwu Sun}
\affiliation{Steward Observatory, University of Arizona, 933 N. Cherry Ave, Tucson, AZ 85721, USA}

\author[0000-0003-2919-7495] {Christina C. Williams}
\affiliation{NSF’s National Optical-Infrared Astronomy Research Laboratory, 950 North Cherry Avenue, Tucson, AZ 85719, USA}

\author[0000-0001-9262-9997] {Christopher N.\ A.\ Willmer}
\affiliation{Steward Observatory, University of Arizona, 933 N. Cherry Ave, Tucson, AZ 85721, USA}

\author[0000-0002-4201-7367]{Chris Willott}
\affiliation{NRC Herzberg, 5071 West Saanich Rd, Victoria, BC V9E 2E7, Canada}

\begin{abstract}
We describe new ultra-deep James Webb Space Telescope (JWST) NIRSpec PRISM and grating spectra for the galaxies JADES-GS-z11-0 ($z_{\mathrm{spec}} = 11.122^{+0.005}_{-0.003}$) and JADES-GS-z13-0 ($z_{\mathrm{spec}} = 13.20^{+0.03}_{-0.04}$), the most distant spectroscopically-confirmed galaxy discovered in the first year of JWST observations. The extraordinary depth of these observations (75 hours and 56 hours, respectively) provides a unique opportunity to explore the redshifts, stellar properties, UV magnitudes, and slopes for these two sources. For JADES-GS-z11-0, we find evidence for multiple emission lines, including [\ion{O}{2}]$\lambda\lambda3726,3729$\AA and [\ion{Ne}{3}$]\lambda3869$\AA, resulting in a spectroscopic redshift we determine with 94\% confidence. We present stringent upper limits on the emission line fluxes and line equivalent widths for JADES-GS-z13-0. At this spectroscopic redshift, the Lyman-$\alpha$ break in JADES-GS-z11-0 can be fit with a damped Lyman-$\alpha$ absorber with $\log{(N_\mathrm{HI}/\mathrm{cm}^{-2})} = 22.42^{+0.093}_{-0.120}$. These results demonstrate how neutral hydrogen fraction and Lyman-damping wings may impact the recovery of spectroscopic redshifts for sources like these, providing insight into the overprediction of the photometric redshifts seen for distant galaxies observed with JWST. In addition, we analyze updated NIRCam photometry to calculate the morphological properties of these resolved sources, and find a secondary source $0.3^{\prime\prime}$ south of JADES-GS-z11-0 at a similar photometric redshift, hinting at how galaxies grow through interactions in the early Universe.
\end{abstract}


\section{Introduction} \label{sec:intro}

The first two years of science from the James Webb Space Telescope (JWST) have completely transformed our understanding of galaxies in the very early Universe. A number of studies have led to spectroscopic confirmations of dozens of ``ultra-high redshift'' galaxies ($z > 10$), where these sources are seen less than $\sim 500$~Myr after the Big Bang \citep{robertson2023, curtislake2023, arrabalharo2023b, fujimoto2023, bunker2023a, bunker2023b, wang2023, hsiao2023, finkelstein2023, deugenio2023, castellano2024, zavala2024}. The spectra of these sources, as observed with the JWST near-infrared spectrograph NIRSpec \citep{jakobsen2022}, are varied: some show nebular UV or optical emission lines, while many are featureless except for the Lyman-$\alpha$ break.

The absence of emission lines in the spectra of ultra high-redshift galaxy spectra is surprising given the large star-formation rates (SFRs) and lack of dust predicted for these sources. Many explanations have been put forth to explain these observations, such as a lower gas-phase metallicity \citep{schaerer2022, curti2023, nakajima2023}, a higher escape fraction of ionizing photons in these sources \citep{curtislake2023, tacchella2023}, or a bursty star formation history with a duty cycle favoring extended periods of low star formation \citep{endsley2023, looser2023}. At such large distances and low observed fluxes, faint emission lines are difficult to discern from the noise in these spectra. 

The very bright ($M_{\mathrm{UV}} = -21.5$) galaxy GN-z11 \citep{oesch2016, tacchella2023b} at $z_{\mathrm{spec}} = 10.6$ was observed using the NIRSpec PRISM and grating dispersers and the resulting UV spectrum shows several strong lines \citep{bunker2023a}. Many of the emission line strengths and flux ratios in this source were dissimilar to those measured in metal-poor star-forming galaxies in the local Universe, which has been ascribed to stellar collisions, tidal disruption events, globular clusters, a top-heavy initial mass function, contributions from Wolf-Rayet and supermassive stars or the effects of a growing supermassive black hole \citep{cameron2023, senchyna2023, kobayashi2023, maiolino2023, bekki2023, dantona2023, isobe2023, watanabe2024}. This variety of physical phenomena demonstrates the complexity of the UV spectra observed in ultra-high redshift galaxies. 

Recently, \citet{deugenio2023} analyzed deep NIRSpec observations of JADES-GS-z12-0 ($z_{\mathrm{spec}} = 12.482 \pm 0.012$), a source first discovered in \citet{robertson2023, curtislake2023}. They found strong evidence for \ion{C}{3}]$\lambda\lambda$1907,1909 nebular emission, making this the highest-redshift detection of an emission line to date. As there is only an upper limit on the detection of the [\ion{O}{3}]$\lambda$1666 emission line, these authors calculate a super-solar C/O ratio for this source ([C/O]$ > 0.15$), in tension with results from JWST at $z = 6-9$ \citep{jones2023, stiavelli2023}. Importantly, the authors present evidence for damped Lyman-$\alpha$ absorption \citep[DLA; e.g., ][]{wolfe2005} in this source, in addition to absorption from the neutral intergalactic medium (IGM) along the line-of-sight. This potential DLA system provides insight into the physics of the gas surrounding sources at high redshift. 

The potential presence of a DLA is important for estimating the redshifts of galaxies without emission or absorption lines, as additional DLA absorption can bias the estimated wavelength of the IGM-driven Lyman-$\alpha$ break at $\sim 1216$~\AA \citep{curtislake2023, umeda2023, heintz2023, willott2023}. Given the high neutral gas fractions in the early Universe \citep{naidu2020, umeda2023}, we expect to observe more DLAs in galaxies at $z > 10$. JADES-GS-z12-0 joins a list of other high-redshift galaxies have been observed with evidence for a DLA, including three galaxies at $z = 9 - 11$ from \citet{heintz2023}. This DLA absorption can lead to an overprediction of the spectroscopic redshift of $\Delta z \sim 0.10 - 0.15$, a bias that can negatively impact the search for emission lines in these sources. Indeed, the spectroscopic redshift derived from the Ly$\alpha$ break by \citet{curtislake2023}, $z_{\mathrm{spec}} = 12.63$, is significantly higher than what \citet{deugenio2023} estimate from the emission line detection. This bias has a larger effect on photometric redshifts, which are used for finding these sources, deriving luminosity functions, and understanding the evolution of the cosmic star-formation rate density. Multiple authors have found that photometric redshifts derived for samples of high-redshift galaxies are systematically shifted to larger $z$ by $\sim 0.2$--$0.3$ \citep{arrabalharo2023b, hainline2023, fujimoto2023, finkelstein2023, willott2023}. As DLA absorption is not accounted for in most popular photometric redshift codes, this would naturally explain why the resulting spectroscopic redshifts are lower than the predicted photometric redshifts, a conclusion supported by the work of \citet{deugenio2023}.

In this paper, we explore two $z > 10$ sources from \citet{robertson2023} and \citet{curtislake2023}, JADES-GS-53.16476-27.77463 (hereafter, JADES-GS-z11-0) and JADES-GS-53.14988-27.7765 (hereafter, JADES-GS-z13-0). JADES-GS-z11-0 was originally discovered in deep Hubble imaging by \citet{bouwens2011} and then further discussed in \citet{ellis2013} and \citet{koekemoer2013}. JADES-GS-z13-0, which lies at a redshift where it was not visible to Hubble, is the highest-redshift spectroscopically confirmed galaxy found in the first year of JWST observations \citep{robertson2023, curtislake2023}. Deeper spectroscopy from NIRSpec taken as part of observations of the JADES Origins Field \citep{eisenstein2023b} allows us to explore the UV properties of these distant galaxies, where we can re-evaluate their redshifts, UV slopes and magnitudes, and the inferred stellar masses, star-formation rates, and metallicities. In JADES-GS-z11-0, we find evidence for multiple weak emission lines, which allows us to refine the spectroscopic redshift estimate for this source. For JADES-GS-z13-0, even with a spectrum with five times the observing time, we do not find evidence for any significant UV emission lines.  

We present the new observations of JADES-GS-z11-0 and JADES-GS-z13-0 in Section \ref{sec:observations} along with the data reduction and spectral extraction approaches. In Section \ref{sec:fits}, we describe the details of the multiple fitting procedures we applied to the observed NIRSpec spectra, and in Section \ref{sec:nircam-photometry}, we introduce updated NIRCam photometry for the sources. In Section \ref{sec:results}, we present the results of these fits: the weak emission lines observed and detected in the JADES-GS-z11-0 spectrum, the possible causes for the lack of emission lines in JADES-GS-z13-0, the potential existence of damped Lyman-$\alpha$ absorption in these sources, and how this affects their photometric redshifts. We discuss these results and conclude in Section \ref{sec:discussion}. Throughout, we assume the \citet{planck2020} cosmology, with $H_0 = 67.4$ km s$^{-1}$ Mpc$^{-1}$, $\Omega_{\mathrm{M}} = 0.315$ and $\Omega_\Lambda = 0.685$. All magnitudes are provided using the AB magnitude system \citep{oke1974, oke1983}.

\section{Observations and Data Reduction} \label{sec:observations}

The NIRSpec spectra that form the basis of this paper were taken as part of two programs: PID 1210 (PI N. L\"utzgendorf) and PID 3215 (PIs D.~Eisenstein and R.~Maiolino). The spectra for PID 1210 are part of JADES, and were presented in \citet{curtislake2023} and \citet{bunker2023b}. The spectra for PID 3215 are part of the JADES Origin Field, as outlined in \citet{eisenstein2023b}. For both JADES-GS-z11-0 and JADES-GS-z13-0, the primary spectra described in this study were observed with the NIRSpec Multi-Shutter Array (MSA) using the PRISM/CLEAR disperser-filter combination. The wavelength range covered by these spectra is 0.6 - 5.3 $\mu$m at a resolution of R $\sim$ 100 \citep{jakobsen2022}. For the PID 1210 data, JADES-GS-z11-0 was observed for a total observing time of 100.8 ksec and JADES-GS-z13-0 was observed for 33.6 ksec \citep{curtislake2023}. For PID 3215, JADES-GS-z11-0 and JADES-GS-z13-0 were both observed for a total observing time of 168.1 ksec ($\sim$ 47 hr) each, and for the present analysis, we sum the spectra from both programs for a total observing time of 268.9 ksec ($\sim$ 75 hr) for JADES-GS-z11-0, and 201.7 ksec ($\sim$ 56 hr) for JADES-GS-z13-0.

We supplement the NIRSpec PRISM/CLEAR observations of these sources with NIRSpec medium-resolution spectra (R $\sim$ 1000) taken in PID 3215 using the G140M/F070LP and G395M/F290LP disperser-filter combinations. For G140M/F070LP, the integration times were 42.1 ksec for both galaxies, while for G395M/F290LP, the integration times were 134.5 ksec for both galaxies. Short-circuits in the NIRSpec MSA \citep{rawle2022} affected one of the five visits, resulting in less integration time on each source than what was requested, 168.1 ksec.

We follow the same data reduction as described in \citet{bunker2023b} and \citet{carniani2023}, reducing the PRISM data from both PID 1210 and 3215 using the pipeline developed by the ESA NIRSpec Science Operations Team (SOT) and Guaranteed Time Observations (GTO) NIRSpec teams, as described in \citet{deugenio2024}. We performed background subtraction using nodding along the 3-slitlet array, and we extracted fluxes using a 3-pixel window. We correct for slit losses by modeling each galaxy as a point-source, and account for the relative intra-shutter location at each nodding position and for each different pointing (and different MSA configuration). We will discuss the updated size properties for JADES-GS-z11-0 and JADES-GS-z13-0 further in Section \ref{sec:nircam-photometry}. To calculate the line-spread function (LSF) for the NIRSpec observations, we followed the method outlined in \citet{degraaff2023}. 

The PRISM data taken in PID 1210 and PID 3215 both employed 1400~s duration (19 frame) PRISM sub-exposures taken in NRSIRS2 readout mode \citep{rauscher2012}. A total of 186 sub-exposures of JADES-GS-z11-0 and 138 sub-exposures JADES-GS-z13-0 were taken between the two programs. These sub-exposures were each reduced separately, and the resulting one-dimensional sub-spectra were combined using a customized algorithm to produce the final spectra. The official reduction pipeline is known to occasionally leave sharp spikes in the extracted spectra due to residual signals from noisy pixels and/or cosmic ray hits that are not properly captured in the ramp fitting. Such spikes are eliminated by performing iterative sigma clipping on the reduced sub-spectra on a wavelength bin by wavelength bin basis prior to their being co-added. However, a closer examination of the large number of sub-spectra available for these sources prompted us to refine the standard approach somewhat. Two additional censoring steps were introduced prior to the sigma-clipping, which served to eliminate obviously spurious flux values that deviated from the median measured flux in each bin by more than five times the median pipeline error estimate for the bin. Similarly, sub-spectra bins whose pipeline error estimate exceeded five times the median error estimate for the bin were eliminated. This was then followed by five passes of iterative sigma-clipping that eliminated any flux values that deviated by more than three times from the sample variance of the surviving entries in the bin. A second change introduced was that instead of co-adding the surviving sub-spectra entries through weighting with the inverse of the square of the pipeline error estimate, a statistically more robust straight averaging of the surviving entries in each wavelength bin was performed. In the same vein, the final propagated pipeline error for the co-added bin was calculated as the rms mean of the pipeline errors of the surviving entries divided by the square root of their number. Altogether, this process eliminated 7.4\% of the wavelength bin entries JADES-GS-z11-0 and 12.3\% of the entries in JADES-GS-z13-0. The resulting final spectra we explore in this work do not differ dramatically from their standard versions, but are clearly devoid of anomalous noise spikes exceeding the actual statistical noise present in the data. 

\begin{figure*}
  \centering
  \includegraphics[width=0.9\linewidth]{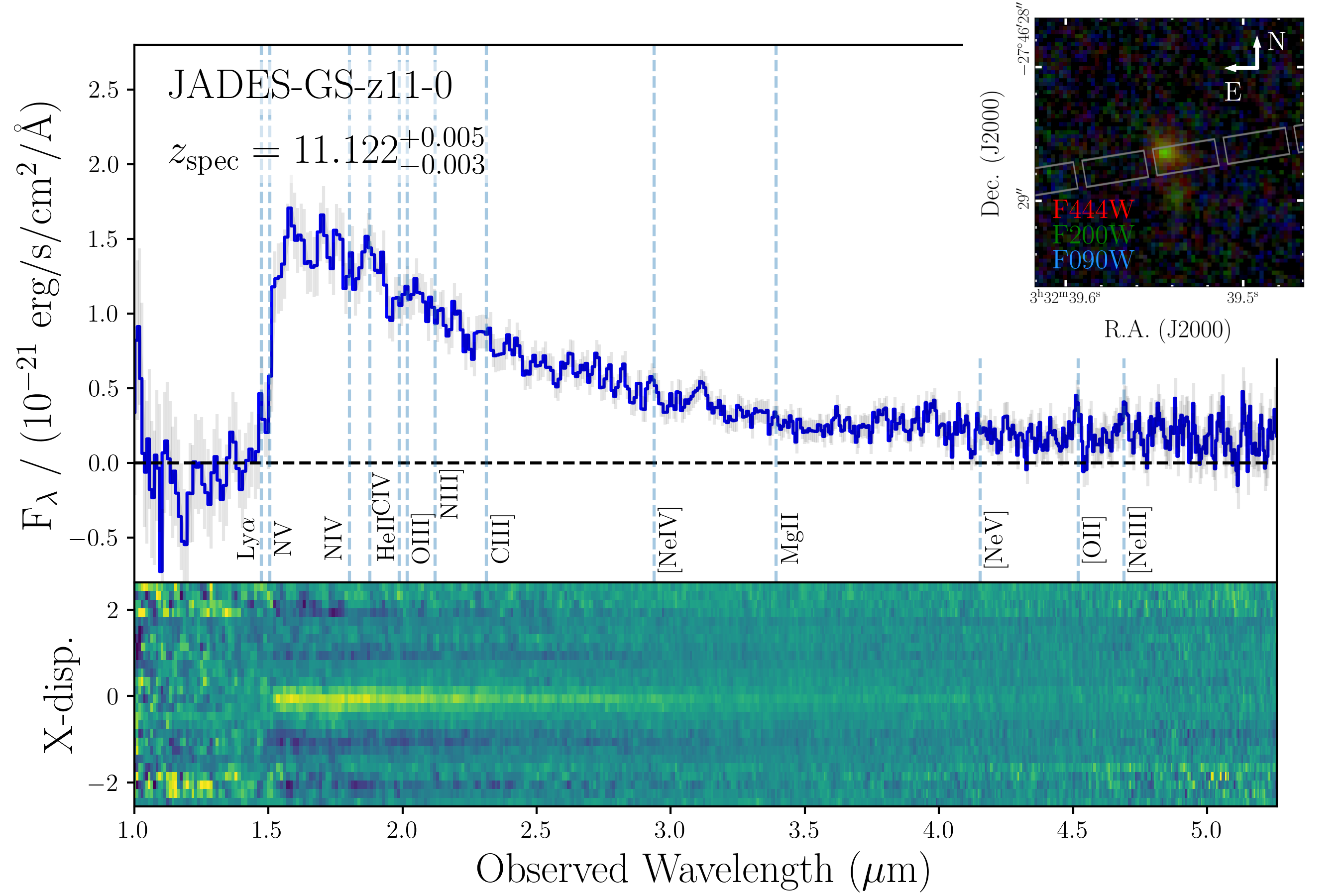}
  \includegraphics[width=0.9\linewidth]{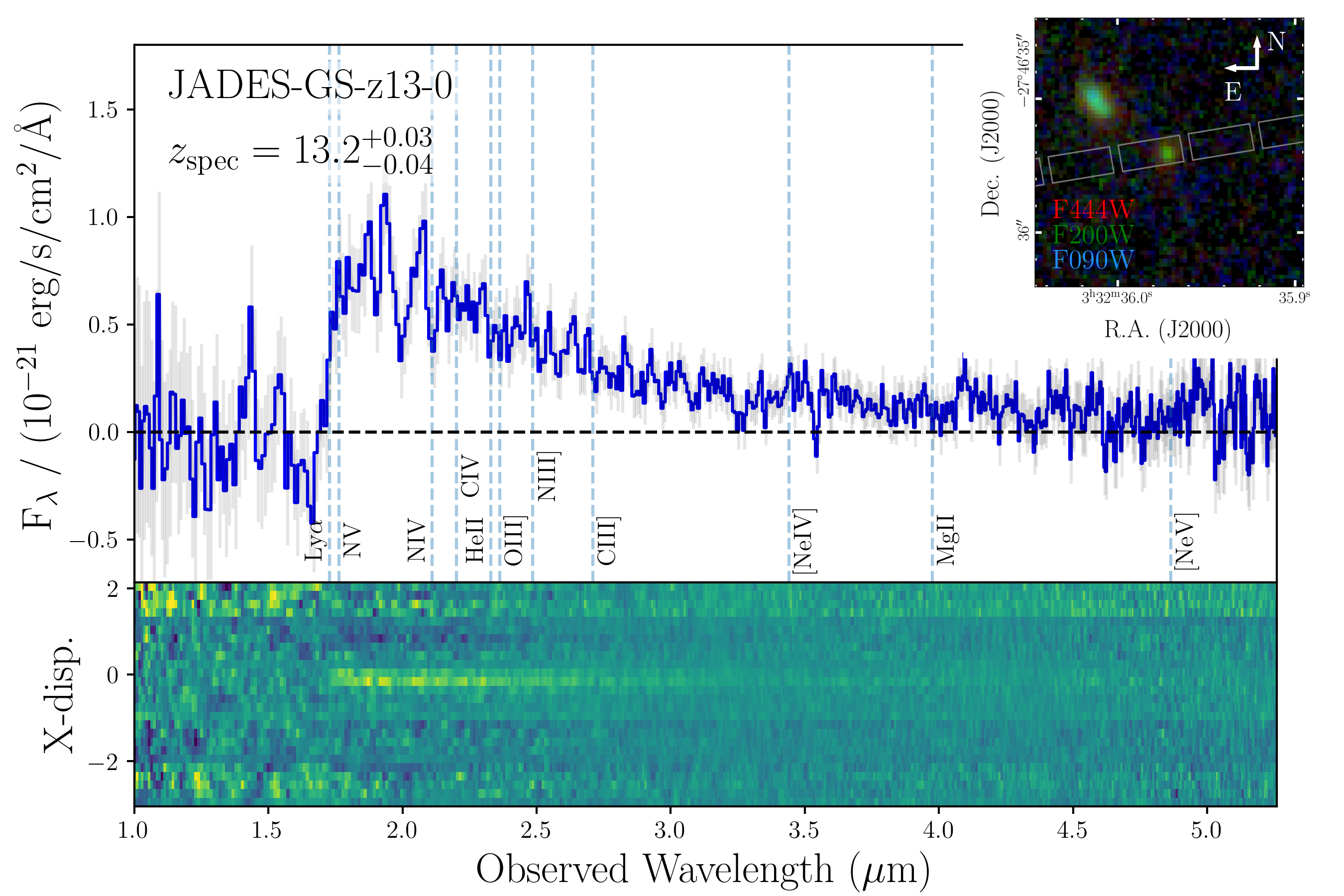}  \caption{2D and 1D NIRSpec PRISM/CLEAR spectra for JADES-GS-z11-0 (top) and JADES-GS-z13-0 (bottom), from the combined spectra from observations under PID 1210 as described in \citet{curtislake2023} and the new, deeper observations under PID 3215. In each panel, we plot the 2D spectrum underneath the sigma-clipped 1D spectrum. For the 1D spectrum, plotted in blue, we also plot uncertainties in light grey. We plot the positions of UV and optical emission lines with dashed lines at the fiducial redshifts estimated for each source. In an insert, we show the $2^{\prime\prime} \times 2^{\prime\prime}$ JADES F444W+F200W+F090W RGB cutout with the MSA slitlets used for PID 3215 overplotted.}
  \label{fig:2d_1d_plots}
\end{figure*}

In Figure \ref{fig:2d_1d_plots} we show the PRISM spectra for both JADES-GS-z11-0 and JADES-GS-z13-0. In the bottom portion of each figure, we plot the 2D NIRSpec PRISM spectrum, with the y-axis depicting the spatial offset along the slitlet. We plot the 1D spectrum in the upper panels along with the $1\sigma$ uncertainties. We also provide our fiducial redshifts we derived for each source (we describe how these are derived in the next section), and we show the wavelengths of prominent ultraviolet and optical emission and absorption features with vertical dashed lines. In an inset for each source, we plot a $2^{\prime\prime} \times 2^{\prime\prime}$ RGB image centered on the source created by combining the JADES NIRCam F444W, F200W, and F090W images, where north is up and east is to the left. In each inset, we show the MSA slitlets used in the PID 3215 observations.

As can be seen from Figure \ref{fig:2d_1d_plots}, the Lyman-$\alpha$ break is very strong in the 2D spectra for both sources, with no significant emission to the blue of the break, a sharp transition, and then smoothly decreasing flux to redder wavelengths. For JADES-GS-z11-0, we see evidence in our deeper spectrum for multiple emission features, including [\ion{O}{2}]$\lambda\lambda3726,3729$\AA, [\ion{Ne}{3}]$\lambda3869$\AA, and possibly \ion{C}{4}$\lambda\lambda1548,1551$\AA. For JADES-GS-z13-0, we do not see any significant emission lines at the fiducial redshift for the source, similar to the results from \citet{curtislake2023}, a topic we explore in Section \ref{sec:gs-z13-no-emission}. 

\section{Spectral Fits and Methodology} \label{sec:fits}

\subsection{Estimating Redshifts} 

We fit the spectra for JADES-GS-z11-0 and JADES-GS-z13-0 with multiple codes and statistical methods to explore the source redshifts and stellar populations. In \citet{curtislake2023}, the JADES-GS-z13-0 spectrum was at a low enough signal-to-noise that the authors fit the spectrum and NIRCam photometry together, so this current work represents the first fit to the spectrum alone, allowing an independent check of the properties as compared to fits to the NIRCam photometry.  

To determine the redshift for each source, we searched each PRISM spectrum for the presence of nebular emission features. To help accomplish this, we developed a novel automated approach designed to ascertain the significance of UV and optical emission features in NIRSpec prism spectra at a given redshift. This method is described in more detail in Appendix \ref{sec:appendix_redshift}. Briefly, we start with the combined, sigma-clipped spectra (itself generated from a number of independent 1400s ``sub-spectra'') for each source, and apply a moving boxcar smoothing to each spectrum to estimate the continuum, which is then subtracted. From this continuum-subtracted spectrum, we create a line flux signal-to-noise ratio array by means of statistical bootstrapping among the sub-exposures making up each spectrum, which allows us to explore the potential significance of emission features found in the spectrum. We plot the signal-to-noise ratio vs wavelength for JADES-GS-z11-0 and JADES-GS-z13-0 in Figures \ref{fig:gsz11_sn} and \ref{fig:gsz13_sn} in the Appendix respectively. We use an ``emission line comb'' to search whether there are redshifts where a significant match of lines is found. The lines used in this search are provided in Table \ref{tab:lines} in the Appendix. We calculate a total probability by combining the individual probabilities for each potential emission line, and we use these total probabilities to find possible values for the systemic redshift for each galaxy.

For JADES-GS-z11-0, this method results in a redshift of $z_{\mathrm{spec}} = 11.122^{+0.005}_{-0.003}$, a value we state with $94$\% confidence (see Appendix \ref{sec:appendix_redshift} for more details). We show the probability vs redshift plot for this galaxy in Figure \ref{fig:gsz11_peak} and describe this redshift and the resulting lines in Section \ref{sec:gs-z11-emission}. For JADES-GS-z13-0, however, the best-fitting redshift resulting from this method, $z_{\mathrm{spec}} = 12.922^{+0.009}_{-0.010}$, is far less likely, and is primarily driven by a potential detection of \ion{N}{4}] emission. We estimate that this solution has $56$\% confidence, and we reject it in favor of the fit to the Lyman-$\alpha$ break for this source. 

\begin{deluxetable}{l c c}
\tabletypesize{\footnotesize}
\tablecolumns{3}
\tablewidth{0pt}
\tablecaption{Best-Fit Spectral Parameters\label{tab:spectrum-fit-parameters}}
\tablehead{
\colhead{Parameter} &  \colhead{JADES-GS-z11-0}  &  \colhead{JADES-GS-z13-0} }
\startdata
  R.A. (degrees) & 53.16476  & 53.14988 \\ 
  DEC (degrees) & -27.77463 & -27.77650 \\
  $z_{\mathrm{spec}}$ & $11.122^{+0.005}_{-0.003}$ & $13.2^{+0.03}_{-0.04}$ \\
  M$_{\mathrm{UV}}$ (spectrum) & -19.32 $\pm$ 0.03 & -18.92 $\pm$ 0.05 \\
  $\beta$ (spectrum) & -2.18 $\pm$ 0.05 & -2.69 $\pm$ 0.10 \\
\hline
  \multicolumn{3}{c}{{\tt BEAGLE}} \\
\hline
   $\log(\Mstar/\Msun)$ & $8.3_{-0.1}^{+0.1}$ & $7.7_{-0.20}^{+0.40}$ \\
   $\log(\SFR / \Msun \yrInv)$ & $0.16_{-0.03}^{+0.03}$ & $0.15_{-0.09}^{+0.16}$ \\
   $\log(\age/\txn{yr})$ & $8.2_{-0.1}^{+0.1}$ & $7.6_{-0.5}^{+0.4}$ \\
   $\log(\Zmetal/\Zsun)$ & $-1.9_{-0.1}^{+0.1}$ & $-1.9_{-0.2}^{+0.3}$ \\
   $\logUs$ & $-2.7_{-0.3}^{+0.2}$ & $-2.9_{-0.7}^{+0.9}$ \\
   $\tauV$ & $0.04_{-0.02}^{+0.02}$ & $0.02_{-0.01}^{+0.03}$ \\
   $\fesc$ & - & $0.91_{-0.1}^{+0.07}$ \\
   $\beta$ & $-2.40_{-0.05}^{+0.05}$ & $-2.76_{-0.07}^{+0.08}$ \\
\hline
\multicolumn{3}{c}{{\tt Prospector}} \\
\hline
   $\log(\Mstar/\Msun)$ & $8.43^{+0.06}_{-0.08}$ & $7.85^{+0.13}_{-0.17}$ \\
   $\log(\SFR / \Msun \yrInv)$ & $0.00^{+0.13}_{-0.10}$ & $0.11^{+0.04}_{-0.03}$ \\
   log(Z$_{\mathrm{stars}}$/Z$_{\odot}$) & $-1.87^{+0.22}_{-0.09}$ & $-1.84^{+0.15}_{-0.12}$ \\
   log(Z$_{\mathrm{gas}}$/Z$_{\odot}$) & $-0.91^{+0.05}_{-0.06}$ & $0.35^{+0.10}_{-0.08}$ \\
   $\logUs$ & $-2.21^{+0.26}_{-0.30}$ & $-1.43^{+0.31}_{-0.32}$ \\
   E(B-V) & $0.010^{+0.001}_{-0.001}$ & $0.004^{+0.002}_{-0.002}$ \\
   $\fesc$ & 0.00 & 0.00 \\
\hline
\multicolumn{3}{c}{Emission Line Fluxes and EWs} \\
\hline
   Flux CIV & $6.2 \pm 4.4$ & $<$ 8.3 \\
   Flux HeII & $< 7.9$ & $<$ 7.6 \\
   Flux OIII] & $< 7.5$ & $<$ 7.2 \\
   Flux NIII]& $< 6.6$ & $<$ 6.3 \\
   Flux CIII] & $< 5.9$ & $<$ 5.4 \\
   Flux [OII] & $4.6 \pm 1.5$ & $<$ 3.9 \\
   Flux [NeIII]  & $3.4 \pm 1.5$ & - \\
 \hline
   EW CIV & $3.9 \pm 2.8$ & $<$ 10.0 \\
   EW HeII & $< 5.7$ & $<$ 10.6 \\
   EW OIII] & $< 5.6$ &$<$ 10.2 \\
   EW NIII] & $< 5.2$ & $<$ 10.1 \\
   EW CIII] & $< 6.0$ & $<$ 11.6 \\
   EW [OII] & $20.8 \pm 6.6$ & - \\
   EW [NeIII] & $14.0 \pm 6.3$ & - \\
\enddata
\tablecomments{Fluxes are provided in units of 10$^{-20}$ erg s$^{-1}$ cm$^{-2}$, while the equivalent width values are units of rest-frame Angstroms. }
\end{deluxetable}

At these systemic redshifts, we estimate emission line fluxes and equivalent widths from the continuum-subtracted spectra using a 5-wavelength bin window. To estimate uncertainties, we repeat the entire process of combining the sub-spectra, estimating and subtracting the continuum, and measuring the line fluxes bootstrapped 2000 times, each time creating a combined spectrum drawn at random from the available sub-spectra. Our estimate of the uncertainties on the fluxes and equivalent widths is calculated from the sample variance derived from this procedure. This is notably different from the method for estimating line fluxes and equivalent widths used in \citet{curtislake2023} and \citet{deugenio2023}, which estimates uncertainties using the NIRSpec reduction pipeline uncertainties and 3-wavelength bin window. The resulting bootstrap errors we estimate agree with those calculated using a covariance matrix measured from the individual sub-spectra for each source. 

\subsection{JADES-GS-z11-0} \label{sec:gs-z11-emission}

With the deeper spectrum for JADES-GS-z11-0, we find evidence for multiple lines in emission in the PRISM spectrum not seen in \citet{curtislake2023}. We detect the [\ion{O}{2}]$\lambda\lambda3726,3729$\AA with flux signal-to-noise ratio (SNR) = 3.1, and observe the [\ion{Ne}{3}]$\lambda3869$\AA with flux SNR = 2.2. In addition, we have tentative evidence for \ion{C}{4}$\lambda\lambda1548,1551$\AA with a flux SNR = 1.4. We list the derived line fluxes and equivalent widths (EWs), and include 2$\sigma$ upper limits for non-detected features, in Table \ref{tab:spectrum-fit-parameters}.

\begin{figure*}
  \centering
  \includegraphics[width=1.0\linewidth]{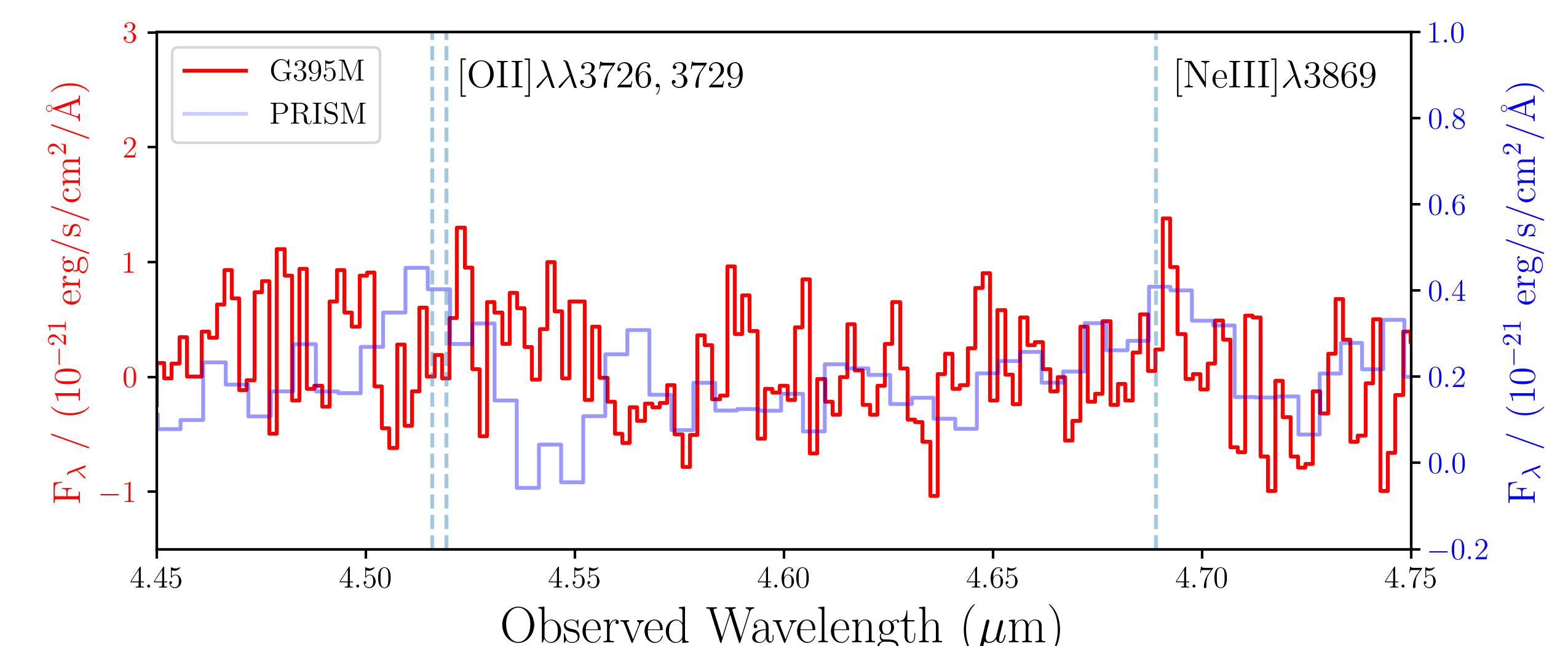}
  \caption{The JADES-GS-z11-0 PRISM spectrum (blue) plotted against the NIRSpec G395M higher-resolution grating spectrum (red). Because of the difference in resolution between the two diffraction modes, we scale the y axes differently for each spectrum, as shown on the left and right sides of the plot. At low significance, the [\ion{O}{2}] and [\ion{Ne}{3}] lines are visible in both spectra.}
  \label{fig:GS_z11_spectrum_z11p13}
\end{figure*}

To further explore the presence of these emission lines, we looked at the higher resolution ($R \sim 1000$) NIRSpec G395M grating spectrum for this source, focusing on the 4.4 - 4.8 $\mu$m region of the observed spectra, which we plot in Figure \ref{fig:GS_z11_spectrum_z11p13}. We see a similar pair of potential emission features at 4.52 $\mu$m (flux SNR = 2.2) and 4.69$\mu$m (flux SNR = 3.11), which correspond to the [\ion{O}{2}] and [\ion{Ne}{3}] lines in the PRISM spectrum. We fit these features and find that the fluxes measured from the grating spectra agree with those measured from the PRISM spectrum within the uncertainties, with similarly low flux SNR = 2 - 3, although we do not see evidence for [\ion{O}{2}]$\lambda$3726 in the grating spectrum. We measure a line width (intrinsic) of $161.07 \pm 70.4$ km/s from this fit.

While we believe that the emission features at 4.52 $\mu$m and 4.69$\mu$m are real, the lack of an observed [\ion{O}{2}]$\lambda$3726 emission line seen in the grating spectrum is curious. The observed [\ion{O}{2}]$\lambda$3729 / [\ion{O}{2}]$\lambda$3726 flux ratio is unphysically high, and implies a very low electron density ($n_e \sim 1-10$ cm$^{-3}$). The large velocity dispersion we measure from the fits to the [\ion{O}{2}] lines arises due to the need to constrain the [\ion{O}{2}]$\lambda$3729 / [\ion{O}{2}]$\lambda$3726 flux ratio to within the physical range. If we allow the [\ion{O}{2}]$\lambda$3726 flux to go to zero for the fit, the measured line width is instead $\sim 98$ km/s. In addition, we observe a positive velocity shift between the observed wavelengths for the putative [\ion{O}{2}] and [\ion{Ne}{3}] features in the G395M grating and PRISM spectra, which is likely a result of the wavelength calibration, and has been discussed for the JADES spectroscopic releases \citep[][D'Eugenio et al. in prep]{bunker2023b}.

To explore the significance of these features, in addition to the G395M grating spectrum reduction described in Section \ref{sec:observations}, we performed a similar sigma clipping and boostrap reduction of the grating spectrum as was done on the PRISM spectrum. The resulting spectrum is consistent with what we present in Figure \ref{fig:GS_z11_spectrum_z11p13}, and we observe both the [\ion{O}{2}] and [\ion{Ne}{3}] emission features. Summing over a three-bin wide box, we observe [\ion{O}{2}] in this spectrum with flux SNR = 2.4 ($p = 0.0124$), and [\ion{Ne}{3}] in this spectrum with flux SNR = 2.79 ($p = 0.00261$), with a Fisher's combined probability $p = 0.000367$.  

\subsection{JADES-GS-z13-0} \label{sec:gs-z13}

For JADES-GS-z13-0, the spectrum shown in the bottom panel of Figure \ref{fig:2d_1d_plots} has a spectral break at $\sim$1.8$\mu$m, with no evidence for flux blueward of this feature, but does not show any significant emission or absorption features. Our fiducial redshift, $z_{\mathrm{spec}} = 13.2^{+0.03}_{-0.04}$, comes from a fit to the spectrum as described in the next section. We calculate 2$\sigma$ upper limits on the line fluxes and equivalent widths at this redshift, and provide these in Table \ref{tab:spectrum-fit-parameters}. These equivalent width values are in agreement with those presented in \citet{curtislake2023}. 

\subsection{SED Fitting} \label{sec:sed-fitting}

We performed SED fitting to the spectra to derive key physical properties of the objects. Even with the exquisite spectroscopic data from NIRSpec, currently we find that systematic uncertainties dominate the inference of galaxy properties like stellar masses and star formation rates, with the star formation history being a dominant source of this uncertainty. We therefore choose to present SED fitting from two different codes with very different prescriptions for the star-formation history to illustrate the magnitude of systematic uncertainties beyond the quoted statistical uncertainties for each code.

The first code that we use to fit the spectra of these sources is the Bayesian galaxy spectral modeling tool {\tt BEAGLE} \citep[BayEsian Analysis of GaLaxy sEds][]{chevallard2016}. For the fits to the spectra, we follow a similar methodology to that adopted in \citet{curtislake2023}. We fit each source using three different models, in which we vary assumptions about the star formation history and escape fraction of ionizing photons. The motivation is that a major challenge in interpreting the spectra of the four $z > 10$ galaxies presented in \citet{curtislake2023} was the absence of detectable emission lines. The new observations presented in this work are significantly deeper than those presented in \citet{curtislake2023}, and yet we only observe and detect tentative emission lines in JADES-GS-z11-0. Explaining the absence of lines in JADES-GS-z13-0 hence requires us to test different model hypotheses.

To estimate the redshift of JADES-GS-z13-0, we fit the spectrum using {\tt BEAGLE} and focus on the observed spectral break at 0.8 -- 1.8 $\mu$m. In this fit, we assume a constant star-formation history, fix the IGM neutral hydrogen fraction ($\hat{x}_{\mathrm{HI}}$) to 0, and let the escape fraction of ionizing photons vary. The resulting redshift, $z_{\mathrm{spec}} = 13.2^{+0.03}_{-0.04}$, agrees with what derived from the Lyman-$\alpha$ break presented in \citet{curtislake2023}, and we adopt this value as the fiducial for this source. 

In all our subsequent modeling, then, we adopt Gaussian priors on the redshift of the sources centered on the spectroscopic redshifts $z_{\mathrm{spec}} = 11.122^{+0.01}_{-0.01}$ for JADES-GS-z11-0, and $z_{\mathrm{spec}} = 13.2^{+0.03}_{-0.03}$ for JADES-GS-z13-0, and with the width of the Gaussian set to the quoted errors. 

We perform a careful fit to each spectrum, pixel-by-pixel, masking the region 1150 -- 1450 \AA\ to prevent biases arising from a potential DLA in this source, and we also include constraints on the measured EWs (including upper limits). We use an updated version of the \citet{Bruzual2003} stellar population synthesis models \citep[see][for details]{Vidal2017}, combined with the (continuum + emission lines) photoionization models of \citet{Gutkin2016}. We assume a \citet{chabrier2003} initial mass function (IMF) with lower and upper mass limits of 0.1 and $300 M_{\odot}$, respectively. The model takes into account the depletion of metals onto dust grains in the photoionized regions of stellar birth clouds, where we fix the dust-to-metal mass ratio to 0.1. We adopt the \citet{charlotfall2000} model for dust attenuation, with the fraction of the attenuation from the diffuse interstellar medium (ISM) fixed at 0.4. The ionization parameter is free to vary, while the interstellar gas-phase metallicity is set to be equal to the stellar metallicity. 

For JADES-GS-z11-0, our fiducial model is based on a constant star formation history, and is defined by six adjustable parameters: the total stellar mass formed \Mtot, age of the oldest stars \age, stellar metallicity \Zstar, gas ionization parameter \logUs, V-band dust attenuation optical depth \tauV, and redshift $z$. Below, we discuss the stellar mass locked into stars \Mstar, which is always lower than the total stellar mass formed \Mtot, since it excludes the mass returned to the ISM by stellar winds and supernovae (SNe) explosions, as well as the mass locked into stellar remnants. Also, we refer to the metallicity \Zmetal, which corresponds to the stellar metallicity \Zstar\ and to the interstellar metallicity \Zism, while the gas abundance of a metal further depends on its dust depletion factor. The star formation rate \SFR\ is computed as the SFR averaged over the last 10 Myr of star formation (although, for a constant SFH, the rate will not change in this time). 

For JADES-GS-z13-0, our fiducial model is the same as for JADES-GS-z11-0, but with the addition of the parameter defining the escape fraction of ionizing photons \fesc. The justification for adopting these models is provided in Section \ref{sec:gs-z13-no-emission} below, where we also discuss the alternative models explored, including evidence for a possible recent cessation of star formation and the tension between the blue UV slope for this source and the lack of observed UV emission lines. 

\begin{figure*}
  \centering
  \includegraphics[width=0.95\linewidth]{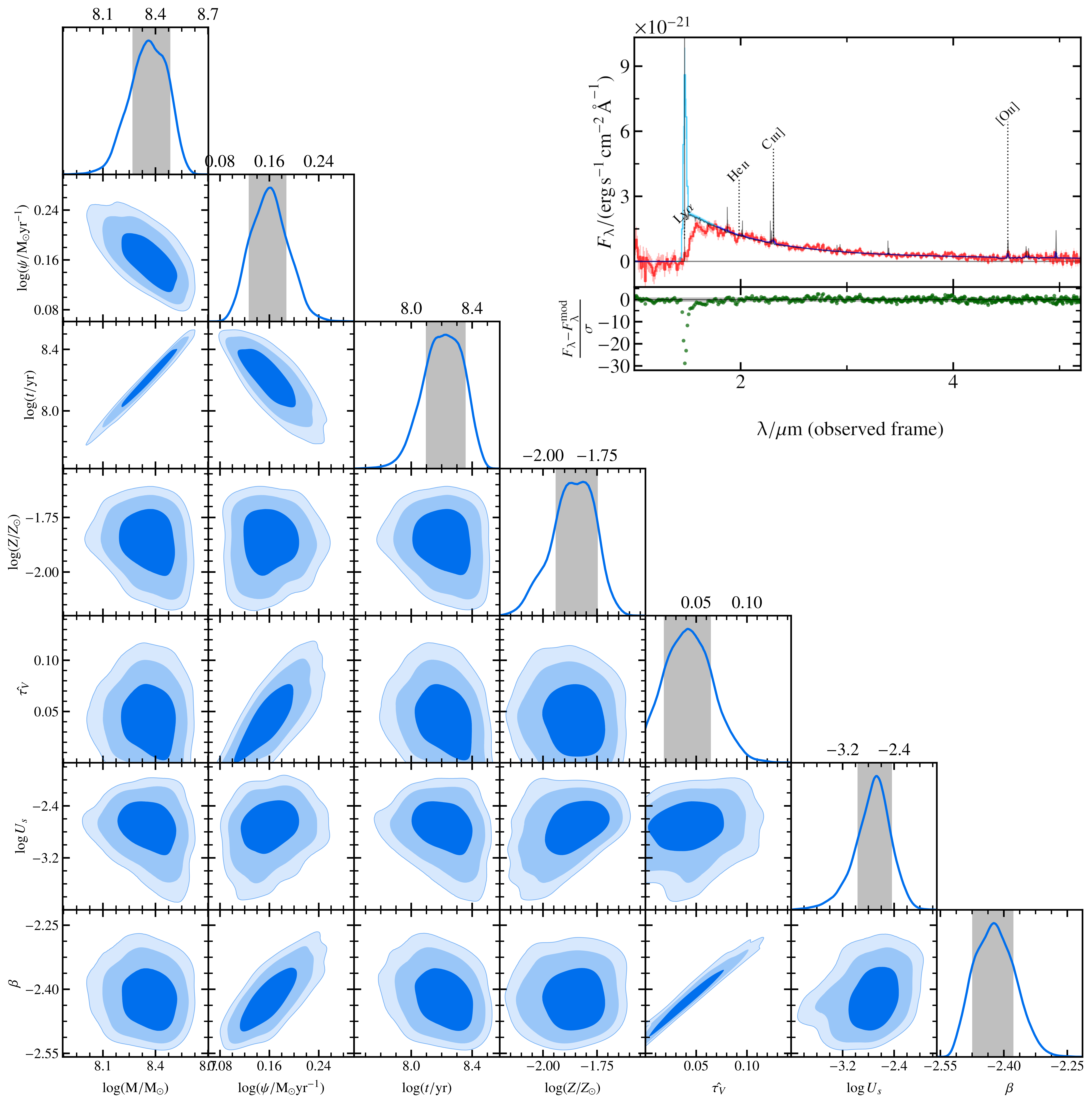}
\caption{Posterior probability distributions obtained with \texttt{BEAGLE} for our fiducial fits, along with the observed spectrum and model prediction, for JADES-GS-z11-0. From left to right, the columns show the stellar mass \Mstar, star formation rate, age of the oldest stars \age, (stellar and interstellar) metallicity \Zmetal, V-band attenuation optical depth \tauV, ionization parameter \logUs, and UV slope $\beta$. The 1D (marginal) posterior distribution of each parameter is plotted along the diagonal, where the shaded gray regions represent the 1$\sigma$ credible interval. The off-diagonal panels show the 2D (joint) posterior distributions, with the shaded blue regions representing the 1, 2, and 3$\sigma$ credible intervals. In the top panel of the inset, we show the observed spectrum (red line), along with the model predictions (dark blue line). The model predictions at $\lambda < 1450$ \AA\ are shown with a cyan line, to indicate that this region was masked during the fitting. In the bottom panel of the inset, we show the residuals in units of observed errors and the $\pm 1 \sigma$ region in grey.}
  \label{fig:beagle_triangle_gsz11}
\end{figure*}

\begin{figure*}
  \centering
  \includegraphics[width=0.95\linewidth]{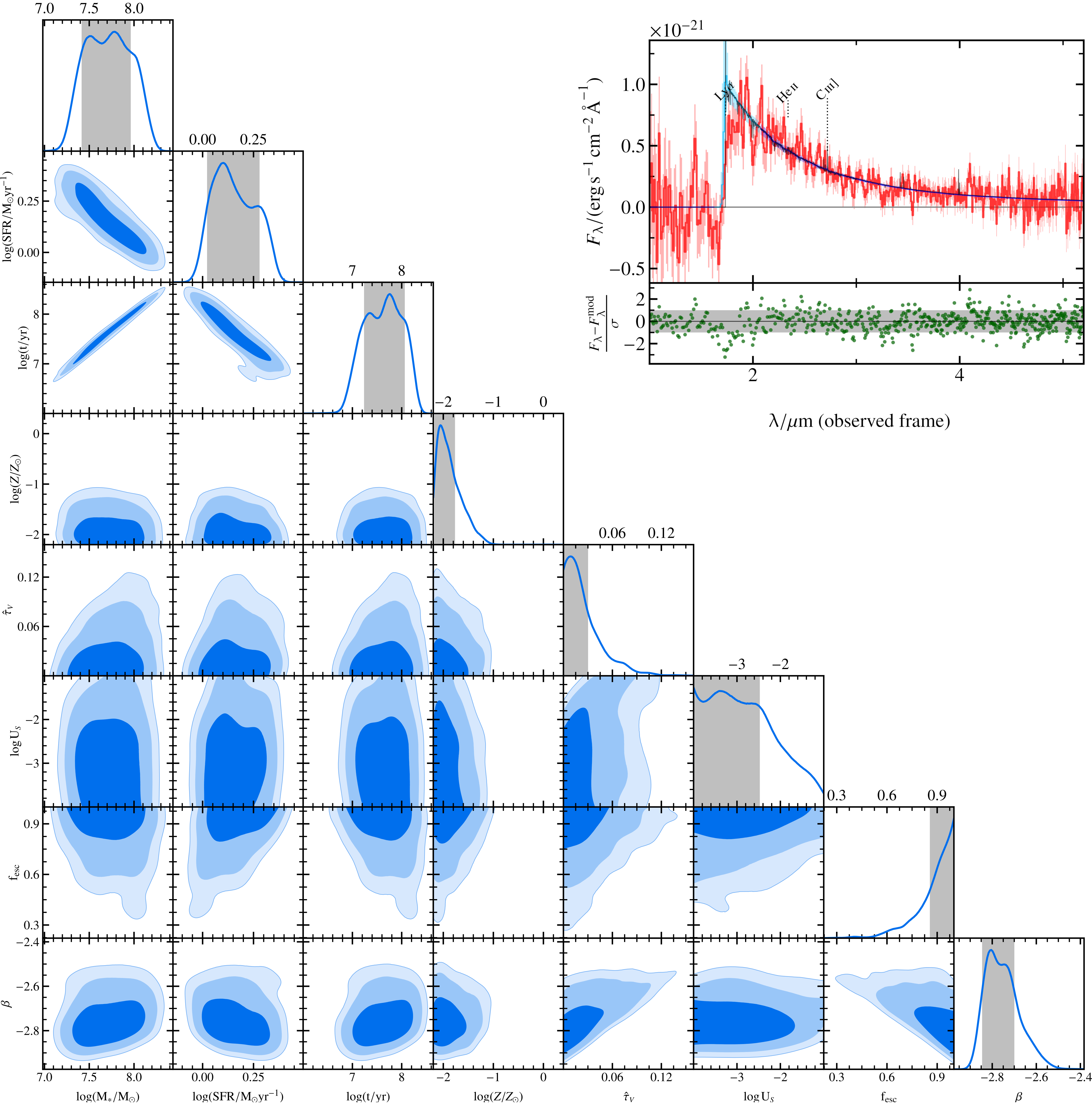}  \caption{Same as in Figure~\ref{fig:beagle_triangle_gsz11}, but for JADES-GS-z13-0 and also including the escape fraction \fesc.}
  \label{fig:beagle_triangle_gsz13}
\end{figure*}

For our fiducial models, we plot in Figures~\ref{fig:beagle_triangle_gsz11} and \ref{fig:beagle_triangle_gsz13} the \texttt{BEAGLE} predictions and posterior probability distributions. We summarize in Table~\ref{tab:spectrum-fit-parameters} the \texttt{BEAGLE} output parameters from these fits. In Table \ref{tab:spectrum-fit-parameters}, we additionally provide observational estimates of both the UV slope $\beta$ and M$_{\mathrm{UV}}$ (which we provide at the top and where we use the word ``spectrum'' to differentiate from the value of $\beta$ from \texttt{BEAGLE}), calculated directly from each spectrum given our fiducial redshifts. To compute $\beta$, we fit the observed flux density of each source over spectral windows defined by \citet{Calzetti1994} in the region 1500 to 3300 \AA, and we use the 1$\sigma$ flux uncertainties to estimate the errors on the derived slope. This wavelength range was chosen so that any additional Lyman-$\alpha$ damping would not affect the calculation of the UV slope. To estimate M$_{\mathrm{UV}}$, we calculate the absolute magnitude for each source through a simulated box hat filter covering the wavelengths 1400 to 1600 \AA. 

In order to explore the range of estimated galaxy parameters for these sources, we also fit the observed spectra with the Bayesian population synthesis code \texttt{Prospector} \citep{johnson2021}. For these fits, we adopt the MIST \citep{choi2016} isochrones and MILES/BaSeL stellar library \citep{lejeune1997, lejeune1998, westera2002, sanchezblazquez2006} as implemented in the Flexible Stellar Population Synthesis (FSPS) package \citep{conroy2010}.  We mask the Lyman-$\alpha$ break region in the same wavelength range as for the BEAGLE fits, 1150 - 1450 \AA. We assume a \citet{kroupa2001} IMF, which results in stellar masses that are larger on average by 6\% from those measured using a \citet{chabrier2003} IMF \citep{speagle2014}. For dust obscuration, we use the \citet{charlotfall2000} dust prescription, where the dust obscuring the nebular emission and stars younger than 10 Myr is modeled using a power-law attenuation, and the additional dust obscuring the older stars is modeled with a modified \citet{calzetti2000} law from \citet{kriekconroy2013}. We assume the \citet{madau1995} model to account for IGM absorption. We allow the stellar- and gas-phase metallicity to be independent and free parameters in the fit, and assume that the escape fraction of ionizing photons $f_{\mathrm{esc}} = 0$. For JADES-GS-z11-0 we restrict redshift to $z = 11.122$, and for JADES-GS-z13-0, we restrict redshift to $z = 13.2$. For our star-formation history, we assume a non-parametric model with 6 bins in lookback time and the \texttt{Prospector} ``continuity prior.'' This parameterization of the SFH is split into multiple bins, with the SFR in each bin being derived from the ratios of those in adjacent bins \citep[see][ for more details]{leja2019, johnson2021}. For modeling the PRISM spectra, we employ the same line-spread function as was used for the \texttt{BEAGLE} fits. 

The total number of lookback time bins ($N_{tot}$) that were chosen for the \texttt{Prospector} fit determined the time resolution of our nonparametric SFH. As demonstrated by \citet{leja2019} using mock observations, the recovered stellar population properties of mock galaxies show large deviations from their intrinsic properties when $N_{tot} < 5$, and they argue that $N_{tot} = 6$ is the smallest value for a stable and unbiased inference of stellar population properties. Several studies of galaxies at $z > 10$ have adopted $N_{tot} = 6$, including \citet{tacchella2022}. This study only included photometric data, while we analyze high-quality spectra over 0.6 - 5$\mu$m.

The \texttt{Prospector} fits potentially suffer from over-fitting problems which are caused by the excessive model flexibility which can lead to overestimated uncertainties. As demonstrated in fits with a flexible SFH done in \citet{leja2019} and \citet{carnall2019}, however, this potential over-fitting issue can be largely mitigated by choosing a prior, like the continuity prior, to weight for physically plausible forms of SFHs.

We show the corner plots, SEDs, and star-formation histories for the \texttt{Prospector} fits in Figures \ref{fig:GS_z11_Prospector} and \ref{fig:GS_z13_Prospector} in the Appendix, and include the stellar population parameters in Table \ref{tab:spectrum-fit-parameters}.

\section{NIRCam Observations and {\tt ForcePho} Fits} \label{sec:nircam-photometry}

The NIRCam photometry for the two sources comes from the JADES data taken as of November 2023, which includes doubling the exposure time in the JADES NIRCam filters: F090W, F115W, F150W, F200W, F277W, F335M, F356W, F410M, and F444W. We add to this the medium-band NIRCam photometry in filters F182M, F210M, F430M, F460M, and F480M from the JWST Extragalactic Medium Survey \citep[JEMS, ][]{williams2023} as well as observations with the filters F182M, F210M and F444W from the First Reionization Epoch Spectroscopic COmplete Survey \citep[FRESCO, ][]{oesch2023} programs. We supplement the NIRCam data observations with those from the Hubble Space Telescope's Advanced Camera for Surveys (HST/ACS), using updated mosaics from the Hubble Legacy Fields program \citep{illingworth2013, whitaker2019}. For our purposes, we use the HST/ACS F435W, F606W, F775W, F814W, and F850LP filters. In total, we have observations in 5 HST/ACS filters and 14 JWST/NIRCam filters, for a total of 19 filters. Compared to the observations described for these sources in \citet{robertson2023}, these data are significantly deeper in both the primary JADES filters and in F182M, F210M, and F444W due to the additional FRESCO observations not included in their analysis. We now reach 5$\sigma$ observational depths of $2.4$ nJy in the F200W mosaic (in an $0.2^{\prime\prime}$ diameter aperture). We can compare this to the first-year depth provided in \citet{hainline2023} with the same aperture of $3.0$ nJy. 

Because of the small sizes of these sources, we extracted fluxes using 0.2$^{\prime\prime}$ diameter circular apertures, and applied an aperture correction assuming they are point sources. In addition, we use the software {\tt ForcePho} (Johnson et al. in prep) to estimate the total fluxes of these two sources. {\tt ForcePho} models the pixel-level fluxes for sources as the sum of PSF-convolved S\'ersic profiles for each galaxy, and fits these models directly to the pixel fluxes of the individual NIRCam exposures in every band. The usage of {\tt ForcePho} on JADES galaxies is described more extensively in \citet{robertson2023} and \citet{baker2023}. For the fits to JADES-GS-z11-0 and JADES-GS-z13-0, we assumed a uniform prior on the S\'ersic index between 0.9 and 1.1 to better constrain the fits. We list the updated NIRCam circular aperture and resulting {\tt ForcePho} fluxes, measured half-light radii and S\'ersic indices for both sources in Table \ref{tab:nircam_fluxes}. We plot the marginalized and joint posterior distributions for the half-light radius and the semiminor to semimajor axis ratios $b/a$ for both objects in Figure \ref{fig:half_light_sersic} in the Appendix.

\begin{figure*}
  \centering
  \includegraphics[width=0.95\linewidth]{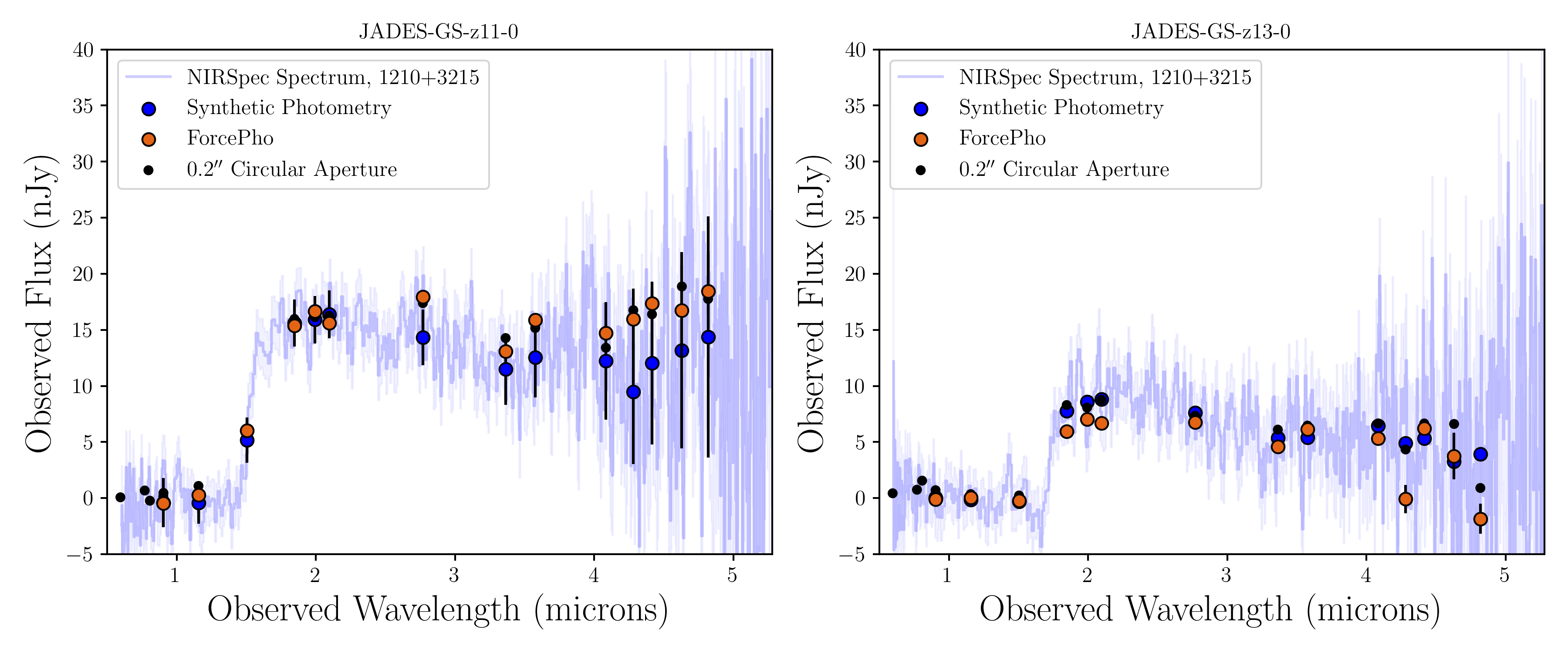}
  \caption{NIRSpec PRISM/CLEAR spectroscopy (light blue lines) plotted against the {\tt ForcePho} (red circles) and 0.2$^{\prime\prime}$ diameter circular aperture (black circles) NIRCam photometry for JADES-GS-z11-0 (left) and JADES-GS-z13-0 (right). In each panel, we compare to synthetic photometry measured from the NIRSpec PRISM data. The aperture photometry agrees well with the spectrum and synthetic photometry, while the {\tt ForcePho} photometry is slightly larger than what's observed in the spectrum for JADES-GS-z11-0 and lower for JADES-GS-z13-0.}
  \label{fig:photometry-comparison}
\end{figure*}

We compare these fluxes, as well as those measured using {\tt ForcePho}, to the observed spectra in Figure \ref{fig:photometry-comparison}. On this plot, for comparison, we also include synthetic photometry measured directly from the spectra for each source, calculated by interpolating the spectra with the HST/ACS and JWST/NIRCam filter curves. For JADES-GS-z11-0, the redshift of the source places the Lyman-$\alpha$ break in the F150W band, while for JADES-GS-z13-0, the break is between the F150W and F182M filters. The circular aperture and {\tt ForcePho} fluxes at $\lambda_{\mathrm{obs}} > 2\mu$m are somewhat higher, but within the uncertainties, than these synthetic photometric points, potentially due to variations in background subtraction at long wavelengths. For JADES-GS-z13-0, the circular aperture fluxes agree quite well, but the {\tt ForcePho} fluxes at $\sim 2 \mu$m are slightly underpredicted compared to the spectrum by $\sim15$\%.

We also measured the $M_{UV}$ and UV slope $\beta$ values from the {\tt ForcePho} and circular aperture photometry directly to compare with the fiducial values from the {\tt BEAGLE} fit. Using the aperture photometry, and only fitting photometry that corresponds to the rest-frame $1500 - 3300$\AA at the fiducial redshifts, we measure $M_{UV} = -19.41 \pm 0.11$ and $\beta = -2.1 \pm 0.1$ for JADES-GS-z11-0, and $M_{UV} = -18.79\pm 0.06$ and $\beta = -2.37 \pm 0.07$ for JADES-GS-z13-0. From the {\tt ForcePho} photometry, we measure $M_{UV} = -19.49\pm 0.12$ and $\beta = -2.10 \pm 0.14$ for JADES-GS-z11-0, and $M_{UV} = -18.79\pm 0.11$ and $\beta = -2.18 \pm 0.13$ for JADES-GS-z13-0. The $M_{UV}$ values are within 2$\sigma$ between the measurements from the spectrum and the photometry, but the slopes disagree, with the photometric slopes being significantly shallower, likely due to the uncertainty in fitting to discrete photometric points. 

\begin{deluxetable}{l c c c c}
\tabletypesize{\footnotesize}
\tablecolumns{5}
\tablewidth{0pt}
\tablecaption{0.2$^{\prime\prime}$ Diameter Circular Aperture and {\tt ForcePho} Photometry \label{tab:nircam_fluxes}}
\tablehead{
 & \multicolumn{2}{c}{JADES-GS-z11-0} & \multicolumn{2}{c}{JADES-GS-z13-0} \\
\colhead{Filter} & \colhead{Aperture} & \colhead{{\tt ForcePho}} & \colhead{Aperture} & \colhead{{\tt ForcePho}} }
\startdata
    F090W & $0.15 \pm 0.54$ & $-0.47 \pm 0.38$ & $0.7 \pm 0.48$ & $-0.12 \pm 0.13$ \\
    F115W & $1.08 \pm 0.43$ & $0.27 \pm 0.34$ & $0.35 \pm 0.35$ & $0.05 \pm 0.08$ \\
    F150W & $5.78 \pm 0.46$ & $6.02 \pm 0.38$ & $0.23 \pm 0.33$ & $-0.21 \pm 0.1$ \\
    F182M & $15.97 \pm 1.0$ & $15.39 \pm 0.68$ & $8.3 \pm 1.03$ & $5.95 \pm 0.28$ \\
    F200W & $16.12 \pm 0.51$ & $16.65 \pm 0.49$ & $8.07 \pm 0.37$ & $7.05 \pm 0.17$ \\
    F210M & $16.26 \pm 1.2$ & $15.62 \pm 0.76$ & $8.76 \pm 1.18$ & $6.7 \pm 0.33$ \\
    F277W & $17.38 \pm 0.41$ & $17.94 \pm 0.47$ & $7.34 \pm 0.28$ & $6.75 \pm 0.12$ \\
    F335M & $14.27 \pm 0.73$ & $13.1 \pm 0.88$ & $6.11 \pm 0.47$ & $4.59 \pm 0.24$ \\
    F356W & $15.17 \pm 0.45$ & $15.88 \pm 0.52$ & $6.46 \pm 0.31$ & $6.14 \pm 0.16$ \\
    F410M & $13.41 \pm 0.71$ & $14.7 \pm 0.84$ & $6.66 \pm 0.49$ & $5.33 \pm 0.26$ \\
    F430M & $16.76 \pm 2.49$ & $15.95 \pm 2.73$ & $4.34 \pm 2.28$ & $-0.08 \pm 1.27$ \\
    F444W & $16.39 \pm 0.59$ & $17.34 \pm 0.74$ & $6.65 \pm 0.4$ & $6.22 \pm 0.24$ \\
    F460M & $18.87 \pm 3.42$ & $16.72 \pm 4.04$ & $6.61 \pm 3.02$ & $3.75 \pm 2.09$ \\
    F480M & $17.76 \pm 2.98$ & $18.43 \pm 3.27$ & $0.9 \pm 2.57$ & $-1.85 \pm 1.33$ \\
    \hline
    $r_{\mathrm{half}}$/$^{\prime\prime}$ & \multicolumn{2}{c}{$0.030^{+0.001}_{-0.001}$} & \multicolumn{2}{c}{$0.017^{+0.001}_{-0.001}$} \\
    $n_{\mathrm{S\acute{e}rsic}}$ & \multicolumn{2}{c}{$1.02^{+0.07}_{-0.06}$} & \multicolumn{2}{c}{$0.99^{+0.04}_{-0.03}$} \\    
    $b/a$ & \multicolumn{2}{c}{$0.75^{+0.06}_{-0.06}$} & \multicolumn{2}{c}{$0.64^{+0.08}_{-0.08}$} \\    
\enddata
\tablecomments{All Fluxes in nJy}
\end{deluxetable}

\section{Results} \label{sec:results}

\subsection{Galaxy Fit and Morphological Properties}
As can be seen from Figures~\ref{fig:beagle_triangle_gsz11} and \ref{fig:beagle_triangle_gsz13}, both the \texttt{BEAGLE} and Prospector fits to the PRISM spectra agree given the uncertainties, with limited evidence for strong emission lines. We stress that the uncertainties we provide from both fitting methods are derived entirely from the flux and model uncertainties, and do not account for any potential systematic uncertainties that arise from deriving galaxy parameters from fits to the UV alone.

Looking at the posterior distributions and the values in Table \ref{tab:spectrum-fit-parameters}, we see that, for JADES-GS-z11-0, the fit results in a stellar mass of log(M$_*$/M$_{\odot}$)$ = 8.3^{+0.1}_{-0.1}$, and for JADES-GS-z13-0, log(M$_*$/M$_{\odot}$)$ = 7.7^{+0.4}_{-0.2}$. This is slightly smaller than what was measured in \citet{curtislake2023} for JADES-GS-z11-0, log(M$_*$/M$_{\odot}$)$ = 8.67^{+0.08}_{-0.13}$, while it agrees with the values presented in that study for JADES-GS-z13-0, log(M$_*$/M$_{\odot}$)$ = 7.95^{+0.19}_{-0.29}$. The likely cause of this difference in stellar mass for JADES-GS-z11-0 is due to the fits to the source at $\lambda_{\mathrm{obs}} > 4 \mu$m, where a potential Balmer break was predicted in the spectra described in \citet{curtislake2023}. In our updated spectra and fits for JADES-GS-z11-0, we do not find evidence for a Balmer break given the uncertainty at $\lambda_{\mathrm{obs}} > 4\mu$m. The SFR estimated from the fit to JADES-GS-z11-0 ($\log(\SFR / \Msun \yrInv) \sim 0.16$) is also slightly smaller than that reported in \citet{curtislake2023} ($\log(\SFR / \Msun \yrInv) \sim 0.34$), while the SFR for JADES-GS-z13-0 ($\log(\SFR / \Msun \yrInv) \sim 0.15$) is very similar to the previous results ($\log(\SFR / \Msun \yrInv) \sim 0.13$). 

\begin{figure*}
  \centering
  \includegraphics[width=0.95\linewidth]{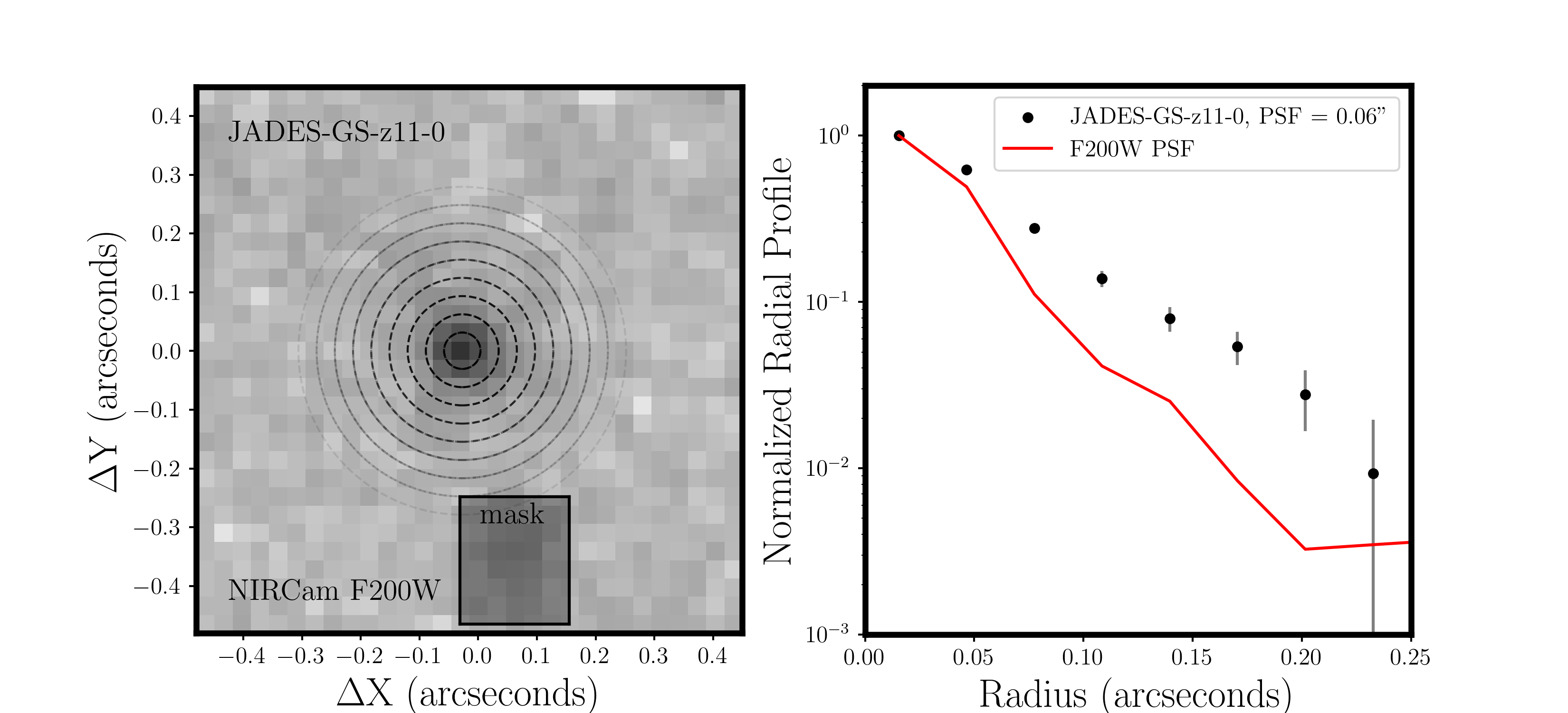}
  \includegraphics[width=0.95\linewidth]{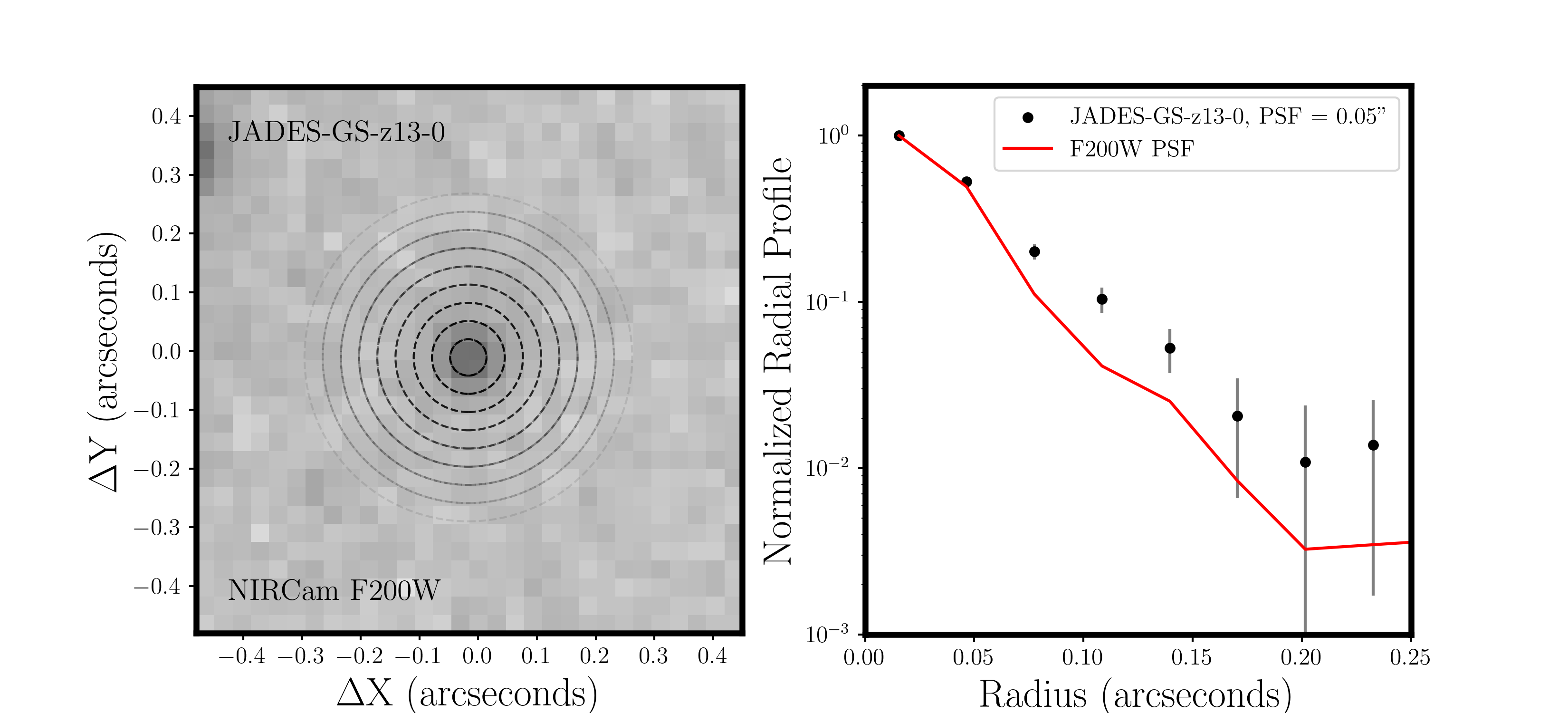}  
  \caption{Radial profile fits to the F200W images for JADES-GS-z11-0 (top row) and JADES-GS-z13-0 (bottom row). In the left column we show the F200W image centered on each source, with circular apertures used in calculating the radial profile. In the right panel, we plot the normalized radial profile for each source with black points and error bars as compared to the F200W mosaic PSF which we plot with a red line. For JADES-GS-z11-0, we mask a nearby source and show this mask with a grey box.}
  \label{fig:radial_profiles}
\end{figure*}

In Figure \ref{fig:radial_profiles}, we additionally plot the F200W radial profiles for JADES-GS-z11-0 and JADES-GS-z13-0. In the left panels we show the sources and the apertures used in deriving the profiles, and in the right panels we show the radial profiles as compared to the measured F200W mosaic PSF. For JADES-GS-z11-0, we mask out the source JADES-GS+53.16474-27.77471, which we discuss further in Section \ref{sec:JADES-GS-z11b}. Both sources are resolved beyond the extent of the PSF in this filter out to $\sim0.25^{\prime\prime}$, where each source is too faint to measure a significant flux. 

The sizes estimated from the {\tt ForcePho} fits are small, with half-light radii of only $0.030^{\prime\prime} \pm 0.001^{\prime\prime}$ for JADES-GS-z11-0 and $0.017^{\prime\prime} \pm 0.001^{\prime\prime}$ for JADES-GS-z13-0. The axis ratio for JADES-GS-z11-0 is $b/a = 0.75^{+0.06}_{-0.05}$ and for JADES-GS-z13-0, $b/a = 0.64^{+0.07}_{-0.08}$. These values for the half-light radii are larger than the sizes presented in \citet{robertson2023}, likely due to the deeper photometry and updated PSF. \citet{robertson2023} were only able to provide an upper limit on the size for JADES-GS-z13-0, as there was substantial probability that the half-light radius for the source was $0.001^{\prime\prime}$, at the lower bound of their fit, and with these updated fits we find strong evidence that the source is resolved. The half-light radii for the sources correspond to 119 pc at the fiducial redshift of JADES-GS-z11-0, and 59 pc at the fiducial redshift of JADES-GS-z13-0, and further demonstrate the very small sizes for these sources. These values are below the FWHM of the NIRCam PSF, demonstrating that due to the dithering from the generation of the mosaic, we are able to resolve the diameter of each source. This is supported by the work of \citet{robertson2023}, where they discuss how {\tt ForcePho} fits to unresolved brown dwarfs in the GOODS-S field indicated the ability to resolve sources of these sizes.

At these sizes, we can use the SFR values measured from {\tt BEAGLE} to estimate the star-formation rate surface densities for these sources, following the definition given in \citet{shibuya2019}:

\begin{equation}
    \Sigma_{\mathrm{SFR}} [\mathrm{M}_{\odot} \mathrm{yr}^{-1} \mathrm{kpc}^{-2}] = \frac{\mathrm{SFR}_{\mathrm{UV}}/2}{\pi r_e^2}
\end{equation}

Where, here, we use the half-light radius as $r_e$. For JADES-GS-z11-0, we calculate $\Sigma_{\mathrm{SFR}} = 16$ M$_{\odot}$ yr$^{-1}$ kpc$^{-2}$, and for JADES-GS-z13-0, we calculate $\Sigma_{\mathrm{SFR}} = 64$ M$_{\odot}$ yr$^{-1}$ kpc$^{-2}$. These values, which are lower than what is presented in \citet{robertson2023} due to the difference in measured sizes, are still above what is seen for most starburst galaxies out to $z \sim 2 - 4$ \citep{genzel2010, reddy2023}, and are more similar to local ultra compact starbursts like the ``green pea'' galaxies \citep{izotov2016a, izotov2016b}. 

\subsection{Ionized Gas Properties in JADES-GS-z11-0} \label{sec:results-jades-gs-z11-ratios}

For JADES-GS-z11-0, we observe three emission lines with SNR $> 1$, \ion{C}{4}, [\ion{O}{2}], and [\ion{Ne}{3}]. We can use these line fluxes, and the upper limits on other strong lines, to investigate the ionization properties of this source. In galaxies, both neon and oxygen are generated as a part of the carbon-burning cycle in stars, and are spread through supernovae explosions \citep[see][for a review]{maiolino2019}. The ratio of the high-ionization line [\ion{Ne}{3}] and low-ionization line [\ion{O}{2}] (commonly known as Ne3O2) traces mainly the ionization state of the gas. For JADES-GS-z11-0, we measure [\ion{Ne}{3}] / [\ion{O}{2}] $= 0.7 \pm 0.4$, a value higher (but consistent within 1$\sigma$) than that measured for Maisie's Galaxy ([\ion{Ne}{3}] / [\ion{O}{2}] $= 0.3$) at $z_{\mathrm{spec}} = 11.42$ or CEERS2\_588 ([\ion{Ne}{3}] / [\ion{O}{2}] $= 0.6$) at $z_{\mathrm{spec}} = 11.04$ \citep{harikane2024}. In addition, the value we estimate is slightly lower (but again, consistent within 1$\sigma$) than the value estimated for JADES-GS-z12-0 ($z_{\mathrm{spec}} = 12.48$), [\ion{Ne}{3}] / [\ion{O}{2}] $= 0.9 \pm 0.3$ \citep{deugenio2023}. These values are similar to a sample of low-redshift, low-metallicity galaxies assembled by \citet{nakajima2022}. 

We can explore other diagnostics of C/O abundance and gas photoionization using \ion{C}{3}] / [\ion{O}{2}] + [\ion{Ne}{3}] and \ion{C}{4} / \ion{C}{3}]. Because we only have a 3$\sigma$ upper limit for the \ion{C}{3}] line, we only report upper limits for JADES-GS-z11-0: \ion{C}{3}] / [\ion{O}{2}] + [\ion{Ne}{3}] $< 1.1$ and \ion{C}{4} / \ion{C}{3}] $ > 0.7$. The upper limit for \ion{C}{3}] / [\ion{O}{2}] + [\ion{Ne}{3}] is not as extreme as JADES-GS-z12-0 \citep{deugenio2023}, but is still consistent with the high end of the values calculated from theoretical models derived from photoionization due to star formation and AGN in \citet{nakajima2022} and \citet{Gutkin2016}. 

\subsection{The Blue UV Slope and Absence of Emission Lines in JADES-GS-z13-0} \label{sec:gs-z13-no-emission}

The JADES-GS-z13-0 spectrum shown in Figure \ref{fig:2d_1d_plots} is notable in that we estimate a very blue UV slope ($\beta = -2.69$) and we do not see any obvious strong emission lines given our fiducial redshifts. This latter point is surprising given the detection of emission lines in other galaxies at $z > 10$, including JADES-GS-z11-0, in multiple galaxies in \citet{arrabalharo2023a}, MACS0647-JD \citep{hsiao2023}, GN-z11 \citep{bunker2023a}, GLASS-z12 \citep{castellano2024, zavala2024}, and JADES-GS-z12-0 \citep{deugenio2023}. The depth of the spectra in this paper puts tight upper limits on the possible flux of any lines, as shown in Table~\ref{tab:spectrum-fit-parameters}.

To interpret the spectrum of JADES-GS-z13-0, we first adopted the same fiducial model as the one used to model JADES-GS-z11-0 and discussed in Section~\ref{sec:sed-fitting} above, i.e. a model with a constant star formation history and no escape fraction of ionising photons. As shown in Figure~\ref{fig:JADES-GS-z13-0-constant-SFH} in Appendix~\ref{sec:supp_figures}, this simple model does not match well to the observed spectrum of JADES-GS-z13-0. The model predicts a UV slope $\beta \sim -2.5$, significantly less steep than what is observed: the nebular continuum reddens the UV slope, preventing the model from reaching values below $\sim 2.5$. Moreover, matching the upper limits on the emission line EWs requires an unlikely combination of parameters, i.e. a very low metallicity $\log(\Zmetal/\Zsun) \lesssim -2$ and a very low ionization parameter $\logUs \lesssim -3$, as this suppresses the UV high-ionization lines. Interestingly, this model predicts significant \OII\ emission, $\txn{EW}(\OII)\sim 25$ \AA, thus more stringent constraints on \OII\ might observationally rule out this model.

We also tested the impact of the assumed star formation history and modeled JADES-GS-z13-0 using a delayed exponential star formation plus a burst of 10 Myr duration. This model thus allows for the separation of the current star formation rate (over the last 10 Myr) from the past star formation history, i.e. decoupling the strength of emission lines (powered by stars younger than 10 Myr) from the UV continuum emission (powered by stars with ages up to few $10^8$ yr). We plot the model predictions and posterior probability distributions in Figure~\ref{fig:JADES-GS-z13-0-delayed-SFH} in Appendix~\ref{sec:supp_figures}. This model provides a formally good fit to the data, reaching $\beta \sim -2.9$, but, again, thanks to an unlikely combination of model parameters: emission lines are suppressed thanks to a very low current (observed) star formation rate ($\log(\SFR/\Msun \yrInv) \lesssim -1.5$), while the blue UV slope is produced by stars covering a narrow age range older than 10 Myr (mass-weighted age between 10 and 20 Myr). This model thus requires a very vigorous star formation that ceased precisely at the time required for the stars emitting ionizing photons to have evolved and died by the time of observation, a conclusion that strains believability. When we run this exact same fit, with a delayed star formation history but without the 10Myr burst, we find a similar result - the resulting fit requires a very low value for for the delayed SFH e-folding time ($\tau \sim 10^7$) with a very low metallicity and ionization parameter, to reproduce the UV slope. This essentially renders the effects of any current star formation in JADES-GS-z13-0 negligible. The luminosity of this fit at a rest-frame $1500$\AA is therefore completely dominated by stars older than 10Myr, as both must be modeled outside of their birth clouds so as to not include a nebular continuum that would redden the UV slope. There is a clear tension btween current star formation (to allow for blue stars and a blue UV slope) and the reddening of a nebular continuum.  

We therefore consider the most physically plausible model to be the one that allows for the escape of ionizing photons from JADES-GS-z13-0. The large predicted escape fraction ($\fesc \gtrsim 0.8$) and low metallicity ($\log(\Zmetal/\Zsun) \lesssim -1.6$) of the fiducial \texttt{BEAGLE} model, which we described in Section \ref{sec:sed-fitting}, enable the suppression of emission lines and a blue UV slope. This model also provides sensible values for the other parameters, namely a very low dust attenuation ($\tauV \lesssim 0.05$), and a mass-weighted age of 8 -- 50 Myr. The high derived escape fraction of this model is reproduced for each of the different star-formation histories we explored with \texttt{BEAGLE}. However, this high value is likely still dependent on our modeling choices. We have not explored density bounded nebulae, or how a more complex SFH in the most recent 10 Myrs since observation might require a less extreme value to explain the blue UV slope and a lack of emission lines. 

We can interpret the observed $\beta$ and escape fraction for JADES-GS-z13-0 in the context of what has been measured for other galaxies at high redshift. Multiple studies have explored the evolution of UV slope $\beta$ estimated from photometry for independent JWST/NIRCam imaging surveys \citep{topping2022, topping2024, morales2024, austin2024, cullen2024}. The slope we measure from the spectrum for JADES-GS-z13-0, $\beta = -2.7$, is only slightly more blue than what has been seen for other galaxies at similar photometric redshifts, but in agreement given the large uncertainties and the small number of sources. The escape fraction that we estimate from the fiducial model for JADES-GS-z13-0, $\fesc \gtrsim 0.8$, is higher than what has been measured for local low-mass Lyman continuum leakers \citep[$\fesc =  0.11 - 0.50$,][]{izotov2021, flury2022}, with more extreme values seen for galaxies at $z = 2 - 3$ \citep[$\fesc \sim 0.6$,][]{vanzella2016, fletcher2019, riverathorsen2019}. More recently, \citet{kim2023} explored the gravitationally-lensed ``Sunburst Arc'' at $z = 2.37$, where they observe a compact region of $ <100$ pc with a UV slope $\beta = -2.9$ and $\fesc \sim 0.3$, estimated from the observed H$\beta$ line flux.
There is a source at lower redshifts with large calculated escape fraction: J1316+2614, at $z = 3.613$, has a UV slope $\beta = -2.59 \pm 0.05$ and escape fraction of $90\%$ \citep{marqueschaves2021, marqueschaves2022, marqueschaves2024}. 

At higher redshift, \citet{topping2024} explore the escape fractions required to achieve extremely blue UV slopes in a sample of JADES sources and conclude that density-bounded photoionization models, such as the ones from \citet{plat2019} can result in such blue UV slopes at escape fractions of only $\fesc \sim 0.5$, with the highest estimated value of $\fesc =  0.86$. Similarly, \citet{menon2024} used numerical radiation
hydrodynamic simulations of dense star clusters at high-redshift to find that a burst of star formation, followed by rapid gas dispersal in ionized outflows that can permit high escape fractions of $\fesc > 0.8$.

Based on the fiducial \texttt{BEAGLE} model, and given the high escape fraction we find for the source, we can estimate the size of the ionized bubble surrounding JADES-GS-z13-0 following the method described in \citet{witstok2023} and \citet{masongronke2020}, where the latter study demonstrates the analytical solution for the evolution of an ionization front. In this method, we assume that the IGM neutral fraction of hydrogen outside the bubble, ($\hat{x}_{\mathrm{HI}}$) is 1, and estimate the source emissivity from the measured $M_{UV}$, UV slope $\beta$, and the slope of the ionizing continuum $\alpha$. For $\alpha$, we assume a value of $\alpha = -2$ based on the NIRSpec spectrum of a $z = 7.3$ Lyman-$\alpha$ emitter in \citet{saxena2023}. The resulting size of the ionized bubble is $\sim 0.1$ pMpc, in agreement with the values measured for high-redshift galaxies in \citet{umeda2023}. We caution that this size is highly uncertain, especially given that it assumes a low hydrogen recombination rate that may not be applicable for galaxies at $z > 8$.

\subsection{DLA Fits} \label{sec:dla-fits}

The redshift of a galaxy can be calculated from either a measurement of one or multiple emission or absorption lines in the spectrum, or it can be estimated from the observed wavelength of the Lyman-$\alpha$ break. This latter technique is uncertain, given the potential for additional UV absorption at high redshift, which serves to push the Lyman-$\alpha$ break to longer wavelengths. In \citet{deugenio2023}, the authors observe the \ion{C}{3}]$\lambda\lambda1907,1909$ emission line in the NIRSpec spectrum for JADES-GS-z12-0, and they propose that strong Lyman damping wing absorption with $\log{(N_\mathrm{HI}/\mathrm{cm}^{-2})} \sim 22$ cm$^{-2}$ is responsible for the observed shift between the Lyman-$\alpha$ rest wavelength and the Lyman-$\alpha$ break for the source. This absorption would result in a slower observed turnover of the Lyman-$\alpha$ break, and similar absorption was observed in a sample of three galaxies at $z = 9 - 11$ by \citep{heintz2023}. We note that for both GN-z11 \citep{bunker2023a} and for one of the two $z > 11$ sources with detected emission lines observed in \citet{arrabalharo2023a}, no additional absorption was necessary in their fit, although these sources are almost a magnitude brighter in $M_{UV}$, and GN-z11 displays Ly$\alpha$ emission.

To explore the potential need for a DLA in both JADES-GS-z11-0 and JADES-GS-z13-0, we followed the fitting approach in \citet{deugenio2023}\footnote{Based on the publicly available python package \texttt{\href{https://github.com/joriswitstok/lymana_absorption}{lymana\_absorption}.}},where we set the redshift for JADES-GS-z11-0 to be at $z = 11.122$, and let redshift be a free parameter for JADES-GS-z13-0. We attenuated the fiducial \texttt{BEAGLE} fits presented in Figures~\ref{fig:beagle_triangle_gsz11} and \ref{fig:beagle_triangle_gsz13} with both a damped Lyman-$\alpha$ system while fixing the IGM neutral hydrogen fraction ($\hat{x}_{\mathrm{HI}}$) to 1. Note that in both sources, the spectra were masked in the wavelength range $\lambda$ = 1150 - 1450 \AA when fitting with \texttt{BEAGLE}, and while the resulting fit for JADES-GS-z11-0 does show Lyman-$\alpha$ in emission, this is not observed in either the PRISM or grating spectra.

For the DLA fit, we tied the redshift of any potential DLA to be at the redshift of the galaxy, and for the IGM, we follow \citet{witstok2023} and assume that the IGM gas is at a mean cosmic density with T = 1K, although raising this temperature has a negligible impact on the results. We estimate the likelihood of the fits over $\lambda$ = 1100 - 1520 \AA (respectively spanning 42 and 45 wavelength bins for JADES-GS-z11-0 and JADES-GS-z13-0) by calculating the inverse-weighted squared residuals between the model (convolved with the effective NIRSpec PRISM line spread function) and the observed spectrum. We assume flat uniform priors on redshift between $z = 12.7 - 13.3$ for JADES-GS-z13-0, and for each source allow $\log{(N_\mathrm{HI}/\mathrm{cm}^{-2})}$ to vary between $\log{(N_\mathrm{HI}/\mathrm{cm}^{-2})} = 19.0 - 24.0$ with a flat prior. For comparison, we also present a fit where we do not include an additional DLA component. The uncertainties we use for calculating $\chi^2$ in the fits were derived from the covariance matrix measured from the individual sub-spectra for each source. 

\begin{figure}
  \centering
  \includegraphics[width=0.98\linewidth]{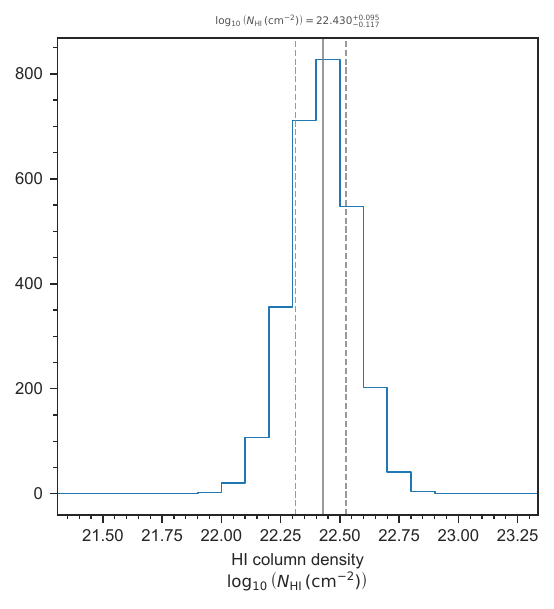}\\
  \includegraphics[width=0.98\linewidth]{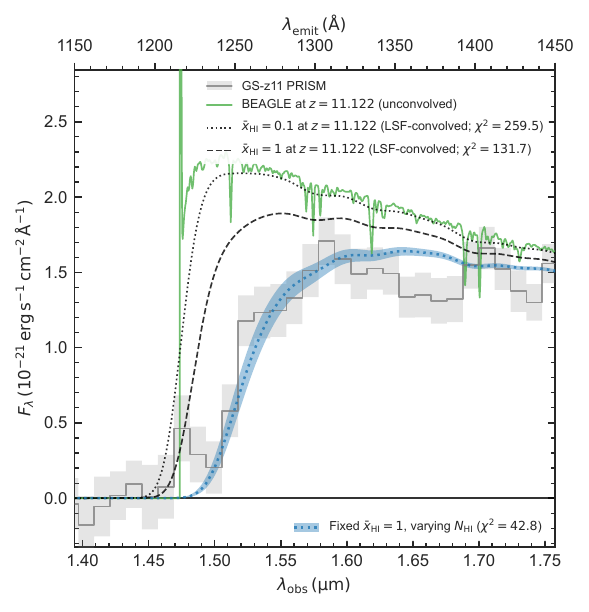}
  \caption{(top) Posterior on $\log{(N_\mathrm{HI}/\mathrm{cm}^{-2})}$ and (bottom) SED fit for JADES-GS-z11-0 where we fix the source at $z_{\mathrm{spec}} = 11.122$, while allowing the DLA HI column density to vary. For this fit, we fix the IGM HI fraction $\hat{x}_{\mathrm{HI}} = 1$. Here, because of the spectroscopic redshift of this source, we find a DLA is required with $\log{(N_\mathrm{HI}/\mathrm{cm}^{-2})} = 22.43^{+0.10}_{-0.12}$ to account for the shape of the Lyman-$\alpha$ break.}
  \label{fig:DLA_GSz11_z11.122}
\end{figure}

For JADES-GS-z11-0, the Lyman-$\alpha$ break implies a significantly higher redshift than we estimate from the observed emission lines. As a result, our fit requires additional absorption, and we estimate $\log{(N_\mathrm{HI}/\mathrm{cm}^{-2})} = 22.43^{+0.10}_{-0.12}$, a column density similar to what was measured for JADES-GS-z12-0 in \citet{deugenio2023}. We plot the posterior on the column density (top), and a fit to the observed spectrum (bottom) for JADES-GS-z11-0 in Figure \ref{fig:DLA_GSz11_z11.122}. 

\begin{figure}
  \centering
  \includegraphics[width=0.98\linewidth]{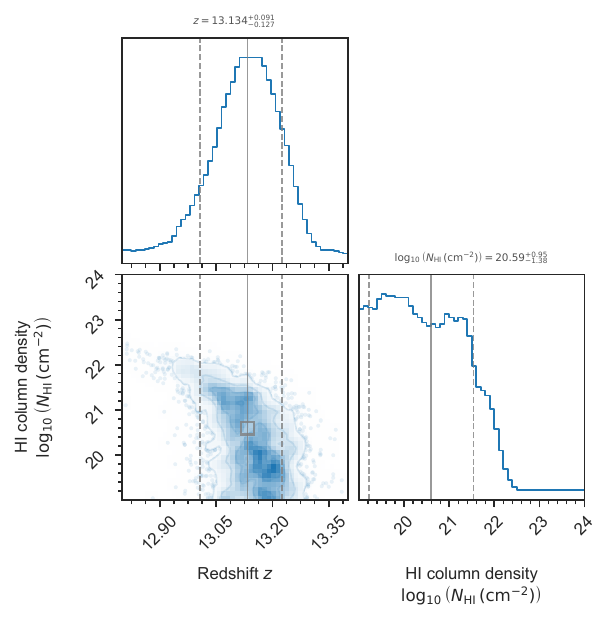}\\
  \includegraphics[width=0.98\linewidth]{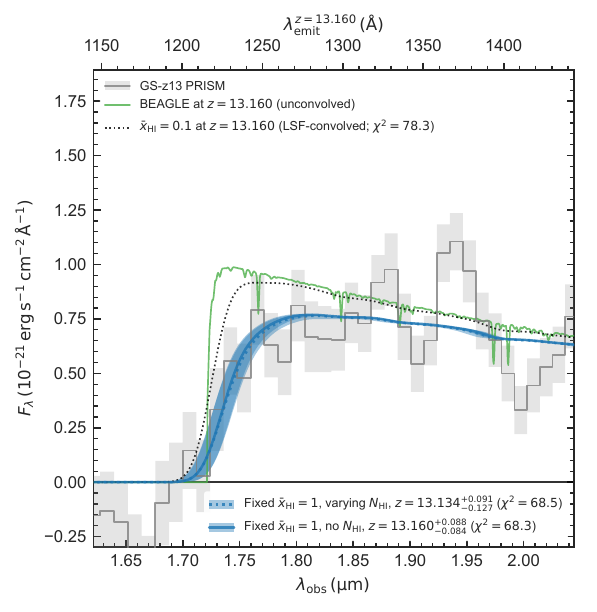}
  \caption{Corner plot (top) and SED fit (bottom) for JADES-GS-z13-0 where we allow the redshift and DLA HI column density to vary while fitting the observed spectrum. We begin with the fiducial BEAGLE fit, shown with the green line in the bottom panel, and then explore the relationship between redshift, IGM absorption, and DLA absorption. The $\chi^2$ does not change when we include a DLA as a free parameter for this source, demonstrating that it is not statistically preferred by the data.}
  \label{fig:DLA_GSz13}
\end{figure}

In Figure \ref{fig:DLA_GSz13} we plot our fit to the JADES-GS-z13-0 spectrum. For this source, we measure a redshift of $z_{\mathrm{spec}} = 13.13^{+0.09}_{-0.13}$ when we allow $\log{(N_\mathrm{HI})}$ to vary, and $z_{\mathrm{spec}} = 13.16^{+0.09}_{-0.08}$ when we don't include a DLA. For this source, we find from the SED fit that we do not need to include a DLA beyond the effects of setting $\hat{x}_{\mathrm{HI}} = 1$, as the best-fit $\chi^2$ is not significantly improved with the addition of a DLA. Most notably in the joint posterior in the bottom left panel of Figure \ref{fig:DLA_GSz13}, we can see how redshift varies with $\log{N_\mathrm{HI}}$ such that at higher DLA column densities, the best-fit redshift is lower. We note that the estimated redshift for JADES-GS-z13-0 is lower than the fiducial redshift from the \texttt{BEAGLE} fit, largely due to the effects of fixing the IGM neutral hydrogen fraction to 1.

These results highlight the uncertainty in estimating the redshifts for these ultra-distant galaxies without observed emission lines. Fitting directly to the observed spectrum without accounting for IGM absorption or a potential DLA may result in artificially high redshifts, as was observed for both JADES-GS-z11-0 and JADES-GS-z12-0 in \citet{curtislake2023}. 

\subsection{Photometric Redshifts} \label{sec:results-photoz}

In \citet{hainline2023}, the authors explore the relationship between photometric redshift and spectroscopic redshift for a large sample of $z > 8$ sources from across the JADES GOODS-S and GOODS-N footprints. They find that, on average, their photometric redshifts overpredict the spectroscopic redshift for these sources by $\langle z_{\mathrm{spec}} - z_{\mathrm{phot}} \rangle = -0.26$, which has been observed for other high-redshift surveys \citep{arrabalharo2023b, fujimoto2023, finkelstein2023, willott2023}. These authors have attributed the offset to potential DLAs and the existence of a 2-photon nuclear continuum that becomes increasingly important at high redshift. We can better explore the origin of this discrepancy using the spectra for JADES-GS-z11-0 and JADES-GS-z13-0.

We fit the synthetic photometry estimated from the NIRSpec PRISM spectra (described in Section \ref{sec:nircam-photometry}) using the template-fitting code \texttt{EAZY} \citep{brammer2008} following the procedure described in \cite{hainline2023}. We let redshift vary between $z_{\mathrm{min}} = 0.01$ to $z_{\mathrm{max}} = 22.0$ in bins of $\Delta z = 0.01$. As we are fitting to synthetic photometry, we don't calculate or use photometric offsets for the fits. The redshift corresponding to the minimum $\chi^2$ ($z_a$) for JADES-GS-z11-0 is $z_a = 11.8$ and for JADES-GS-z13-0 is $z_a = 14.0$, both in excess of our fiducial spectroscopic redshifts.

\begin{figure*}
  \centering
  \includegraphics[width=0.98\linewidth]{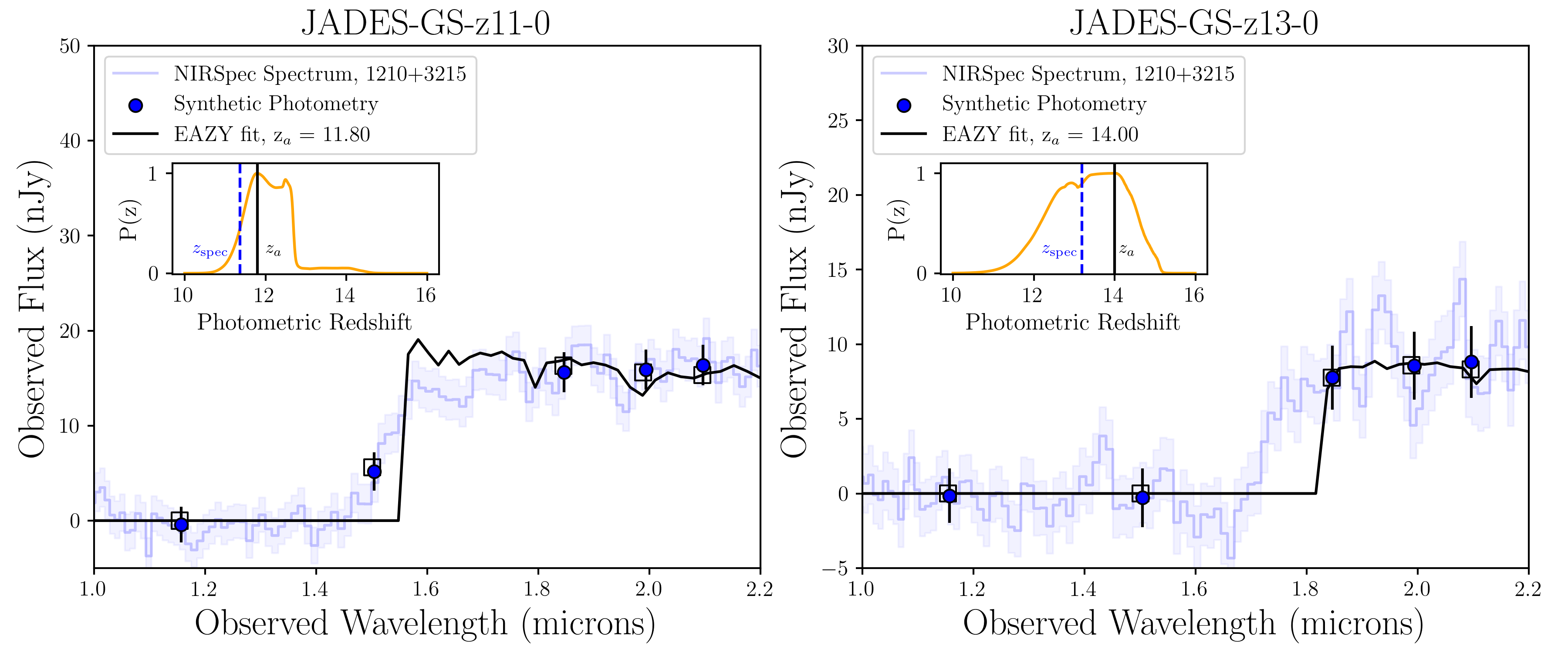}
  \caption{EAZY fit to JADES-GS-z11-0 (left) and JADES-GS-z13-0 (right). In each panel, the NIRSpec PRISM spectrum and uncertainty are plotted in blue, and synthetic photometry estimated from the spectrum through the JADES NIRCam bands are given with blue circles. In black we plot the \texttt{EAZY} SED corresponding to the minimum $\chi^2$ of the fit (the redshift of each fit is given in the legend), and with black squares are the \texttt{EAZY} photometry derived from that SED. In the inset in each panel, we show the \texttt{EAZY} $P(z)$ in orange, and plot with a black vertical line the photometric redshift $z_a$ and a blue dashed vertical line the fiducial spectroscopic redshift. For each source, the photometric redshift is biased high compared to our fiducial photometric redshifts due to the shallower Lyman-$\alpha$ break in the observed spectrum, which is not simulated based on the IGM model in the \texttt{EAZY} fit.}
  \label{fig:LyA_EAZY}
\end{figure*}

In Figure \ref{fig:LyA_EAZY}, we plot the \texttt{EAZY} SED corresponding to the minimum $\chi^2$ and the PRISM spectra and synthetic photometry, focusing on the region around the Lyman-$\alpha$ break for each source. In this Figure, we plot the synthetic photometry with black points, and the \texttt{EAZY} template photometry with black squares. For both sources, the fit is excellent, with $\chi^2 = 0.77$ for JADES-GS-z11-0 and $\chi^2 = 0.36$ for JADES-GS-z13-0, but in each case, the observed Lyman-$\alpha$ break from the spectrum falls off to the blue more gradually than the \texttt{EAZY} SED. In addition, for JADES-GS-z13-0, the gap between the F150W and F182M filters makes determining a precise photometric redshift more difficult. For galaxies at $z > 12$, deep images taken with the NIRCam F162M filter, similar to those obtained for the JADES Origins Field \citep{eisenstein2023b}, would help with this issue. The larger problem, however, is the possibility that DLA absorption, or a 2-photon nebular continuum becoming increasingly important at higher redshifts, which is not currently simulated in most photometric redshift codes, leading to photometric redshift estimates that are biased high.

\subsection{JADES-GS+53.16474-27.77471} \label{sec:JADES-GS-z11b}

\begin{figure}
  \centering
  \includegraphics[width=0.98\linewidth]{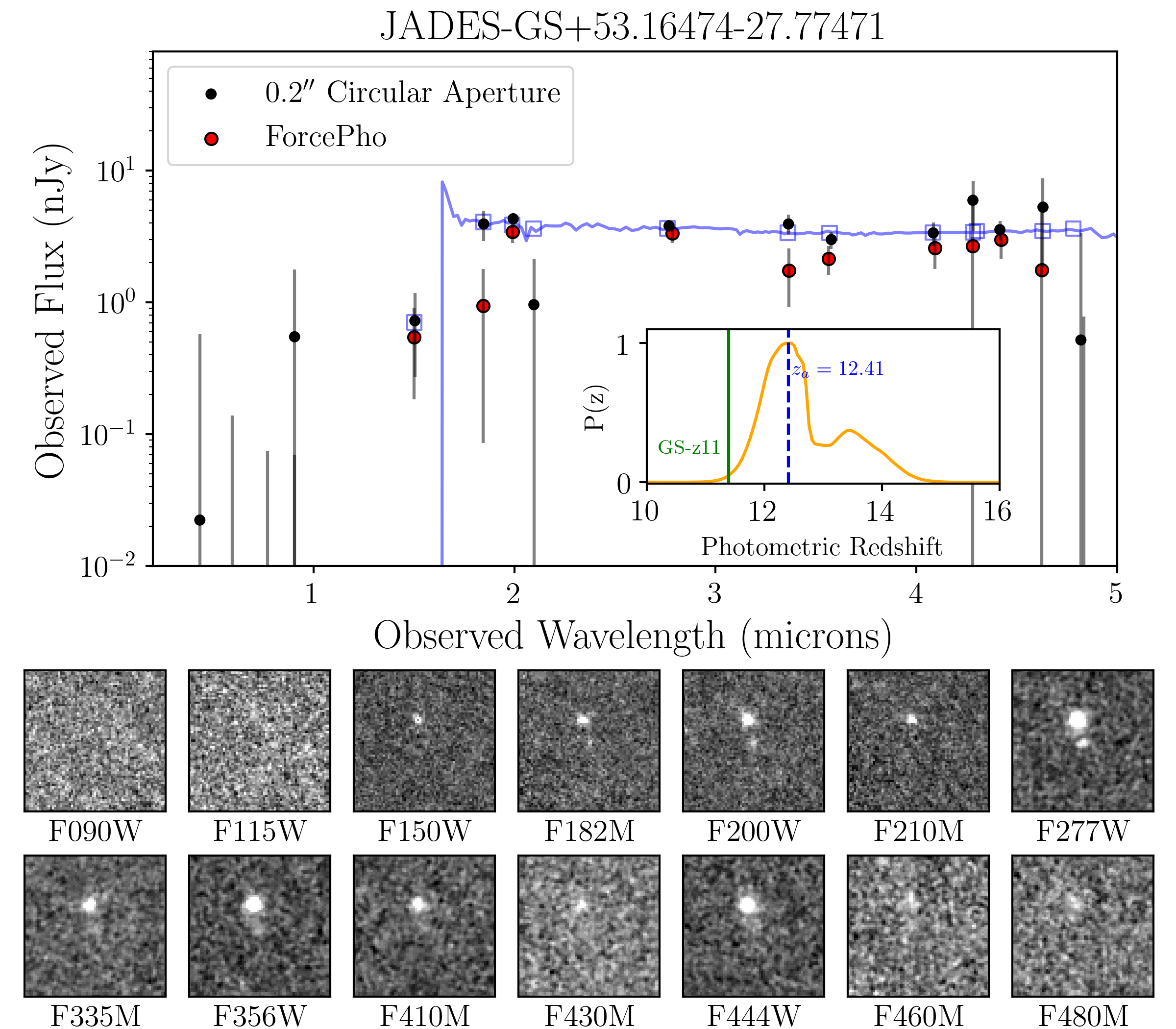}
  \caption{Photometry and \texttt{EAZY} fit (top panel), and NIRCam $2^{\prime\prime} \times 2^{\prime\prime}$ thumbnails for JADES-GS+53.16474-27.77471. This distinct source, which can be most easily seen south of JADES-GS-z11-0 in the F277W thumbnail, has a best-fit photometric redshift $z_a = 12.41$ when fitting to the 0.2$^{\prime\prime}$ circular aperture photometry (black points), as represented by the blue fit in the top panel. In the inset, we show the P(z) surface from \texttt{EAZY}, where we indicate our spectroscopic redshift for JADES-GS-z11-0 with a green vertical line. The NIRCam data for F182M, F210M, F430M, F460M, and F480M are taken from the JEMS data and are shallower than the other filters, and non-detections are not surprising in these filters given the flux levels of this source.}
  \label{fig:JADES-GS-z11b}
\end{figure}

Our updated, deeper NIRCam imaging provides stronger evidence of a secondary source $\sim 0.3^{\prime\prime}$ south ($\sim 1.2$kpc at $z = 11.39$) of JADES-GS-z11-0, which can be seen in the thumbnail in Figure \ref{fig:2d_1d_plots}. This object, JADES-GS+53.16474-27.77471, appears to be an F150W dropout potentially associated with JADES-GS-z11-0, although it did not fall onto the NIRSpec MSA. In Figure \ref{fig:JADES-GS-z11b}, we plot the SED and thumbnails for this source, where we show both the 0.2$^{\prime\prime}$ circular aperture and \texttt{ForcePho} photometry. We fit both sets of photometry with \texttt{EAZY} following the procedure in Section \ref{sec:nircam-photometry}, and the minimum $\chi^2$ redshift is $z_a = 12.41$ for the fit to the circular aperture photometry, and $z_a = 12.31$ for the fit to the \texttt{ForcePho} photometry. At this redshift, using a fit to the \texttt{ForcePho} photometry, we calculate M$_{\mathrm{UV}} = -17.8 \pm 0.5$ for this source. While this redshift is potentially biased high for the same reasons as are described in Section \ref{sec:dla-fits}, there is some probability of the source being at the spectroscopic redshift of JADES-GS-z11-0 as shown in the P(z) plot inset of the figure. One of the primary reasons for the difference in photometric redshifts is the redder F150W - F200W color for JADES-GS+53.16474-27.77471 ($m_{\mathrm{F150W}} - m_{\mathrm{F200W}} = 1.9$) as compared to JADES-GS-z11-0 ($m_{\mathrm{F150W}} - m_{\mathrm{F200W}} = 1.1$). 

From the \texttt{ForcePho} fit to the source, we calculate a half-light radii of only $0.03^{\prime\prime} \pm 0.01^{\prime\prime}$ for JADES-GS+53.16474-27.77471, which is 109 pc at $z_a = 12.41$ (116 pc at the redshift of JADES-GS-z11-0), a similar size to JADES-GS-z11-0 given the uncertainties. We fit the \texttt{ForcePho} photometry for this source with \texttt{Prospector} to estimate the stellar mass of this potential satellite, and calculate a stellar mass of $\log{(M_*/M_\odot)} = 8.0^{+0.4}_{-0.6}$. 

In \citet{hainline2023}, the authors find a number of sources from across the JADES survey at $z > 8$ with complex morphologies. Many of these galaxies have multiple knots of a similar brightness, or, like JADES-GS-z11-0 and JADES-GS+53.16474-27.77471, a bright central knot with a fainter satellite. Multiple galaxies at $z \sim 7 - 8$ have been targeted with JWST/NIRSpec as part of JADES \citep{bunker2023b}, which also show potential satellite galaxies similar to JADES-GS+53.16474-27.77471. The central galaxies are bright and compact, with markedly redder observed colors as compared to their smaller satellites. The UV+optical spectra for these sources show evidence for strong line emission. It would be of interest to target these satellites directly to understand the complex interactions of high-redshift galaxies. 

\section{Discussion and Conclusions} \label{sec:discussion}

These new spectra for JADES-GS-z11-0 and JADES-GS-z13-0 confirm that these are among the farthest galaxies observed in the first two years after the launch of JWST. We can compare the properties we derive from these deep spectra with the properties for other spectroscopically-confirmed galaxies at high redshift observed with JWST/NIRSpec. 

With these new, deeper observations, JADES-GS-z11-0 joins other sources at $z > 10$ with observed emission lines: MACS0647–JD \citep[$z_{\mathrm{spec}} = 10.17$, ][]{hsiao2023}, GN-z11 \citep[$z_{spec} = 10.6$, ][]{bunker2023a}, Maisie's Galaxy and CEERS2\_588 \citep[$z_{spec} = 11.42$ and $11.04$ respectively, ][]{arrabalharo2023a}, GLASS-z12 \citep[$z_{spec} = 12.34$,][]{castellano2024, zavala2024}, and JADES-GS-z12-0 \citep[$z_{spec} = 12.48$, ][]{deugenio2023}. MACS0647–JD, which was originally detected using HST \citep{coe2013}, is triply-lensed, and the NIRSpec spectrum for this source shows multiple emission lines. \citet{hsiao2023} estimate $M_{UV} = -20.3$ for this source. Due to the gravitational lensing, this source has a brighter apparent magnitude than GN-z11, a source that is more luminous with $M_{UV} = -21.5$ \citep{bunker2023a}. Maisie's Galaxy and CEERS2\_588 are estimated to have $M_{UV} = -20.1$ and $M_{UV} = -20.3$ \citet{heintz2023}, and GLASS-z12, at $M_{UV} = -20.49$ \citep{castellano2024}, is also quite bright. JADES-GS-z12-0 \citep[$M_{UV} = -18.23$,][]{curtislake2023} and JADES-GS-z11-0 ($M_{UV} = -19.22$) are currently the least luminous galaxies at $z > 10$ observed with emission lines. 

Among the sources at $z > 10$ with emission line detections, there are a variety of Lyman-$\alpha$ profiles. The spectra for GN-z11 and CEERS2\_588 do not show evidence for additional DLA absorption. Fits to the MACS0647–JD spectrum, however, do demonstrate a need for a damping wing, and can be explained with a high neutral fraction ($\hat{x}_{\mathrm{HI}} = 0.9$) and a small ionized bubble around the source with a radius smaller than 1 physical Mpc \citep{hsiao2023}. In \citet{heintz2023}, they fit the spectrum of this source with an absorber with $\log{(N_\mathrm{HI}/\mathrm{cm}^{-2})} = 22.4$, which, given the redshift and bright $M_{UV}$ value for this source, is puzzling when compared to GN-z11 and CEERS2\_588. \citet{heintz2023} also require an absorber with $\log{(N_\mathrm{HI}/\mathrm{cm}^{-2})} = 22.2$ for Maisie's Galaxy. There is no discussion of any DLA in the fits to the GLASS-z12 NIRspec spectrum shown in \citet{castellano2024}, and the spectrum for the source shows a very sharp Lyman-$\alpha$ break, consistent with the fact that the spectroscopic redshift for this very bright source is very similar to those presented in \citet{castellano2022} and \cite{naidu2022}. Fits to the significantly fainter JADES-GS-z12-0 spectrum in \citet{deugenio2023} require $\log{(N_\mathrm{HI}/\mathrm{cm}^{-2})} = 22.1$ to explain the observed Lyman-$\alpha$ profile. The value for the hydrogen column density we derive for JADES-GS-z11-0 is similar, and well in excess of what is seen for lower-redshift analogues as assembled in \citet{heintz2023}. Although there are still limited sources thus far found with a need for such extreme column densities, it appears that they are preferentially found in less luminous sources.

We can compare JADES-GS-z13-0 with other galaxies in the literature that have been observed to have spectra devoid of emission lines. In \citet{wang2023}, the authors present JWST/NIRSpec PRISM spectra for two galaxies (UNCOVER-z12 and UNCOVER-z13) at $z > 12$ selected from the JWST Treasure Cycle 1 UNCOVER survey \citep{bezanson2022}. Emission lines are not detected in either of these spectra, and the redshifts are derived from fits to the Lyman-$\alpha$ break, with one source at $z_{\mathrm{spec}} = 12.393^{+0.004}_{-0.001}$ and the other at $z_{\mathrm{spec}} = 13.079^{+0.013}_{-0.001}$. While the authors do not provide estimates of $M_{UV}$ for these sources, they do indicate the rest-frame absolute magnitude in the F200W filter at the spectroscopic redshifts, $M_{\mathrm{F200W}} = -19.2 \pm 0.5$ for UNCOVER-z12 and $M_{\mathrm{F200W}} = -19.4 \pm 1.8$ for UNCOVER-z13. In addition, the stellar masses measured by \citet{wang2023} for these sources with \texttt{Prospector} are similar to what we measure for JADES-GS-z11-0 and JADES-GS-z13-0 using the same code, but with higher SFRs. Most notably, however, the gravitational lensing-corrected sizes estimated for these two sources (300 - 400 pc) are 3 - 8$\times$ larger than what we measure for JADES-GS-z11-0 and JADES-GS-z13-0. The lower-redshift source in the \citet{wang2023} sample, UNCOVER-z12, shows evidence for multiple clumps, similar to what we observe with JADES-GS-z11-0 and JADES-GS+53.16474-27.77471. 

The NIRSpec PRISM spectrum for JADES-GS-z10-0 shown in \citet{curtislake2023} has marginal evidence for a Lyman-$\alpha$ emission line at 1.44$\mu$m, which would put this source at $z = 10.84$, higher than the fiducial redshift those authors provide of $z_{\mathrm{spec}} = 10.38$. They provide upper limits on the equivalent widths of \ion{C}{3}], \ion{He}{2}, and [\ion{O}{2}] at this redshift, and these values are in agreement with the EW values we measure for JADES-GS-z11-0, indicating that perhaps this is an effect of the shallower depth of their observations. 

For JADES-GS-z13-0, our modeling indicates that the lack of emission lines is likely due to the high escape fraction of ionizing photons. Our results show that low-metallicity models still result in detectable emission lines, even down to $\log\,Z/Z_\odot \approx -2$, the lowest values explored in our grids. Observations of other galaxies at similar redshifts which do show emission lines reveal that these sources are metal-enriched, although still at significantly sub-solar values \citep[$<0.17 Z/Z_{\odot}$, ][]{bunker2023a, deugenio2023, zavala2024}, so we cannot fully rule out that lower metallicity outside the range of our models plays a part for JADES-GS-z13-0. Additional key information could possibly come from observing these high-redshift sources at even longer wavelengths, for instance with JWST/MIRI, to seek evidence for H$\beta$ and/or [\ion{O}{3}]$\lambda$5007 emission. In addition, it will take significantly larger populations of galaxies spectroscopically confirmed at these redshifts to understand whether the lack of lines is due to their SFH having a relative lull at observation, similar to the mini-quenched galaxy seen at $z = 7.3$ in \citet{looser2023}. 

Our results, along with those in the literature with significant predicted DLA absorption, demonstrate the uncertainties in estimating redshifts from UV spectra without emission lines. The DLA fits shown for JADES-GS-z13-0 in Figure \ref{fig:DLA_GSz13} demonstrate the degeneracy between HI column density and redshift, such that redshifts for this source could vary as much as 0.2-0.3. For samples of sources with photometric redshifts, or those selected entirely by colors spanning the Lyman-$\alpha$ break, this would have an effect of moving objects to higher redshift bins, significantly affecting any recovered evolution of the cosmic star-formation rate density or luminosity function. In addition, these absorbers can strongly affect the recovery of the UV slope in high-redshift objects. Indeed, new or updated codes should be developed to help account for this absorption. 

We also find evidence that both JADES-GS-z11-0 and JADES-GS-z13-0 are spatially resolved, from both the \texttt{ForcePho} fits and from the radial profiles plotted in Figure \ref{fig:radial_profiles}. These results stand in contrast to the speculation that these sources are supermassive ``dark stars'' presented in \citet{ilie2023}, as this model would require these sources to be unresolved. Our results indicate that both JADES-GS-z11-0 and JADES-GS-z13-0 have radial profiles significantly in excess of the PSF, and are unlikely to be unresolved stars. 

In conclusion, we present significantly deeper spectra and updated photometry for the ultra-high redshift galaxies JADES-GS-z11-0 and JADES-GS-z13-0. We find:

\begin{enumerate}
    \item The PRISM spectrum for JADES-GS-z11-0, combined with a careful assessment of the spectral errors, reveals multiple weak emission lines that indicate a redshift of $z_{\mathrm{spec}} = 11.122^{+0.005}_{-0.003}$. This redshift derived from emission lines is lower than what would be predicted from fitting the Lyman-$\alpha$ break alone.
    \item The PRISM spectrum for JADES-GS-z13-0 does not show any emission features, and we estimate a redshift of $z_{\mathrm{spec}} = 13.20^{+0.03}_{-0.04}$ from a fit to the Lyman-$\alpha$ break.  
    \item We use both {\tt BEAGLE} and {\tt Prospector} to fit the spectra for these sources, and find stellar masses that range from $\log{(M_*/M_\odot)} = 7.8 - 8.4$, with low stellar and gas phase metallicities and little to no dust content. The UV slopes and SFRs we derive indicate that both sources are actively star-forming, consistent with previous results for the objects in \citet{curtislake2023} and \citet{robertson2023}.
    \item Updated NIRCam photometry and fits provide further evidence of the small sizes of these galaxies (half-light radii of 119 pc for JADES-GS-z11-0 and 59 pc for JADES-GS-z13-0). Both are resolved above the PSF. 
    \item We demonstrate that additional damped Lyman-$\alpha$ absorption ($\log{(N_\mathrm{HI}/\mathrm{cm}^{-2})} = 22.43^{+0.10}_{-0.12}$) can explain the shape of the Lyman-$\alpha$ break at the fiducial redshift for JADES-GS-z11-0. However, we don't find that a damped Lyman-$\alpha$ absorber is necessary for fitting the spectrum for JADES-GS-z13-0, as its Lyman-$\alpha$ profiles can be fit with a high neutral fraction of hydrogen gas.
    \item The photometric redshifts we derive are systematically high for both of these sources because of the treatment of the Lyman-$\alpha$ profile in the photometric redshift code used. Higher neutral fractions and potential DLA absorption can produce smoother Lyman-$\alpha$ break profiles that are not accounted for, pushing the photometric redshifts higher. 
    \item We uncover evidence for a secondary source $\sim 0.3^{\prime\prime}$ south of JADES-GS-z11-0 with a similar photometric redshift. Both sources may be part of an interacting pair similar to others seen in the early Universe in \citet{hainline2023}. 
\end{enumerate}

These results demonstrate the highly complex nature of star formation and its effect on the observed UV spectra within galaxies from the first few hundred million years after the Big Bang. It is vital to obtain additional deep spectra of sources at these redshifts to understand the evolution of DLA absorption in these systems, especially given the need to update photometric redshift and stellar population synthesis codes to account for these effects. Future deep NIRSpec/MOS campaigns following up on samples of high-redshift candidates will go a long way towards helping understand the way in which galaxies drove cosmic reionization.

%\begin{acknowledgments}
$ $\\

K.H., Z.J., B.J., J.H., M.R., F.W., and C.N.A.W are funded in part by the JWST/NIRCam contract to the University of Arizona NAS5-02015.
F.D.E., R.M., J.W., and J.S. acknowledge support by the Science and Technology Facilities Council (STFC), by the ERC through Advanced Grant 695671 “QUENCH”, and by the UKRI Frontier Research grant RISEandFALL.
The Cosmic Dawn Center (DAWN) is funded by the Danish National Research Foundation under grant DNRF140.
S.C acknowledges support by European Union’s HE ERC Starting Grant No. 101040227 - WINGS.
ECL acknowledges support of an STFC Webb Fellowship (ST/W001438/1)
BER acknowledges support from the NIRCam Science Team contract to the University of Arizona, NAS5-02015, and JWST Program 3215.
ST acknowledges support by the Royal Society Research Grant G125142.
S.A. acknowledges support from Grant PID2021-127718NB-I00 funded by the Spanish Ministry of Science and Innovation/State Agency of Research (MICIN/AEI/ 10.13039/501100011033). 
AJB, AJC, and JC acknowledge funding from the "FirstGalaxies" Advanced Grant from the European Research Council (ERC) under the European Union’s Horizon 2020 research and innovation programme (Grant agreement No. 789056)
DJE is supported as a Simons Investigator and by JWST/NIRCam contract to the University of Arizona, NAS5-02015
Funding for this research was provided by the Johns Hopkins University, Institute for Data Intensive Engineering and Science (IDIES)
RM also acknowledges funding from a research professorship from the Royal Society.
PGP-G acknowledges support from grant PID2022-139567NB-I00 funded by Spanish Ministerio de Ciencia e Innovaci\'on MCIN/AEI/10.13039/501100011033, FEDER, UE.
The research of CCW is supported by NOIRLab, which is managed by the Association of Universities for Research in Astronomy (AURA) under a cooperative agreement with the National Science Foundation.

%\end{acknowledgments}

\appendix
\section{Statistical Redshift Determination for NIRSpec Prism Observations}\label{sec:appendix_redshift}

The NIRSpec spectra from the JWST PID 1210 and 3215 programs are notable for their very long exposure times, and the fact that they are constructed from a large number of nominally identical 19 frame (1400~s) NRSIRS2 mode sub-exposures that are reduced separately by the GTO pipeline and then co-added to produce the final spectrum. The JADES-GS-z11-0 PRISM spectrum presented in this paper was created from of a total of 72 sub-spectra taken in PID 1210 and 114 taken in PID 3215, resulting in a total of 186 sub-spectra and a combined exposure time of 72.3 h. The equivalent total number of sub-spectra of JADES-GS-z13-0 is 138 (53.6 h), with 24 from PID 1210 and 114 from PID 3215. 

Having this many independent sub-spectra available provides a unique opportunity to directly measure the actual level of statistical noise present in NIRSpec spectra and quantify the significant correlation occurring between adjacent wavelength bins (Jakobsen et al. 2024, in preparation). In particular it allowed us to determine the covariance matrix of the two prism spectra used as the noise model in the two detailed fits of Section~\ref{sec:dla-fits}. Here we exploit these multiple sub-spectra to perform an automated search for the systemic redshifts of JADES-GS-z11-0 and JADES-GS-z13-0 by looking for statistically significant redshift matches among any weak emission lines present in the spectra. 

Starting with a 674 wavelength-bin combined NIRSpec PRISM spectrum with wavelength $\lambda(i)$ and flux $F_\lambda(i)$, the first step of the process is to determine the reference continuum level $F_\lambda^c(i)$. Since there are no obvious strong emission lines visible in the spectra of our targets, we calculated the continuum level by boxcar-smoothing $F_\lambda(i)$ with a variable box width that follows the uneven dispersion of the PRISM spectra and varies between 9 and 87 wavelength bins across the spectrum. Starting at the wavelength of the onset of the Gunn-Peterson trough, a fixed five-pixel-wide window was then shifted across the spectrum, and the strength of any narrow emission line present at the central wavelength $\lambda(i)$ of the window was quantified by the total summed excess flux above the continuum contained within the window, and the equivalent width of this excess signal:
\begin{equation}
F_l(i) = \sum_{j=i-2}^{i+2} (F_\lambda(j)-F_\lambda^c(j))\Delta \lambda(j)
\end{equation}
and 
\begin{equation}
W_\lambda(i) = \sum_{j=i-2}^{i+2} \frac{(F_\lambda(j)-F_\lambda^c(j))}{F_\lambda^c(j)}\Delta \lambda(j)
\end{equation}
where $\Delta \lambda(i)$ is the width of wavelength bin $i$ in the spectrum. A fixed-sized extraction window could be used since the PRISM spectra are unevenly sampled at a wavelength binning that closely mirrors the uneven dispersion and native pixel sampling on the detector. As a consequence, the Line Spread Function of NIRSpec PRISM spectra has a FWHM corresponding to around 3.0 local wavelength bins at all wavelengths such that a fixed 5 wavelength bin wide window is adequate to capture narrow emission lines at all wavelengths. 

The statistical error on the measured line flux and equivalent width at each wavelength was then determined by repeating the same measurements on 2000 bootstrapped versions of the combined spectrum $F_\lambda(i)$ drawn from the 186 and 138 available sub-spectra (with replacement), and determining the sample scatter seen in $F_l(i)$ and $W_\lambda(i)$ at each wavelength bin $i$ \citep[cf.][]{Efron21}. The outcomes are the error arrays $\sigma F_l(i)$ and $\sigma W_\lambda(i)$. It should be noted that these empirical bootstrapped statistical errors should be considered as more reliable than the errors on $F_l(i)$ and $W_\lambda(i)$ calculated from the estimated error spectrum output by the pipeline processing, in that they measure the actual statistical fluctuations in the quantities $F_l(i)$ and $W_\lambda(i)$ due to all sources of noise, and implicitly take into account the significant inter-bin correlation present in NIRSpec spectra. 

\begin{deluxetable}{l l l l l l}
\tabletypesize{\footnotesize}
\tablecolumns{6}
\tablewidth{0pt}
\tablecaption{Emission lines used in the redshift search, with their `effective' vacuum wavelength, i.e., averaging over spectrally unresolved multiplets.\label{tab:lines} }
\tablehead{
\colhead{Emission line(s)} & \colhead{$\lambda_l$[\AA]} & \colhead{Emission line(s)} & \colhead{$\lambda_l$[\AA]} & \colhead{Emission line(s)} & \colhead{$\lambda_l$[\AA]}
}
\startdata
\ion{N}{4}]$\lambda\lambda$1483,1486 &  1486 &   \ion{Mg}{2}$\lambda\lambda$2796,2803  &  2799  & H$\gamma$                   &  4342\\
\ion{C}{4}$\lambda\lambda$1548,1551  &  1549 &   [\ion{O}{2}]$\lambda\lambda$3726,3729 &  3728  & [\ion{O}{3}]$\lambda$4363 &  4364\\
\ion{He}{2}$\lambda\,$1640             &  1640 &   [\ion{Ne}{3}]$\lambda$3869            &  3870  & H$\beta$                    &  4863\\
\ion{O}{3}]$\lambda\lambda$1661,1666 &  1663 &   [\ion{Ne}{3}]$\lambda$3968            &  3969  & [\ion{O}{3}]$\lambda$4959 &  4960\\
\ion{N}{3}]$\lambda\lambda$1747--1754 &  1754 &   H$\epsilon$                             &  3970  & [\ion{O}{3}]$\lambda$5007 &  5008\\
\ion{C}{3}]$\lambda\lambda$1907,1909 &  1909 &   H$\delta$                               &  4103  & H$\alpha$                   &  6565\\
\enddata
\end{deluxetable}

The arrays $\lambda(i)$, $F_l(i)$, $\sigma F_l(i)$ together capture the signal strength and statistical significance of any narrow emission line potentially detected anywhere in the spectrum, and form the basis of the line strengths reported in Section \ref{sec:fits}. A statistical search for the systemic redshifts of JADES-GS-z11-0 and JADES-GS-z13-0 was performed by passing a ``comb'' of the most common emission lines seen in high-redshift galaxy spectra though the (signed) signal-to-noise array $SN(i) = F_l(i)/\sigma F_l(i)$, and quantifying the statistical significance of the coincidences occurring among the emission lines searched for as a function of redshift. That is, for a given probed redshift $z$, for each line listed in Table~\ref{tab:lines} the signal-to-noise ratio $SN_l(j)$ of any line possibly present at its redshifted wavelength $\lambda_l(j)(1+z)$ was determined through interpolation in the $\lambda(i)$, $SN(i)$ arrays. The statistical significance of each line being present in the spectrum was then assigned the one-sided $p$-value $p_l(j) = 1-\Phi(SN_l(j))$ where $\Phi(x)$ is the cumulative normal distribution. $p_l(j)$ gives the probability that the value of $SN_l(j)$ or greater is reached in the spectrum at the redshifted location of the emission line under the null hypothesis that there are no emission lines present in the spectrum. A line reaching $SN_l=2.0$ is therefore assigned a $p$-value of $p_l=0.0227$, while a line achieving a negative value of $SN_l=-2.0$ is assigned $p_l=0.9772$. Since we are only interested in searching for narrow emission lines that are not all required to be physically present in the spectrum, such asymmetric one-sided $p$-values are appropriate.

In practice, the search for weak emission lines is carried out at wavelengths between the onset of the Gunn-Peterson trough and out to the $\lambda=5.3$~$\mu$m red cutoff of the PRISM spectra. If we find $k$ redshifted line candidates from Table~\ref{tab:lines} falling in this wavelength interval, their individual $p$-values are combined into a single statistic, $X_T$, using Fisher's method:
\begin{equation}
X_T= -2 \sum_{j=1}^k \ln  p_l(j)   \sim \chi^2_{2k}
\end{equation}
Under the null hypothesis of no lines, $X_T$ will be Chi-squared distributed with $2k$ degrees of freedom \citep{Fisher50}.

Possible values for the systemic redshift will reveal themselves as statistically significant peaks in $X_T(z)$ when evaluated over a continuous range of plausible redshifts spanning $z=z_{GP}\pm0.5$, where $z_{GP}$ is the redshift of Lyman-$\alpha$ at the midpoint of the onset of the Gunn-Peterson trough. Note that the search needs to be extended to both sides of $z_{GP}$ since the systemic redshift of the galaxy may lie below $z_{GP}$ if there is a local Damped Lyman Alpha absorber present, and above $z_{GP}$ if the galaxy resides in a local ionized bubble which shifts the onset of the intergalactic absorption to shorter wavelengths.

\begin{figure}
\centering
  \includegraphics[width=0.7\linewidth]{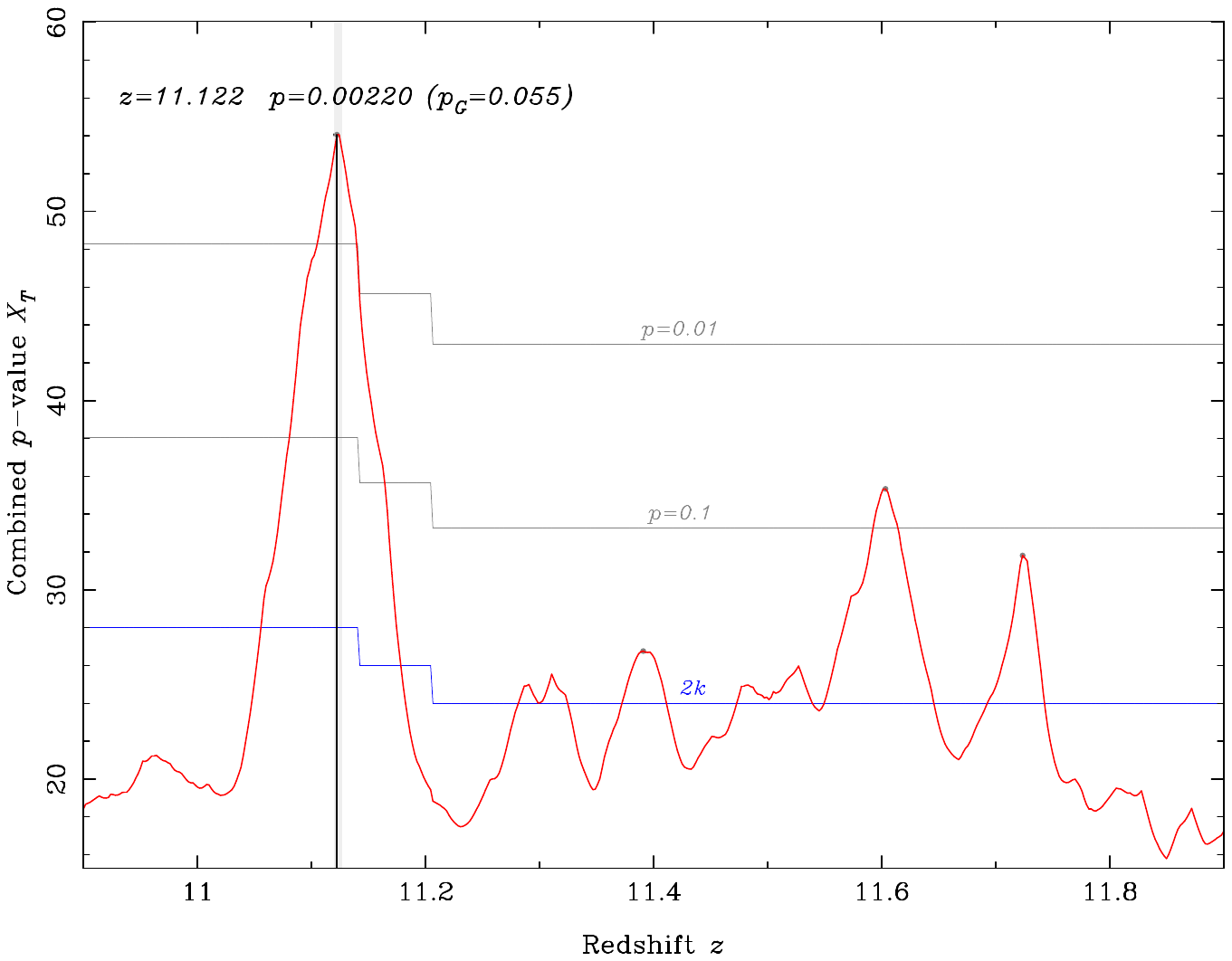}\\
\caption{The combined emission line detection $p$-value $X_T$ as a function of redshift for JADES-GS-z11-0. The prominent peak at $z=11.122$ is evident.}
  \label{fig:gsz11_peak}
\end{figure}

\begin{figure}
\centering
  \includegraphics[width=0.7\linewidth]{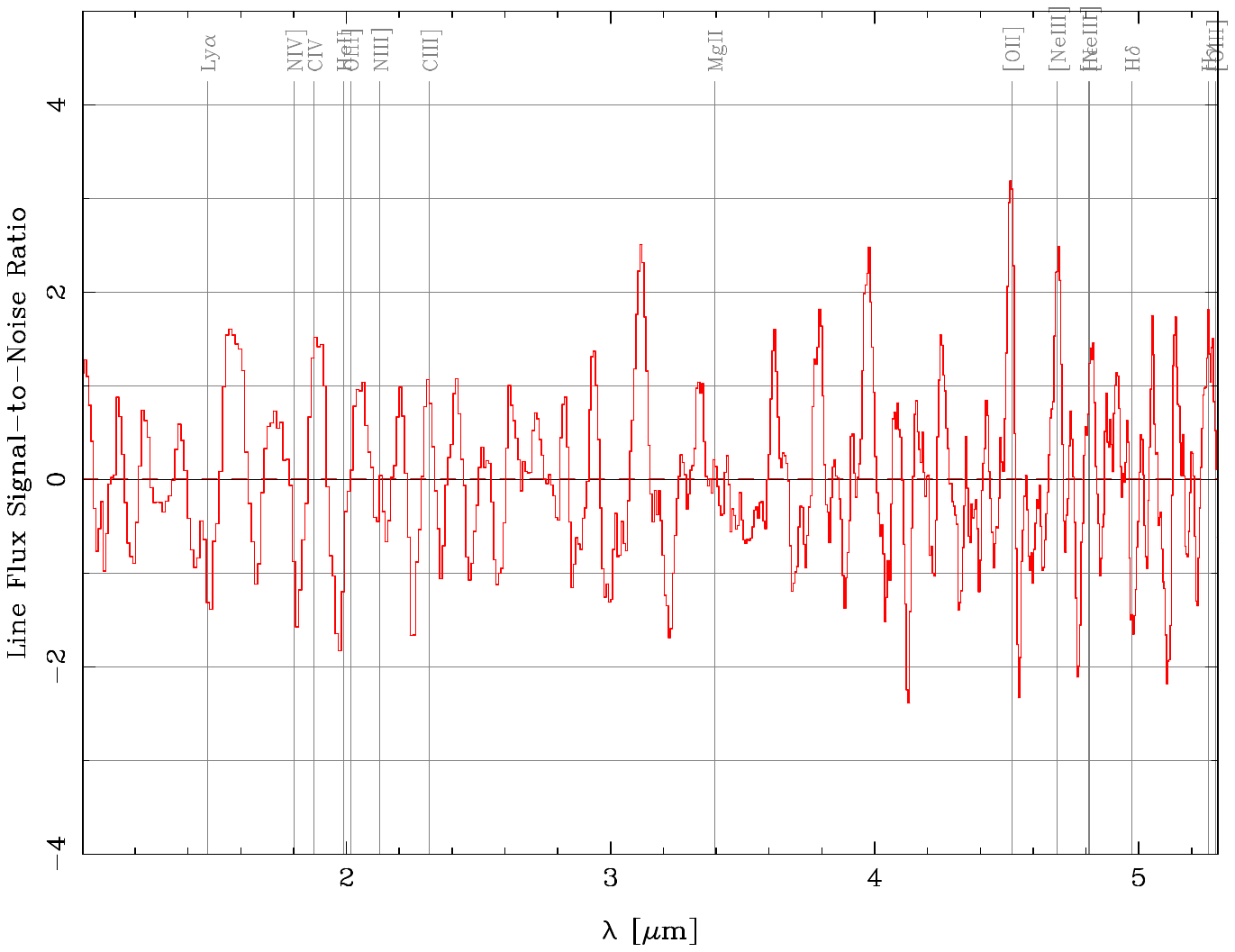}\\
\caption{The potential line flux signal-to-noise ratio versus wavelength for JADES-GS-z11-0. The overlay shows the locations of the emission lines searched for at the peak redshift of $z=11.122$.}
  \label{fig:gsz11_sn}
\end{figure}

The outcome of this redshift sweep of the JADES-GS-z11-0 spectrum is shown in Figure~\ref{fig:gsz11_peak}. It is evident that the combined $p$-value displays a prominent peak at $z=11.122$, reaching an amplitude of $X_T=54.06 \sim \chi^2$ for 26 degrees of freedom. Figure~\ref{fig:gsz11_sn} plots $SN(i)=F_l(i)/\sigma F_l(i)$ as a function of $\lambda(i)$, with the search lines overlaid for this peak redshift. It is seen that the $z=11.122$ peak in $X_T$ is made up of a set of matching weak emission lines consisting of an [\ion{O}{2}] line measured at 3.11$\sigma$, a pair of [\ion{Ne}{3}] lines measured at 2.21$\sigma$ and 0.95$\sigma$, H$\gamma$ and H$\epsilon$ measured  at 1.81$\sigma$ and 1.15$\sigma$, \ion{C}{4} measured at 1.41$\sigma$ and \ion{C}{3}] measured at 0.92$\sigma$. While these lines are not overwhelmingly significant when considered individually, when taken together they do provide good evidence for JADES-GS-z11-0 having a systemic redshift of $z=11.122$. Given the measured noise level in the spectrum, the probability of such a coincidence of erroneous noise spikes giving rise to a peak reaching $X_T=54.06$ occurring purely by chance is only $p_L=0.0022$.

\begin{figure}
\centering
  \includegraphics[width=0.7\linewidth]{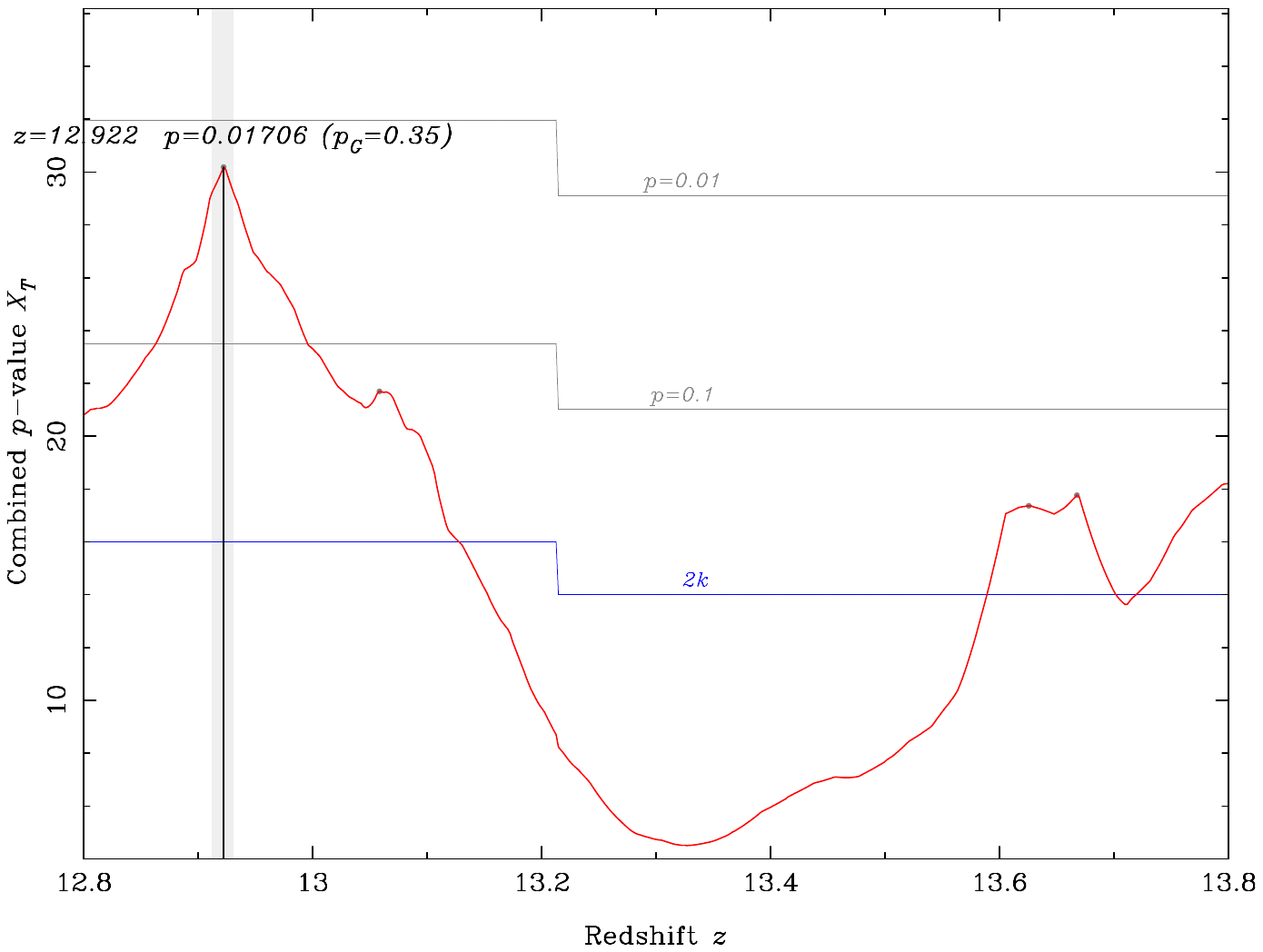}\\
\caption{The combined emission line detection $p$-value $X_T$ as a function of redshift for JADES-GS-z13-0. The peak seen at $z=12.922$ is likely a chance occurrence.}
  \label{fig:gsz13_peak}
\end{figure}

\begin{figure}
    \centering
  \includegraphics[width=0.7\linewidth]{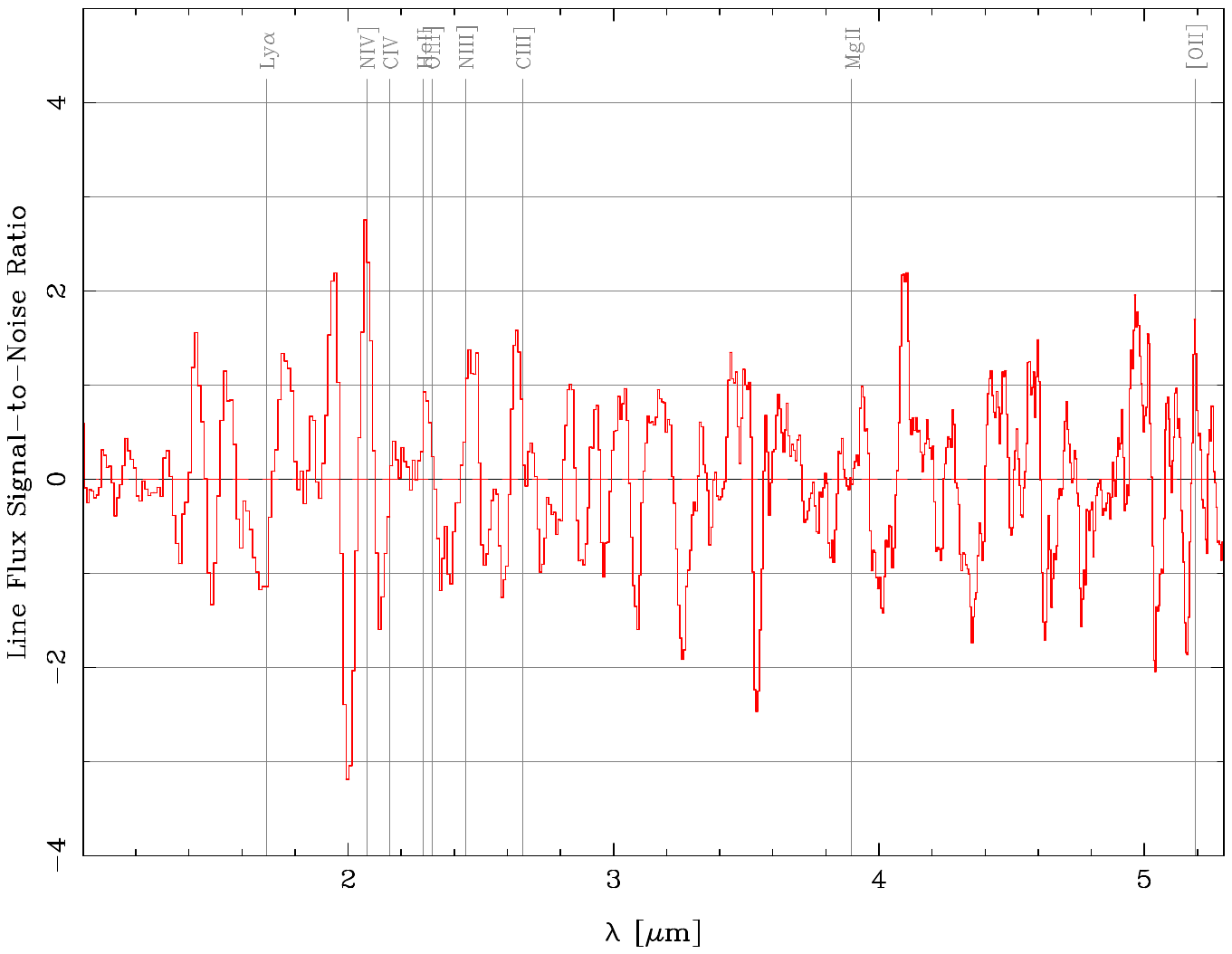}\\
  \caption{The line flux signal-to-noise ratio versus  wavelength for JADES-GS-z13-0. The overlay shows the locations of the emission lines searched for at the peak redshift of $z=12.922$.}
  \label{fig:gsz13_sn}
\end{figure}

However, a well-known issue with this type of automated systematic search is that it probes a range in redshift and therefore samples the $X_T$ statistic multiple times when searching for peaks, thereby increasing the probability that an extreme random excursion may be encountered by chance. This issue is commonly referred to as the ``Look-elsewere'' effect. If $N$ denotes the effective number of independent probes of $X_T$ carried out in the trial, the relevant probability to consider is the global probability of the trial encountering one or more peaks reaching a local $p$-value less or equal to the observed peak value $p_L$:
\begin{equation}
p_G = 1 - (1-p_L)^{N} \simeq N p_L 
\end{equation}
We can attempt to estimate the number of independent probes $N$ as follows. The line-matching in JADES-GS-z11-0 takes place at wavelengths between the onset of the Gunn-Peterson trough at 1.5~$\mu$m and out to 5.3~$\mu$m. The maximum spectral resolution of the NIRSpec PRISM occurs at the red end of this range where it reaches $R=\lambda/\delta\lambda\simeq 300$ \citep{jakobsen2022}. Consequently, at $z_{GP}=11.4$ a change in redshift of $\delta z \simeq \delta\lambda/\lambda (1+z) = 0.041$ will cause the reddest candidate lines contributing to $X_T$ to move off any features present in the spectrum. This implies that the effective number of independent redshifts sampled  over the total $\Delta z=1.0$ range probed is approximately $N\simeq \Delta z/\delta z = 24.4$. As an alternative more stringent approach, \cite{Bayer21} describe a method for self-calibrating trials such as ours for $N$ by comparing the amplitude of the highest peak to those of the second, third and so forth highest peaks. Applying their recipe to the three highest secondary peaks in Figure~\ref{fig:gsz11_peak} yields the values $N=23.6$, $N=25.1$ and $N=27.6$. The average value of $N=25.4$ is in remarkably good agreement with the rough estimate above. In view of this agreement, we can be reasonably confident that the local $p$-value of $p_L=0.0022$ for the peak at $z=11.122$, corresponds to a still significant global $p$-value of $p_G=N p_L = 0.056$. This is the basis for our concluding that JADES-GS-z11-0 lies at this redshift with 94\% confidence.  

We tested the robustness of this finding by carrying out five test trials in which the 186 available sub-spectra of JADES-GS-z11-0 were randomly split into two halves, and each set of 93 sub-exposures was processed in exactly the same way as the full data set. Eight out of the ten redshift scans performed on the resulting half data sets displayed primary peaks well within the anticipated overall accuracy of our approach of $\delta z\simeq \pm \Delta z/N=\pm0.039$ of $z=11.122$ . The two exceptions both displayed prominent secondary peaks in that redshift range, but their slightly lower $p-values$ had been edged out by the algorithm having ascribed a redshift to a single clearly spurious strong emission line that was not evident in any of the other trials. Since such single line redshifts cannot be excluded apriori, they need to be identified and assessed through visual inspection.

Our findings for JADES-GS-z13-0 are less fruitful. The $X_T(z)$ plot derived from its PRISM data is shown in Figure~\ref{fig:gsz13_peak}. In this object, the strongest peak at $z=12.922$ has an amplitude of $X_T=30.19$ for 16 degrees of freedom, corresponding to a local $p$-value of $p_L=0.017$. The corresponding overlaid signal-to-noise plot in Figure~\ref{fig:gsz13_sn} reveals that this peak is dominated by a match between a 2.54$\sigma$ detection of \ion{N}{4}] and a 1.67$\sigma$ detection of [\ion{O}{2}]. The physical plausibility of this match aside, for $N=25.4$ the local $p$-value of this peak corresponds to a global $p$-value of $p_G=0.35$, indicating that this match is likely a chance coincidence. We therefore conclude that our search has failed to determine the systemic redshift of JADES-GS-z13-0 on the basis of weak absorption lines in its PRISM spectrum. 

\section{Supplemental Figures}\label{sec:supp_figures}

\begin{figure}
  \centering
  \includegraphics[width=0.98\linewidth]{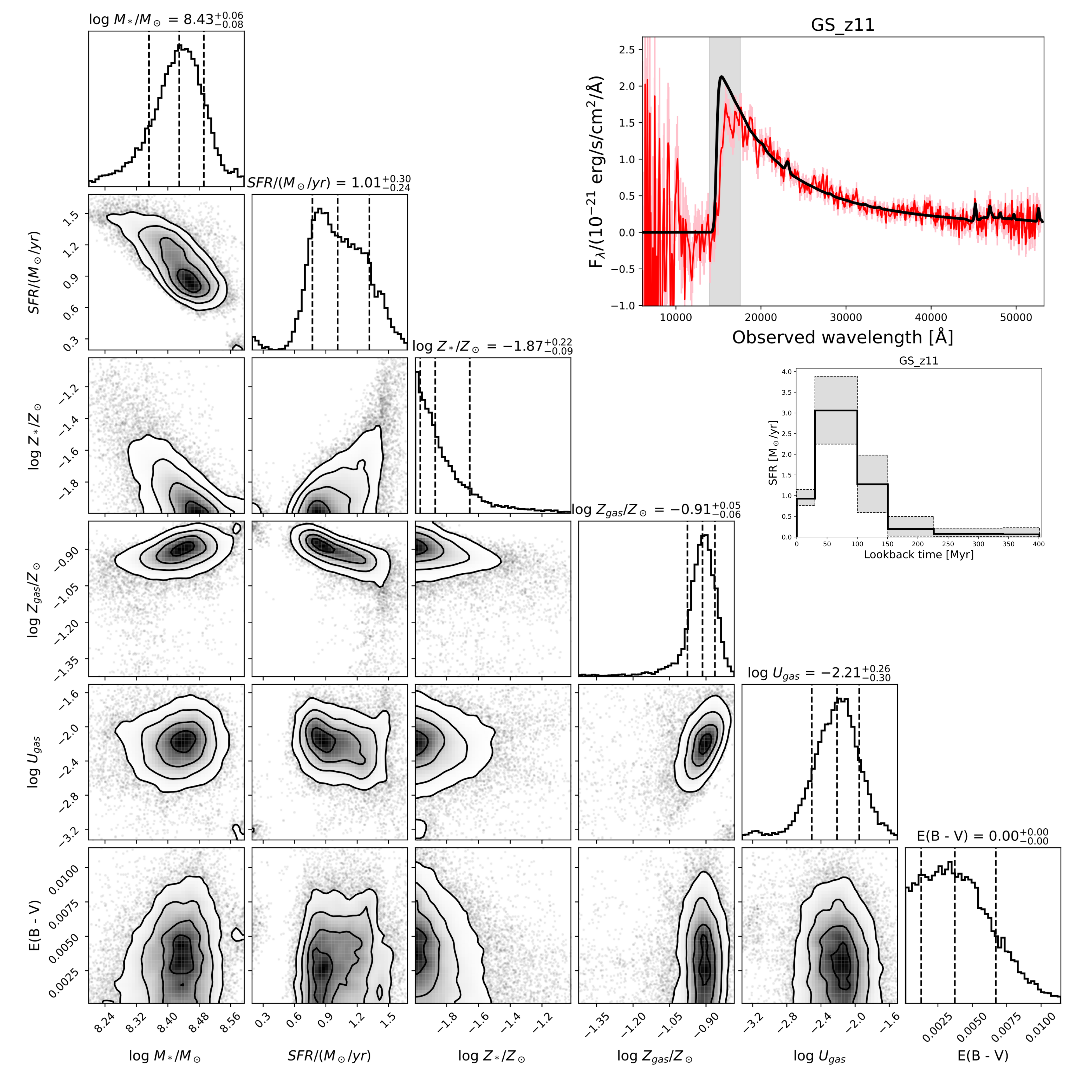}\\
  \caption{Corner plot (left), SED fit (top), and star-formation history (middle) for JADES-GS-z11-0 from \texttt{Prospector}. In the corner plot, the first column from the left is galaxy redshift, the second column is stellar mass, the third column is the stellar metallicity, the fourth column is gas-phase metallicity, the fifth column is the ionization parameter of the gas, and the sixth column is the V-band optical depth for the older ($>10$Myr) stellar population.}
  \label{fig:GS_z11_Prospector}
\end{figure}

\begin{figure}
  \centering
  \includegraphics[width=0.98\linewidth]{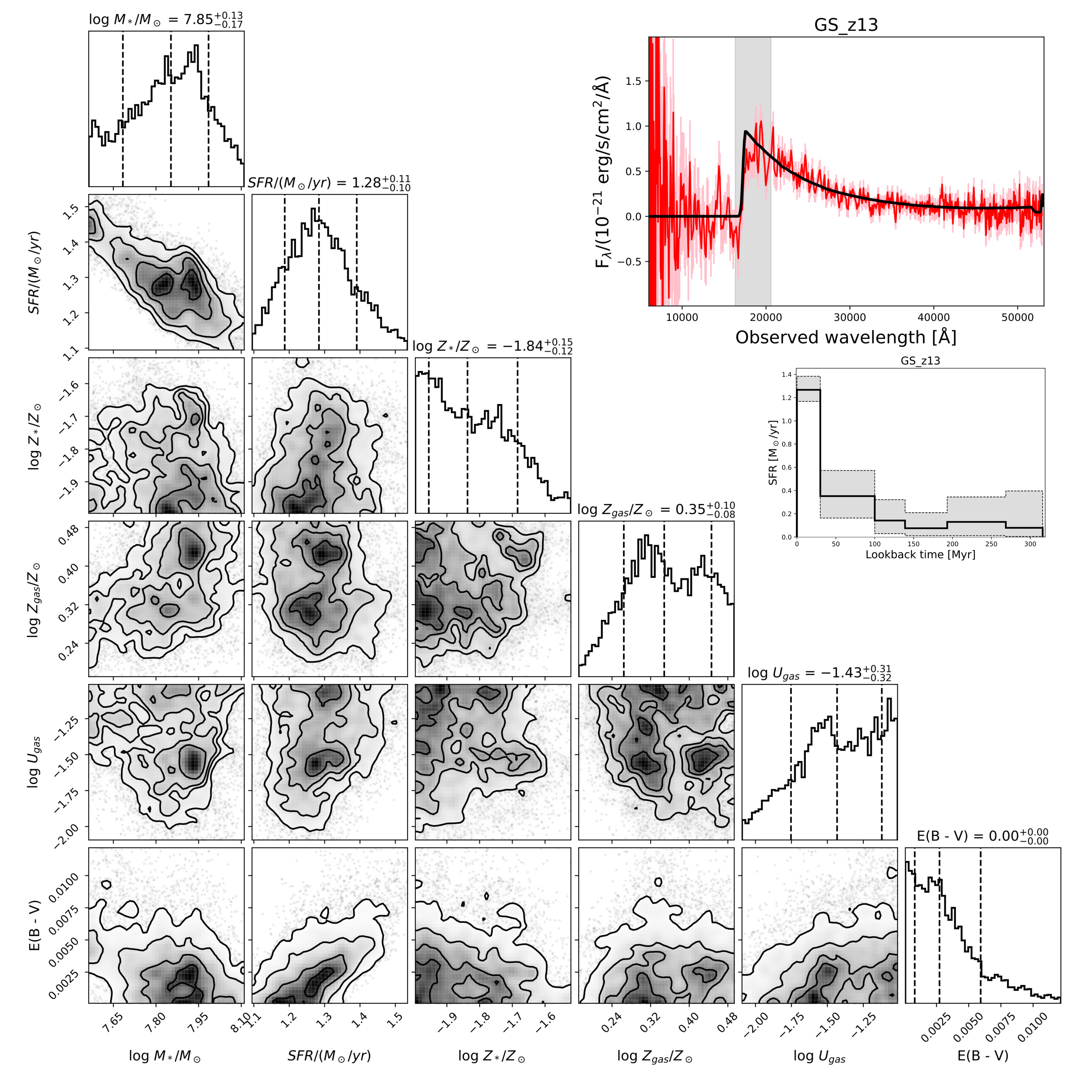}\\
  \caption{Corner plot (left), SED fit (top), and star-formation history (middle) for JADES-GS-z13-0 from Prospector, with columns and description as in Figure \ref{fig:GS_z11_Prospector}.}
  \label{fig:GS_z13_Prospector}
\end{figure}

\begin{figure}
  \centering
  \includegraphics[width=0.4\linewidth]{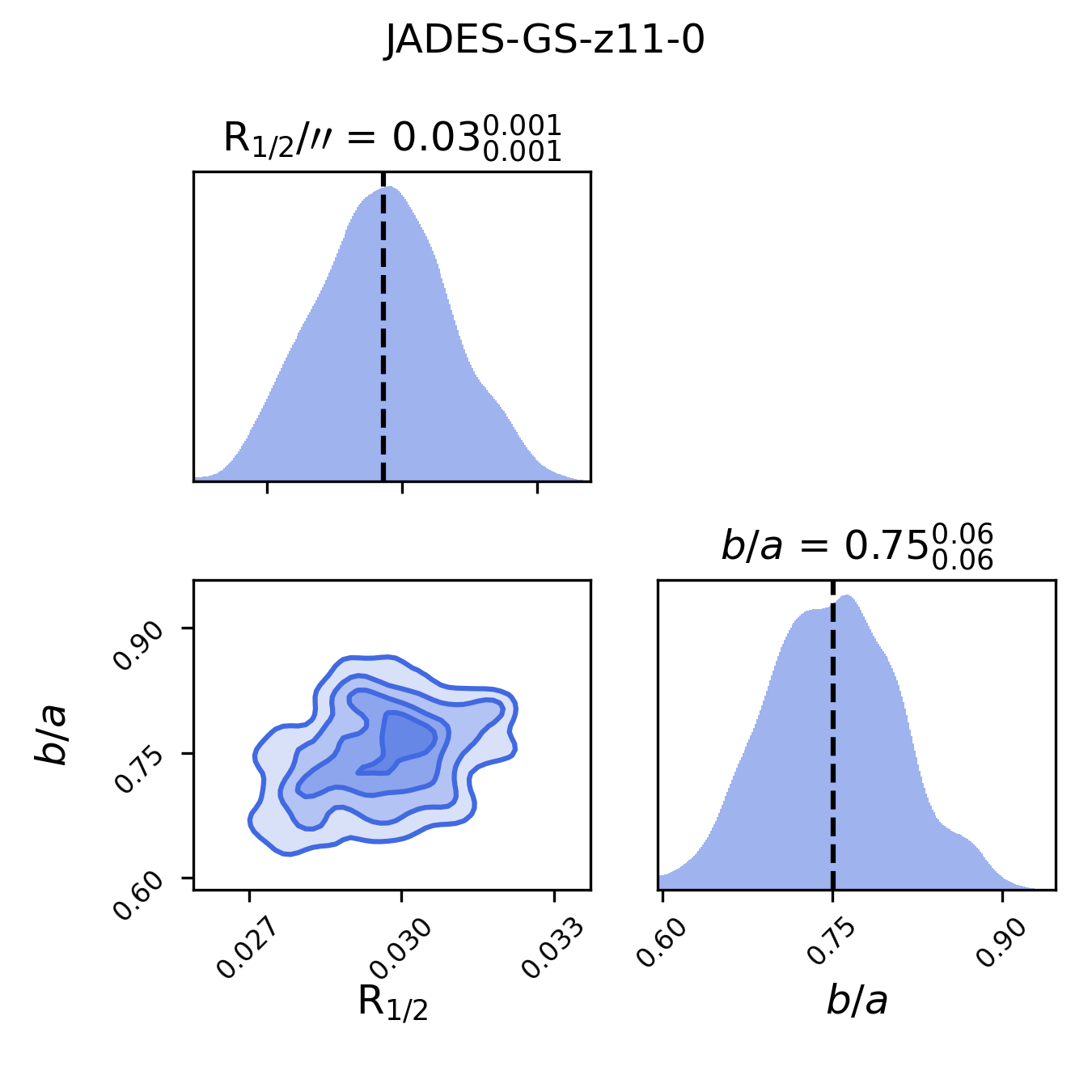}
  \includegraphics[width=0.4\linewidth]{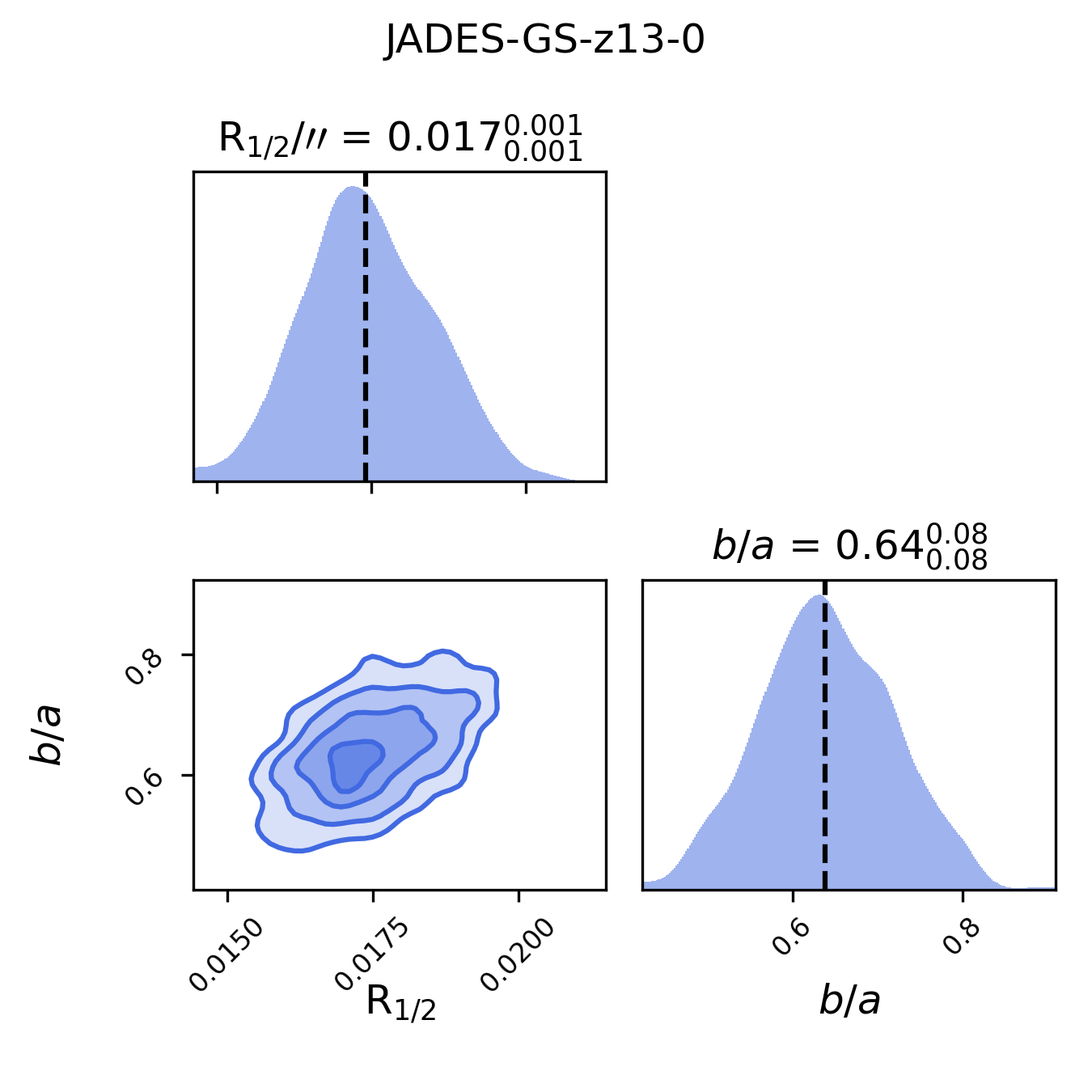}\\
  \caption{\texttt{ForcePho} marginalized and joint posterior distribution for the half-light radius and semiminor to semimajor axis ratios $b/a$ for JADES-GS-z11-0 (left) and JADES-GS-z13-0 (right).}
  \label{fig:half_light_sersic}
\end{figure}

\begin{figure*}
  \centering
  \includegraphics[width=0.95\linewidth]{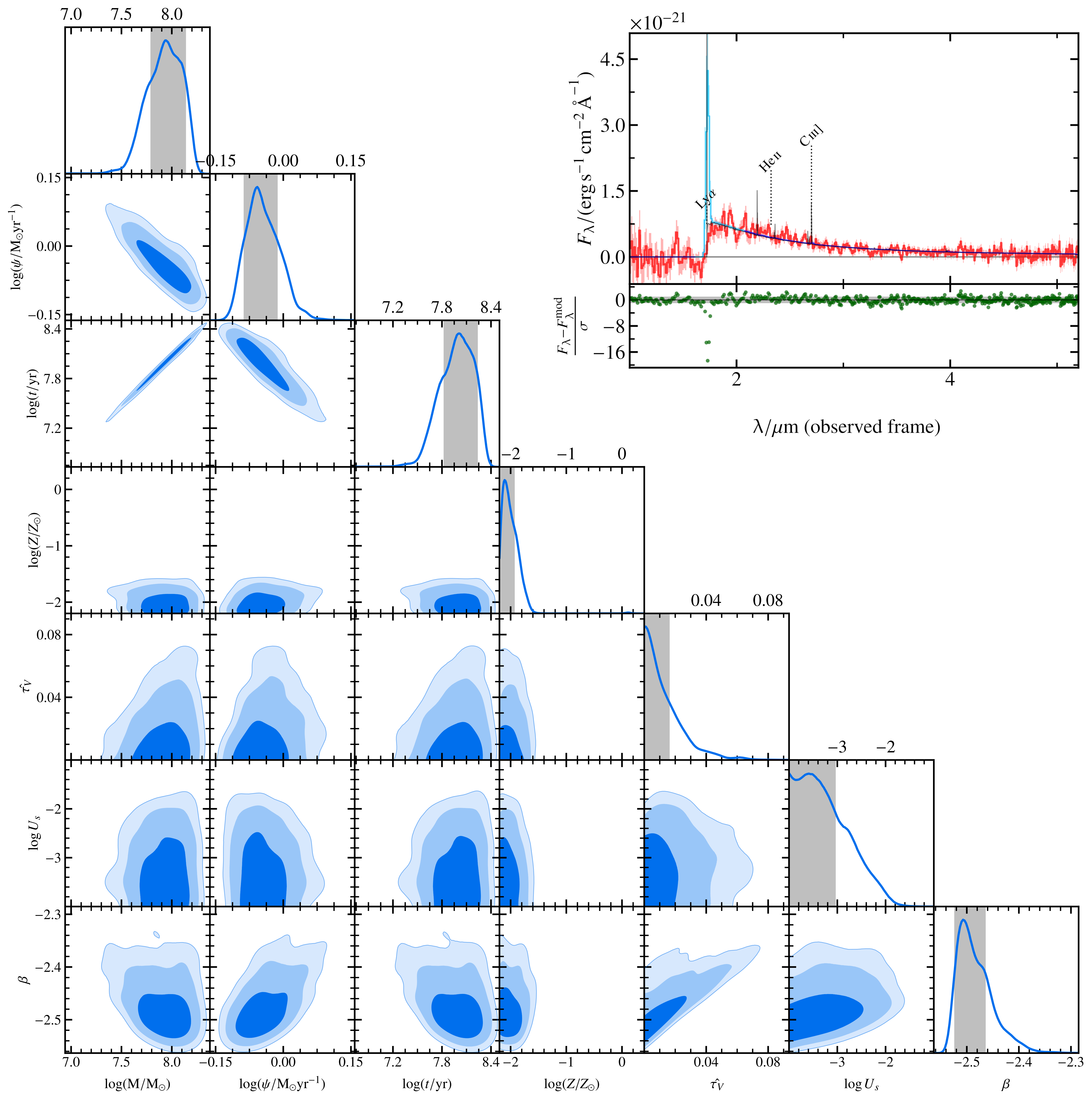}  
  \caption{As in Figure~\ref{fig:beagle_triangle_gsz13}, \texttt{BEAGLE} predictions and posterior probability distributions for JADES-GS-z13-0, but assuming a model with constant star formation history and no escape of ionizing photons.}
  \label{fig:JADES-GS-z13-0-constant-SFH}
\end{figure*}

\begin{figure*}
  \centering
  \includegraphics[width=0.95\linewidth]{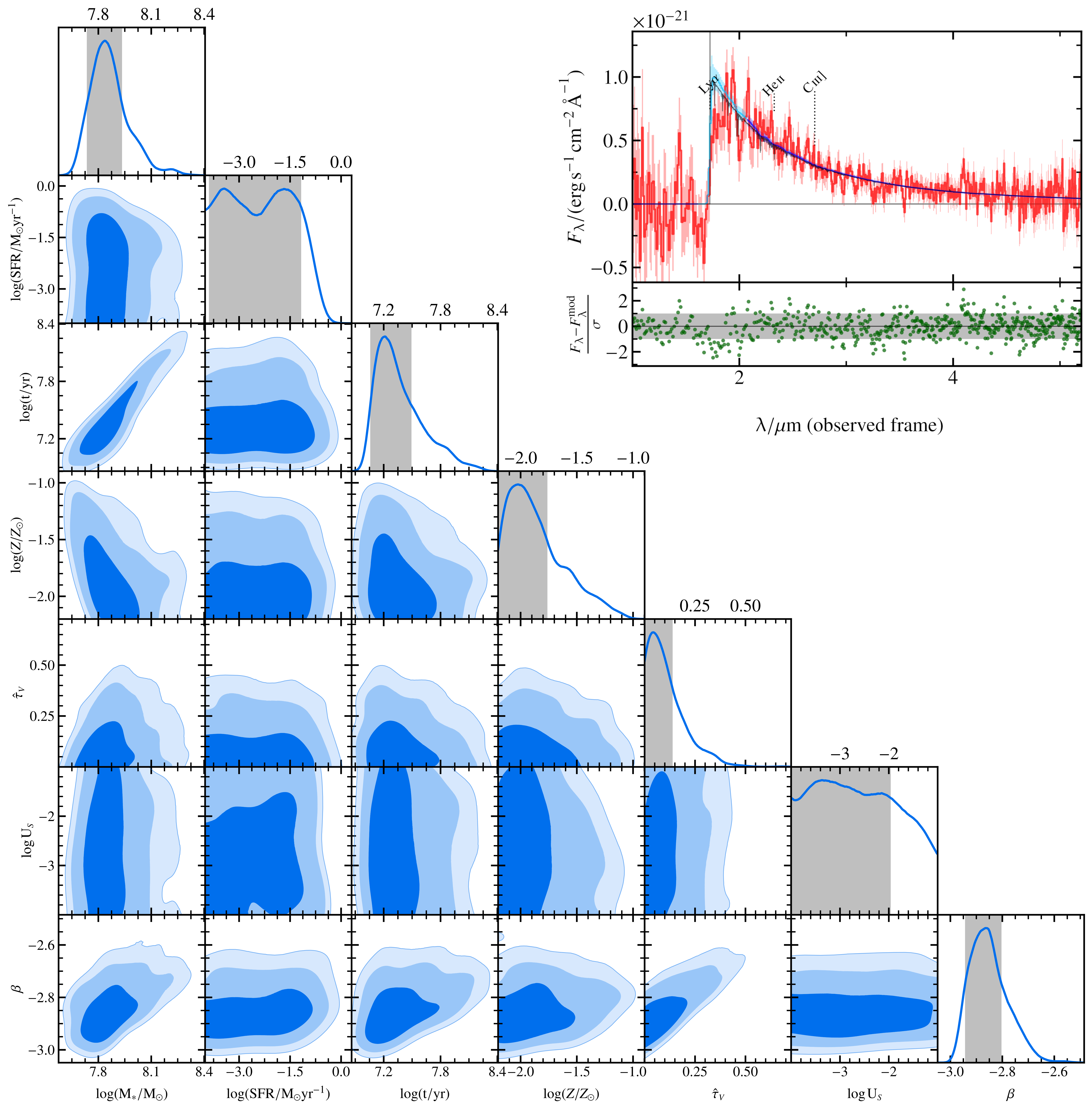}  
  \caption{As in Figure~\ref{fig:beagle_triangle_gsz13}, \texttt{BEAGLE} predictions and posterior probability distributions for JADES-GS-z13-0, but assuming a model with a delayed exponential star formation history plus a 10 Myr burst, and no escape of ionizing photons.}
  \label{fig:JADES-GS-z13-0-delayed-SFH}
\end{figure*}

In this Section we provide figures described in the text that supplement the analysis. In Figures  \ref{fig:GS_z11_Prospector} and \ref{fig:GS_z13_Prospector} we plot the posteriors, SED plots, and SFHs for JADES-GS-z11-0 and JADES-GS-z13-0 derived from \texttt{Prospector} as discussed in Section \ref{sec:sed-fitting}. In Figure \ref{fig:half_light_sersic} we show the marginalized and joint posterior plots for the {\tt ForcePho}-derived half-light radii and axis ratios for JADES-GS-z11-0 and JADES-GS-z13-0, as discussed in Section \ref{sec:nircam-photometry}. In Figures \ref{fig:JADES-GS-z13-0-constant-SFH} and \ref{fig:JADES-GS-z13-0-delayed-SFH} we plot corner plots and SED fits from \texttt{BEAGLE} for JADES-GS-z13-0, which we discuss in Section \ref{sec:gs-z13-no-emission}. 

\vspace{5mm}
\facilities{JWST(NIRCam, NIRSpec), HST(ACS)}

\clearpage

\bibliography{gsz11_z13}{}

\begin{thebibliography}{}
\expandafter\ifx\csname natexlab\endcsname\relax\def\natexlab#1{#1}\fi
\providecommand{\url}[1]{\href{#1}{#1}}
\providecommand{\dodoi}[1]{doi:~\href{http://doi.org/#1}{\nolinkurl{#1}}}
\providecommand{\doeprint}[1]{\href{http://ascl.net/#1}{\nolinkurl{http://ascl.net/#1}}}
\providecommand{\doarXiv}[1]{\href{https://arxiv.org/abs/#1}{\nolinkurl{https://arxiv.org/abs/#1}}}

\bibitem[{{Arrabal Haro} {et~al.}(2023{\natexlab{a}}){Arrabal Haro},
  {Dickinson}, {Finkelstein}, {Fujimoto}, {Fern{\'a}ndez}, {Kartaltepe},
  {Jung}, {Cole}, {Burgarella}, {Chworowsky}, {Hutchison}, {Morales},
  {Papovich}, {Simons}, {Amor{\'\i}n}, {Backhaus}, {Bagley}, {Bisigello},
  {Calabr{\`o}}, {Castellano}, {Cleri}, {Dav{\'e}}, {Dekel}, {Ferguson},
  {Fontana}, {Gawiser}, {Giavalisco}, {Harish}, {Hathi}, {Hirschmann},
  {Holwerda}, {Huertas-Company}, {Koekemoer}, {Larson}, {Lucas}, {Mobasher},
  {P{\'e}rez-Gonz{\'a}lez}, {Pirzkal}, {Rose}, {Santini}, {Trump}, {de la
  Vega}, {Wang}, {Weiner}, {Wilkins}, {Yang}, {Yung}, \&
  {Zavala}}]{arrabalharo2023b}
{Arrabal Haro}, P., {Dickinson}, M., {Finkelstein}, S.~L., {et~al.}
  2023{\natexlab{a}}, \apjl, 951, L22, \dodoi{10.3847/2041-8213/acdd54}

\bibitem[{{Arrabal Haro} {et~al.}(2023{\natexlab{b}}){Arrabal Haro},
  {Dickinson}, {Finkelstein}, {Kartaltepe}, {Donnan}, {Burgarella}, {Carnall},
  {Cullen}, {Dunlop}, {Fern{\'a}ndez}, {Fujimoto}, {Jung}, {Krips}, {Larson},
  {Papovich}, {P{\'e}rez-Gonz{\'a}lez}, {Amor{\'\i}n}, {Bagley}, {Buat},
  {Casey}, {Chworowsky}, {Cohen}, {Ferguson}, {Giavalisco}, {Huertas-Company},
  {Hutchison}, {Kocevski}, {Koekemoer}, {Lucas}, {McLeod}, {McLure}, {Pirzkal},
  {Seill{\'e}}, {Trump}, {Weiner}, {Wilkins}, \& {Zavala}}]{arrabalharo2023a}
---. 2023{\natexlab{b}}, \nat, 622, 707, \dodoi{10.1038/s41586-023-06521-7}

\bibitem[{{Austin} {et~al.}(2024){Austin}, {Conselice}, {Adams}, {Harvey},
  {Duan}, {Trussler}, {Li}, {Juodzbalis}, {Ormerod}, {Ferreira}, {Westcott},
  {Harris}, {Wilkins}, {Bhatawdekar}, {Caruana}, {Coe}, {Cohen}, {Driver},
  {D'Silva}, {Frye}, {Furtak}, {Grogin}, {Hathi}, {Holwerda}, {Jansen},
  {Koekemoer}, {Marshall}, {Nonino}, {Ortiz}, {Pirzkal}, {Robotham}, {Ryan},
  {Summers}, {Willmer}, {Windhorst}, {Yan}, \& {Zackrisson}}]{austin2024}
{Austin}, D., {Conselice}, C.~J., {Adams}, N.~J., {et~al.} 2024, arXiv
  e-prints, arXiv:2404.10751, \dodoi{10.48550/arXiv.2404.10751}

\bibitem[{{Baker} {et~al.}(2023){Baker}, {Tacchella}, {Johnson}, {Nelson},
  {Suess}, {D'Eugenio}, {Curti}, {de Graaff}, {Ji}, {Maiolino}, {Robertson},
  {Scholtz}, {Alberts}, {Arribas}, {Boyett}, {Bunker}, {Carniani}, {Charlot},
  {Chen}, {Chevallard}, {Curtis-Lake}, {Danhaive}, {DeCoursey}, {Egami},
  {Eisenstein}, {Endsley}, {Hausen}, {Helton}, {Kumari}, {Looser}, {Maseda},
  {Pusk{\'a}s}, {Rieke}, {Sandles}, {Sun}, {{\"U}bler}, {Williams}, {Willmer},
  \& {Witstok}}]{baker2023}
{Baker}, W.~M., {Tacchella}, S., {Johnson}, B.~D., {et~al.} 2023, arXiv
  e-prints, arXiv:2306.02472, \dodoi{10.48550/arXiv.2306.02472}

\bibitem[{{Bayer} {et~al.}(2021){Bayer}, {Seljak}, \& {Robnik}}]{Bayer21}
{Bayer}, A.~E., {Seljak}, U., \& {Robnik}, J. 2021, \mnras, 508, 1346,
  \dodoi{10.1093/mnras/stab2331}

\bibitem[{{Bekki} \& {Tsujimoto}(2023)}]{bekki2023}
{Bekki}, K., \& {Tsujimoto}, T. 2023, \mnras, 526, L26,
  \dodoi{10.1093/mnrasl/slad108}

\bibitem[{{Bezanson} {et~al.}(2022){Bezanson}, {Labbe}, {Whitaker}, {Leja},
  {Price}, {Franx}, {Brammer}, {Marchesini}, {Zitrin}, {Wang}, {Weaver},
  {Furtak}, {Atek}, {Coe}, {Cutler}, {Dayal}, {van Dokkum}, {Feldmann},
  {Forster Schreiber}, {Fujimoto}, {Geha}, {Glazebrook}, {de Graaff}, {Greene},
  {Juneau}, {Kassin}, {Kriek}, {Khullar}, {Maseda}, {Mowla}, {Muzzin},
  {Nanayakkara}, {Nelson}, {Oesch}, {Pacifici}, {Pan}, {Papovich}, {Setton},
  {Shapley}, {Smit}, {Stefanon}, {Taylor}, \& {Williams}}]{bezanson2022}
{Bezanson}, R., {Labbe}, I., {Whitaker}, K.~E., {et~al.} 2022, arXiv e-prints,
  arXiv:2212.04026, \dodoi{10.48550/arXiv.2212.04026}

\bibitem[{{Bouwens} {et~al.}(2011){Bouwens}, {Illingworth}, {Labbe}, {Oesch},
  {Trenti}, {Carollo}, {van Dokkum}, {Franx}, {Stiavelli}, {Gonz{\'a}lez},
  {Magee}, \& {Bradley}}]{bouwens2011}
{Bouwens}, R.~J., {Illingworth}, G.~D., {Labbe}, I., {et~al.} 2011, \nat, 469,
  504, \dodoi{10.1038/nature09717}

\bibitem[{{Brammer} {et~al.}(2008){Brammer}, {van Dokkum}, \&
  {Coppi}}]{brammer2008}
{Brammer}, G.~B., {van Dokkum}, P.~G., \& {Coppi}, P. 2008, \apj, 686, 1503,
  \dodoi{10.1086/591786}

\bibitem[{{Bruzual} \& {Charlot}(2003)}]{Bruzual2003}
{Bruzual}, G., \& {Charlot}, S. 2003, \mnras, 344, 1000,
  \dodoi{10.1046/j.1365-8711.2003.06897.x}

\bibitem[{{Bunker} {et~al.}(2023{\natexlab{a}}){Bunker}, {Saxena}, {Cameron},
  {Willott}, {Curtis-Lake}, {Jakobsen}, {Carniani}, {Smit}, {Maiolino},
  {Witstok}, {Curti}, {D'Eugenio}, {Jones}, {Ferruit}, {Arribas}, {Charlot},
  {Chevallard}, {Giardino}, {de Graaff}, {Looser}, {L{\"u}tzgendorf}, {Maseda},
  {Rawle}, {Rix}, {Del Pino}, {Alberts}, {Egami}, {Eisenstein}, {Endsley},
  {Hainline}, {Hausen}, {Johnson}, {Rieke}, {Rieke}, {Robertson}, {Shivaei},
  {Stark}, {Sun}, {Tacchella}, {Tang}, {Williams}, {Willmer}, {Baker}, {Baum},
  {Bhatawdekar}, {Bowler}, {Boyett}, {Chen}, {Circosta}, {Helton}, {Ji},
  {Kumari}, {Lyu}, {Nelson}, {Parlanti}, {Perna}, {Sandles}, {Scholtz},
  {Suess}, {Topping}, {{\"U}bler}, {Wallace}, \& {Whitler}}]{bunker2023a}
{Bunker}, A.~J., {Saxena}, A., {Cameron}, A.~J., {et~al.} 2023{\natexlab{a}},
  \aap, 677, A88, \dodoi{10.1051/0004-6361/202346159}

\bibitem[{{Bunker} {et~al.}(2023{\natexlab{b}}){Bunker}, {Cameron},
  {Curtis-Lake}, {Jakobsen}, {Carniani}, {Curti}, {Witstok}, {Maiolino},
  {D'Eugenio}, {Looser}, {Willott}, {Bonaventura}, {Hainline}, {Uebler},
  {Willmer}, {Saxena}, {Smit}, {Alberts}, {Arribas}, {Baker}, {Baum},
  {Bhatawdekar}, {Bowler}, {Boyett}, {Charlot}, {Chen}, {Chevallard},
  {Circosta}, {DeCoursey}, {de Graaff}, {Egami}, {Eisenstein}, {Endsley},
  {Ferruit}, {Giardino}, {Hausen}, {Helton}, {Hviding}, {Ji}, {Johnson},
  {Jones}, {Kumari}, {Laseter}, {Luetzgendorf}, {Maseda}, {Nelson}, {Parlanti},
  {Perna}, {Rawle}, {Rix}, {Rieke}, {Robertson}, {Rodriguez Del Pino},
  {Sandles}, {Scholtz}, {Sharpe}, {Skarbinski}, {Stark}, {Sun}, {Tacchella},
  {Topping}, {Villanueva}, {Wallace}, {Williams}, \& {Woodrum}}]{bunker2023b}
{Bunker}, A.~J., {Cameron}, A.~J., {Curtis-Lake}, E., {et~al.}
  2023{\natexlab{b}}, arXiv e-prints, arXiv:2306.02467,
  \dodoi{10.48550/arXiv.2306.02467}

\bibitem[{{Calzetti} {et~al.}(2000){Calzetti}, {Armus}, {Bohlin}, {Kinney},
  {Koornneef}, \& {Storchi-Bergmann}}]{calzetti2000}
{Calzetti}, D., {Armus}, L., {Bohlin}, R.~C., {et~al.} 2000, \apj, 533, 682,
  \dodoi{10.1086/308692}

\bibitem[{{Calzetti} {et~al.}(1994){Calzetti}, {Kinney}, \&
  {Storchi-Bergmann}}]{Calzetti1994}
{Calzetti}, D., {Kinney}, A.~L., \& {Storchi-Bergmann}, T. 1994, \apj, 429,
  582, \dodoi{10.1086/174346}

\bibitem[{{Cameron} {et~al.}(2023){Cameron}, {Katz}, {Rey}, \&
  {Saxena}}]{cameron2023}
{Cameron}, A.~J., {Katz}, H., {Rey}, M.~P., \& {Saxena}, A. 2023, \mnras, 523,
  3516, \dodoi{10.1093/mnras/stad1579}

\bibitem[{{Carnall} {et~al.}(2019){Carnall}, {Leja}, {Johnson}, {McLure},
  {Dunlop}, \& {Conroy}}]{carnall2019}
{Carnall}, A.~C., {Leja}, J., {Johnson}, B.~D., {et~al.} 2019, \apj, 873, 44,
  \dodoi{10.3847/1538-4357/ab04a2}

\bibitem[{{Carniani} {et~al.}(2023){Carniani}, {Venturi}, {Parlanti}, {de
  Graaff}, {Maiolino}, {Arribas}, {Bonaventura}, {Boyett}, {Bunker}, {Cameron},
  {Charlot}, {Chevallard}, {Curti}, {Curtis-Lake}, {Eisenstein}, {Giardino},
  {Hausen}, {Kumari}, {Maseda}, {Nelson}, {Perna}, {Rix}, {Robertson},
  {Rodr{\'\i}guez Del Pino}, {Sandles}, {Scholtz}, {Simmonds}, {Smit},
  {Tacchella}, {{\"U}bler}, {Williams}, {Willott}, \& {Witstok}}]{carniani2023}
{Carniani}, S., {Venturi}, G., {Parlanti}, E., {et~al.} 2023, arXiv e-prints,
  arXiv:2306.11801, \dodoi{10.48550/arXiv.2306.11801}

\bibitem[{{Castellano} {et~al.}(2022){Castellano}, {Fontana}, {Treu},
  {Santini}, {Merlin}, {Leethochawalit}, {Trenti}, {Vanzella}, {Mestric},
  {Bonchi}, {Belfiori}, {Nonino}, {Paris}, {Polenta}, {Roberts-Borsani},
  {Boyett}, {Brada{\v{c}}}, {Calabr{\`o}}, {Glazebrook}, {Grillo}, {Mascia},
  {Mason}, {Mercurio}, {Morishita}, {Nanayakkara}, {Pentericci}, {Rosati},
  {Vulcani}, {Wang}, \& {Yang}}]{castellano2022}
{Castellano}, M., {Fontana}, A., {Treu}, T., {et~al.} 2022, \apjl, 938, L15,
  \dodoi{10.3847/2041-8213/ac94d0}

\bibitem[{{Castellano} {et~al.}(2024){Castellano}, {Napolitano}, {Fontana},
  {Roberts-Borsani}, {Treu}, {Vanzella}, {Zavala}, {Arrabal Haro},
  {Calabr{\`o}}, {Llerena}, {Mascia}, {Merlin}, {Paris}, {Pentericci},
  {Santini}, {Bakx}, {Bergamini}, {Cupani}, {Dickinson}, {Filippenko},
  {Glazebrook}, {Grillo}, {Kelly}, {Malkan}, {Mason}, {Morishita},
  {Nanayakkara}, {Rosati}, {Sani}, {Wang}, \& {Yoon}}]{castellano2024}
{Castellano}, M., {Napolitano}, L., {Fontana}, A., {et~al.} 2024, arXiv
  e-prints, arXiv:2403.10238, \dodoi{10.48550/arXiv.2403.10238}

\bibitem[{{Chabrier}(2003)}]{chabrier2003}
{Chabrier}, G. 2003, \pasp, 115, 763, \dodoi{10.1086/376392}

\bibitem[{{Charlot} \& {Fall}(2000)}]{charlotfall2000}
{Charlot}, S., \& {Fall}, S.~M. 2000, \apj, 539, 718, \dodoi{10.1086/309250}

\bibitem[{{Chevallard} \& {Charlot}(2016)}]{chevallard2016}
{Chevallard}, J., \& {Charlot}, S. 2016, \mnras, 462, 1415,
  \dodoi{10.1093/mnras/stw1756}

\bibitem[{{Choi} {et~al.}(2016){Choi}, {Dotter}, {Conroy}, {Cantiello},
  {Paxton}, \& {Johnson}}]{choi2016}
{Choi}, J., {Dotter}, A., {Conroy}, C., {et~al.} 2016, \apj, 823, 102,
  \dodoi{10.3847/0004-637X/823/2/102}

\bibitem[{{Coe} {et~al.}(2013){Coe}, {Zitrin}, {Carrasco}, {Shu}, {Zheng},
  {Postman}, {Bradley}, {Koekemoer}, {Bouwens}, {Broadhurst}, {Monna}, {Host},
  {Moustakas}, {Ford}, {Moustakas}, {van der Wel}, {Donahue}, {Rodney},
  {Ben{\'\i}tez}, {Jouvel}, {Seitz}, {Kelson}, \& {Rosati}}]{coe2013}
{Coe}, D., {Zitrin}, A., {Carrasco}, M., {et~al.} 2013, \apj, 762, 32,
  \dodoi{10.1088/0004-637X/762/1/32}

\bibitem[{{Conroy} \& {Gunn}(2010)}]{conroy2010}
{Conroy}, C., \& {Gunn}, J.~E. 2010, {FSPS: Flexible Stellar Population
  Synthesis}, Astrophysics Source Code Library, record ascl:1010.043

\bibitem[{{Cullen} {et~al.}(2024){Cullen}, {McLeod}, {McLure}, {Dunlop},
  {Donnan}, {Carnall}, {Keating}, {Magee}, {Arellano-Cordova}, {Bowler},
  {Begley}, {Flury}, {Hamadouche}, \& {Stanton}}]{cullen2024}
{Cullen}, F., {McLeod}, D.~J., {McLure}, R.~J., {et~al.} 2024, \mnras, 531,
  997, \dodoi{10.1093/mnras/stae1211}

\bibitem[{{Curti} {et~al.}(2023){Curti}, {Maiolino}, {Curtis-Lake},
  {Chevallard}, {Carniani}, {D'Eugenio}, {Looser}, {Scholtz}, {Charlot},
  {Cameron}, {{\"U}bler}, {Witstok}, {Boyett}, {Laseter}, {Sandles}, {Arribas},
  {Bunker}, {Giardino}, {Maseda}, {Rawle}, {Rodr{\'\i}guez Del Pino}, {Smit},
  {Willott}, {Eisenstein}, {Hausen}, {Johnson}, {Rieke}, {Robertson},
  {Tacchella}, {Williams}, {Willmer}, {Baker}, {Bhatawdekar}, {Egami},
  {Helton}, {Ji}, {Kumari}, {Perna}, {Shivaei}, \& {Sun}}]{curti2023}
{Curti}, M., {Maiolino}, R., {Curtis-Lake}, E., {et~al.} 2023, arXiv e-prints,
  arXiv:2304.08516, \dodoi{10.48550/arXiv.2304.08516}

\bibitem[{{Curtis-Lake} {et~al.}(2023){Curtis-Lake}, {Carniani}, {Cameron},
  {Charlot}, {Jakobsen}, {Maiolino}, {Bunker}, {Witstok}, {Smit}, {Chevallard},
  {Willott}, {Ferruit}, {Arribas}, {Bonaventura}, {Curti}, {D'Eugenio},
  {Franx}, {Giardino}, {Looser}, {L{\"u}tzgendorf}, {Maseda}, {Rawle}, {Rix},
  {Rodr{\'\i}guez del Pino}, {{\"U}bler}, {Sirianni}, {Dressler}, {Egami},
  {Eisenstein}, {Endsley}, {Hainline}, {Hausen}, {Johnson}, {Rieke},
  {Robertson}, {Shivaei}, {Stark}, {Tacchella}, {Williams}, {Willmer},
  {Bhatawdekar}, {Bowler}, {Boyett}, {Chen}, {de Graaff}, {Helton}, {Hviding},
  {Jones}, {Kumari}, {Lyu}, {Nelson}, {Perna}, {Sandles}, {Saxena}, {Suess},
  {Sun}, {Topping}, {Wallace}, \& {Whitler}}]{curtislake2023}
{Curtis-Lake}, E., {Carniani}, S., {Cameron}, A., {et~al.} 2023, Nature
  Astronomy, 7, 622, \dodoi{10.1038/s41550-023-01918-w}

\bibitem[{{D'Antona} {et~al.}(2023){D'Antona}, {Vesperini}, {Calura},
  {Ventura}, {D'Ercole}, {Caloi}, {Marino}, {Milone}, {Dell'Agli}, \&
  {Tailo}}]{dantona2023}
{D'Antona}, F., {Vesperini}, E., {Calura}, F., {et~al.} 2023, \aap, 680, L19,
  \dodoi{10.1051/0004-6361/202348240}

\bibitem[{{de Graaff} {et~al.}(2023){de Graaff}, {Rix}, {Carniani}, {Suess},
  {Charlot}, {Curtis-Lake}, {Arribas}, {Baker}, {Boyett}, {Bunker}, {Cameron},
  {Chevallard}, {Curti}, {Eisenstein}, {Franx}, {Hainline}, {Hausen}, {Ji},
  {Johnson}, {Jones}, {Maiolino}, {Maseda}, {Nelson}, {Parlanti}, {Rawle},
  {Robertson}, {Tacchella}, {{\"U}bler}, {Williams}, {Willmer}, \&
  {Willott}}]{degraaff2023}
{de Graaff}, A., {Rix}, H.-W., {Carniani}, S., {et~al.} 2023, arXiv e-prints,
  arXiv:2308.09742, \dodoi{10.48550/arXiv.2308.09742}

\bibitem[{{D'Eugenio} {et~al.}(2023){D'Eugenio}, {Maiolino}, {Carniani},
  {Curtis-Lake}, {Witstok}, {Chevallard}, {Charlot}, {Baker}, {Arribas},
  {Boyett}, {Bunker}, {Curti}, {Eisenstein}, {Hainline}, {Ji}, {Johnson},
  {Looser}, {Nakajima}, {Nelson}, {Rieke}, {Robertson}, {Scholtz}, {Smit},
  {Venturi}, {Tacchella}, {Uebler}, {Willmer}, \& {Willott}}]{deugenio2023}
{D'Eugenio}, F., {Maiolino}, R., {Carniani}, S., {et~al.} 2023, arXiv e-prints,
  arXiv:2311.09908, \dodoi{10.48550/arXiv.2311.09908}

\bibitem[{{D'Eugenio} {et~al.}(2024){D'Eugenio}, {Cameron}, {Scholtz},
  {Carniani}, {Willott}, {Curtis-Lake}, {Bunker}, {Parlanti}, {Maiolino},
  {Willmer}, {Jakobsen}, {Robertson}, {Johnson}, {Tacchella}, {Cargile},
  {Rawle}, {Arribas}, {Chevallard}, {Curti}, {Egami}, {Eisenstein}, {Kumari},
  {Looser}, {Rieke}, {Rodr{\'\i}guez Del Pino}, {Saxena}, {{\"U}bler},
  {Venturi}, {Witstok}, {Baker}, {Bhatawdekar}, {Bonaventura}, {Boyett},
  {Charlot}, {Danhaive}, {Hainline}, {Hausen}, {Helton}, {Ji}, {Ji}, {Jones},
  {Joud{\v{z}}balis}, {Maseda}, {P{\'e}rez-Gonz{\'a}lez}, {Perna},
  {Pusk{\'a}s}, {Shivaei}, {Silcock}, {Simmonds}, {Smit}, {Sun}, {Villanueva},
  {Williams}, \& {Zhu}}]{deugenio2024}
{D'Eugenio}, F., {Cameron}, A.~J., {Scholtz}, J., {et~al.} 2024, arXiv
  e-prints, arXiv:2404.06531, \dodoi{10.48550/arXiv.2404.06531}

\bibitem[{{Efron} \& {Hastie}(2021)}]{Efron21}
{Efron}, B., \& {Hastie}, T. 2021, {Computer Age Statistical Inference}
  (Cambridge University Press), p155

\bibitem[{{Eisenstein} {et~al.}(2023){Eisenstein}, {Johnson}, {Robertson},
  {Tacchella}, {Hainline}, {Jakobsen}, {Maiolino}, {Bonaventura}, {Bunker},
  {Cameron}, {Cargile}, {Curtis-Lake}, {Hausen}, {Pusk{\'a}s}, {Rieke}, {Sun},
  {Willmer}, {Willott}, {Alberts}, {Arribas}, {Baker}, {Baum}, {Bhatawdekar},
  {Carniani}, {Charlot}, {Chen}, {Chevallard}, {Curti}, {DeCoursey},
  {D'Eugenio}, {de Graaff}, {Egami}, {Helton}, {Ji}, {Jones}, {Kumari},
  {L{\"u}tzgendorf}, {Laseter}, {Looser}, {Lyu}, {Maseda}, {Nelson},
  {Parlanti}, {Rauscher}, {Rawle}, {Rieke}, {Rix}, {Rujopakarn}, {Sandles},
  {Saxena}, {Scholtz}, {Sharpe}, {Shivaei}, {Simmonds}, {Smit}, {Topping},
  {{\"U}bler}, {Venturi}, {Williams}, {Witstok}, \&
  {Woodrum}}]{eisenstein2023b}
{Eisenstein}, D.~J., {Johnson}, B.~D., {Robertson}, B., {et~al.} 2023, arXiv
  e-prints, arXiv:2310.12340, \dodoi{10.48550/arXiv.2310.12340}

\bibitem[{{Ellis} {et~al.}(2013){Ellis}, {McLure}, {Dunlop}, {Robertson},
  {Ono}, {Schenker}, {Koekemoer}, {Bowler}, {Ouchi}, {Rogers}, {Curtis-Lake},
  {Schneider}, {Charlot}, {Stark}, {Furlanetto}, \& {Cirasuolo}}]{ellis2013}
{Ellis}, R.~S., {McLure}, R.~J., {Dunlop}, J.~S., {et~al.} 2013, \apjl, 763,
  L7, \dodoi{10.1088/2041-8205/763/1/L7}

\bibitem[{{Endsley} {et~al.}(2023){Endsley}, {Stark}, {Whitler}, {Topping},
  {Johnson}, {Robertson}, {Tacchella}, {Alberts}, {Baker}, {Bhatawdekar},
  {Boyett}, {Bunker}, {Cameron}, {Carniani}, {Charlot}, {Chen}, {Chevallard},
  {Curtis-Lake}, {Danhaive}, {Egami}, {Eisenstein}, {Hainline}, {Helton}, {Ji},
  {Looser}, {Maiolino}, {Nelson}, {Pusk{\'a}s}, {Rieke}, {Rieke}, {Rix},
  {Sandles}, {Saxena}, {Simmonds}, {Smit}, {Sun}, {Williams}, {Willmer},
  {Willott}, \& {Witstok}}]{endsley2023}
{Endsley}, R., {Stark}, D.~P., {Whitler}, L., {et~al.} 2023, arXiv e-prints,
  arXiv:2306.05295, \dodoi{10.48550/arXiv.2306.05295}

\bibitem[{{Finkelstein} {et~al.}(2023){Finkelstein}, {Leung}, {Bagley},
  {Dickinson}, {Ferguson}, {Papovich}, {Akins}, {Arrabal Haro}, {Dave},
  {Dekel}, {Kartaltepe}, {Kocevski}, {Koekemoer}, {Pirzkal}, {Somerville},
  {Yung}, {Amorin}, {Backhaus}, {Behroozi}, {Bisigello}, {Bromm}, {Casey},
  {Chavez Ortiz}, {Cheng}, {Chworowsky}, {Cleri}, {Cooper}, {Davis}, {de la
  Vega}, {Elbaz}, {Franco}, {Fontana}, {Fujimoto}, {Giavalisco}, {Grogin},
  {Holwerda}, {Huertas-Company}, {Hirschmann}, {Iyer}, {Jogee}, {Jung},
  {Larson}, {Lucas}, {Mobasher}, {Morales}, {Morley}, {Mukherjee},
  {Perez-Gonzalez}, {Ravindranath}, {Rodighiero}, {Rowland}, {Tacchella},
  {Taylor}, {Trump}, \& {Wilkins}}]{finkelstein2023}
{Finkelstein}, S.~L., {Leung}, G. C.~K., {Bagley}, M.~B., {et~al.} 2023, arXiv
  e-prints, arXiv:2311.04279, \dodoi{10.48550/arXiv.2311.04279}

\bibitem[{{Fisher}(1950)}]{Fisher50}
{Fisher}, R.~A. 1950, {Statistical Methods for Research Workers, 11th Edition}
  (Oliver and Boyd), p99

\bibitem[{{Fletcher} {et~al.}(2019){Fletcher}, {Tang}, {Robertson}, {Nakajima},
  {Ellis}, {Stark}, \& {Inoue}}]{fletcher2019}
{Fletcher}, T.~J., {Tang}, M., {Robertson}, B.~E., {et~al.} 2019, \apj, 878,
  87, \dodoi{10.3847/1538-4357/ab2045}

\bibitem[{{Flury} {et~al.}(2022){Flury}, {Jaskot}, {Ferguson}, {Worseck},
  {Makan}, {Chisholm}, {Saldana-Lopez}, {Schaerer}, {McCandliss}, {Wang},
  {Ford}, {Heckman}, {Ji}, {Giavalisco}, {Amorin}, {Atek}, {Blaizot},
  {Borthakur}, {Carr}, {Castellano}, {Cristiani}, {De Barros}, {Dickinson},
  {Finkelstein}, {Fleming}, {Fontanot}, {Garel}, {Grazian}, {Hayes}, {Henry},
  {Mauerhofer}, {Micheva}, {Oey}, {Ostlin}, {Papovich}, {Pentericci},
  {Ravindranath}, {Rosdahl}, {Rutkowski}, {Santini}, {Scarlata}, {Teplitz},
  {Thuan}, {Trebitsch}, {Vanzella}, {Verhamme}, \& {Xu}}]{flury2022}
{Flury}, S.~R., {Jaskot}, A.~E., {Ferguson}, H.~C., {et~al.} 2022, \apjs, 260,
  1, \dodoi{10.3847/1538-4365/ac5331}

\bibitem[{{Fujimoto} {et~al.}(2023){Fujimoto}, {Arrabal Haro}, {Dickinson},
  {Finkelstein}, {Kartaltepe}, {Larson}, {Burgarella}, {Bagley}, {Behroozi},
  {Chworowsky}, {Hirschmann}, {Trump}, {Wilkins}, {Yung}, {Koekemoer},
  {Papovich}, {Pirzkal}, {Ferguson}, {Fontana}, {Grogin}, {Grazian}, {Kewley},
  {Kocevski}, {Lotz}, {Pentericci}, {Ravindranath}, {Somerville}, {Wilkins},
  {Amor{\'\i}n}, {Backhaus}, {Calabr{\`o}}, {Casey}, {Cooper}, {Fern{\'a}ndez},
  {Franco}, {Giavalisco}, {Hathi}, {Harish}, {Hutchison}, {Iyer}, {Jung},
  {Lucas}, \& {Zavala}}]{fujimoto2023}
{Fujimoto}, S., {Arrabal Haro}, P., {Dickinson}, M., {et~al.} 2023, \apjl, 949,
  L25, \dodoi{10.3847/2041-8213/acd2d9}

\bibitem[{{Genzel} {et~al.}(2010){Genzel}, {Tacconi}, {Gracia-Carpio},
  {Sternberg}, {Cooper}, {Shapiro}, {Bolatto}, {Bouch{\'e}}, {Bournaud},
  {Burkert}, {Combes}, {Comerford}, {Cox}, {Davis}, {F{\"o}rster Schreiber},
  {Garcia-Burillo}, {Lutz}, {Naab}, {Neri}, {Omont}, {Shapley}, \&
  {Weiner}}]{genzel2010}
{Genzel}, R., {Tacconi}, L.~J., {Gracia-Carpio}, J., {et~al.} 2010, \mnras,
  407, 2091, \dodoi{10.1111/j.1365-2966.2010.16969.x}

\bibitem[{{Gutkin} {et~al.}(2016){Gutkin}, {Charlot}, \&
  {Bruzual}}]{Gutkin2016}
{Gutkin}, J., {Charlot}, S., \& {Bruzual}, G. 2016, \mnras, 462, 1757,
  \dodoi{10.1093/mnras/stw1716}

\bibitem[{{Hainline} {et~al.}(2023){Hainline}, {Johnson}, {Robertson},
  {Tacchella}, {Helton}, {Sun}, {Eisenstein}, {Simmonds}, {Topping}, {Whitler},
  {Willmer}, {Rieke}, {Suess}, {Hviding}, {Cameron}, {Alberts}, {Baker},
  {Bhatawdekar}, {Boyett}, {Bunker}, {Carniani}, {Charlot}, {Chen}, {Curti},
  {Curtis-Lake}, {D'Eugenio}, {Egami}, {Endsley}, {Hausen}, {Ji}, {Looser},
  {Lyu}, {Maiolino}, {Nelson}, {Puskas}, {Rawle}, {Sandles}, {Saxena}, {Smit},
  {Stark}, {Williams}, {Willott}, \& {Witstok}}]{hainline2023}
{Hainline}, K.~N., {Johnson}, B.~D., {Robertson}, B., {et~al.} 2023, arXiv
  e-prints, arXiv:2306.02468, \dodoi{10.48550/arXiv.2306.02468}

\bibitem[{{Harikane} {et~al.}(2024){Harikane}, {Nakajima}, {Ouchi}, {Umeda},
  {Isobe}, {Ono}, {Xu}, \& {Zhang}}]{harikane2024}
{Harikane}, Y., {Nakajima}, K., {Ouchi}, M., {et~al.} 2024, \apj, 960, 56,
  \dodoi{10.3847/1538-4357/ad0b7e}

\bibitem[{{Heintz} {et~al.}(2023){Heintz}, {Watson}, {Brammer}, {Vejlgaard},
  {Hutter}, {Strait}, {Matthee}, {Oesch}, {Jakobsson}, {Tanvir}, {Laursen},
  {Naidu}, {Mason}, {Killi}, {Jung}, {Hsiao}, {Abdurro'uf}, {Coe}, {Arrabal
  Haro}, {Finkelstein}, \& {Toft}}]{heintz2023}
{Heintz}, K.~E., {Watson}, D., {Brammer}, G., {et~al.} 2023, arXiv e-prints,
  arXiv:2306.00647, \dodoi{10.48550/arXiv.2306.00647}

\bibitem[{{Hsiao} {et~al.}(2023){Hsiao}, {Abdurro'uf}, {Coe}, {Larson}, {Jung},
  {Mingozzi}, {Dayal}, {Kumari}, {Kokorev}, {Vikaeus}, {Brammer}, {Furtak},
  {Adamo}, {Andrade-Santos}, {Antwi-Danso}, {Bradac}, {Bradley}, {Broadhurst},
  {Carnall}, {Conselice}, {Diego}, {Donahue}, {Eldridge}, {Fujimoto}, {Henry},
  {Hernandez}, {Hutchison}, {James}, {Norman}, {Park}, {Pirzkal}, {Postman},
  {Ricotti}, {Rigby}, {Vanzella}, {Welch}, {Wilkins}, {Windhorst}, {Xu},
  {Zackrisson}, \& {Zitrin}}]{hsiao2023}
{Hsiao}, T. Y.-Y., {Abdurro'uf}, {Coe}, D., {et~al.} 2023, arXiv e-prints,
  arXiv:2305.03042, \dodoi{10.48550/arXiv.2305.03042}

\bibitem[{{Ilie} {et~al.}(2023){Ilie}, {Paulin}, \& {Freese}}]{ilie2023}
{Ilie}, C., {Paulin}, J., \& {Freese}, K. 2023, Proceedings of the National
  Academy of Science, 120, e2305762120, \dodoi{10.1073/pnas.2305762120}

\bibitem[{{Illingworth} {et~al.}(2013){Illingworth}, {Magee}, {Oesch},
  {Bouwens}, {Labb{\'e}}, {Stiavelli}, {van Dokkum}, {Franx}, {Trenti},
  {Carollo}, \& {Gonzalez}}]{illingworth2013}
{Illingworth}, G.~D., {Magee}, D., {Oesch}, P.~A., {et~al.} 2013, \apjs, 209,
  6, \dodoi{10.1088/0067-0049/209/1/6}

\bibitem[{{Isobe} {et~al.}(2023){Isobe}, {Ouchi}, {Tominaga}, {Watanabe},
  {Nakajima}, {Umeda}, {Yajima}, {Harikane}, {Fukushima}, {Xu}, {Ono}, \&
  {Zhang}}]{isobe2023}
{Isobe}, Y., {Ouchi}, M., {Tominaga}, N., {et~al.} 2023, \apj, 959, 100,
  \dodoi{10.3847/1538-4357/ad09be}

\bibitem[{{Izotov} {et~al.}(2016{\natexlab{a}}){Izotov}, {Orlitov{\'a}},
  {Schaerer}, {Thuan}, {Verhamme}, {Guseva}, \& {Worseck}}]{izotov2016a}
{Izotov}, Y.~I., {Orlitov{\'a}}, I., {Schaerer}, D., {et~al.}
  2016{\natexlab{a}}, \nat, 529, 178, \dodoi{10.1038/nature16456}

\bibitem[{{Izotov} {et~al.}(2016{\natexlab{b}}){Izotov}, {Schaerer}, {Thuan},
  {Worseck}, {Guseva}, {Orlitov{\'a}}, \& {Verhamme}}]{izotov2016b}
{Izotov}, Y.~I., {Schaerer}, D., {Thuan}, T.~X., {et~al.} 2016{\natexlab{b}},
  \mnras, 461, 3683, \dodoi{10.1093/mnras/stw1205}

\bibitem[{{Izotov} {et~al.}(2021){Izotov}, {Worseck}, {Schaerer}, {Guseva},
  {Chisholm}, {Thuan}, {Fricke}, \& {Verhamme}}]{izotov2021}
{Izotov}, Y.~I., {Worseck}, G., {Schaerer}, D., {et~al.} 2021, \mnras, 503,
  1734, \dodoi{10.1093/mnras/stab612}

\bibitem[{{Jakobsen} {et~al.}(2022){Jakobsen}, {Ferruit}, {Alves de Oliveira},
  {Arribas}, {Bagnasco}, {Barho}, {Beck}, {Birkmann}, {B{\"o}ker}, {Bunker},
  {Charlot}, {de Jong}, {de Marchi}, {Ehrenwinkler}, {Falcolini}, {Fels},
  {Franx}, {Franz}, {Funke}, {Giardino}, {Gnata}, {Holota}, {Honnen}, {Jensen},
  {Jentsch}, {Johnson}, {Jollet}, {Karl}, {Kling}, {K{\"o}hler}, {Kolm},
  {Kumari}, {Lander}, {Lemke}, {L{\'o}pez-Caniego}, {L{\"u}tzgendorf},
  {Maiolino}, {Manjavacas}, {Marston}, {Maschmann}, {Maurer}, {Messerschmidt},
  {Moseley}, {Mosner}, {Mott}, {Muzerolle}, {Pirzkal}, {Pittet}, {Plitzke},
  {Posselt}, {Rapp}, {Rauscher}, {Rawle}, {Rix}, {R{\"o}del}, {Rumler},
  {Sabbi}, {Salvignol}, {Schmid}, {Sirianni}, {Smith}, {Strada}, {te Plate},
  {Valenti}, {Wettemann}, {Wiehe}, {Wiesmayer}, {Willott}, {Wright}, {Zeidler},
  \& {Zincke}}]{jakobsen2022}
{Jakobsen}, P., {Ferruit}, P., {Alves de Oliveira}, C., {et~al.} 2022, \aap,
  661, A80, \dodoi{10.1051/0004-6361/202142663}

\bibitem[{{Johnson} {et~al.}(2021){Johnson}, {Leja}, {Conroy}, \&
  {Speagle}}]{johnson2021}
{Johnson}, B.~D., {Leja}, J., {Conroy}, C., \& {Speagle}, J.~S. 2021, \apjs,
  254, 22, \dodoi{10.3847/1538-4365/abef67}

\bibitem[{{Jones} {et~al.}(2023){Jones}, {Sanders}, {Chen}, {Wang},
  {Morishita}, {Roberts-Borsani}, {Treu}, {Dressler}, {Merlin}, {Paris},
  {Santini}, {Bergamini}, {Henry}, {Huntzinger}, {Nanayakkara}, {Boyett},
  {Bradac}, {Brammer}, {Calabr{\'o}}, {Glazebrook}, {Grasha}, {Mascia},
  {Pentericci}, {Trenti}, \& {Vulcani}}]{jones2023}
{Jones}, T., {Sanders}, R., {Chen}, Y., {et~al.} 2023, \apjl, 951, L17,
  \dodoi{10.3847/2041-8213/acd938}

\bibitem[{{Kim} {et~al.}(2023){Kim}, {Bayliss}, {Rigby}, {Gladders},
  {Chisholm}, {Sharon}, {Dahle}, {Rivera-Thorsen}, {Florian}, {Khullar},
  {Mahler}, {Mainali}, {Napier}, {Navarre}, {Owens}, \& {Roberson}}]{kim2023}
{Kim}, K.~J., {Bayliss}, M.~B., {Rigby}, J.~R., {et~al.} 2023, \apjl, 955, L17,
  \dodoi{10.3847/2041-8213/acf0c5}

\bibitem[{{Kobayashi} \& {Ferrara}(2023)}]{kobayashi2023}
{Kobayashi}, C., \& {Ferrara}, A. 2023, arXiv e-prints, arXiv:2308.15583,
  \dodoi{10.48550/arXiv.2308.15583}

\bibitem[{{Koekemoer} {et~al.}(2013){Koekemoer}, {Ellis}, {McLure}, {Dunlop},
  {Robertson}, {Ono}, {Schenker}, {Ouchi}, {Bowler}, {Rogers}, {Curtis-Lake},
  {Schneider}, {Charlot}, {Stark}, {Furlanetto}, {Cirasuolo}, {Wild}, \&
  {Targett}}]{koekemoer2013}
{Koekemoer}, A.~M., {Ellis}, R.~S., {McLure}, R.~J., {et~al.} 2013, \apjs, 209,
  3, \dodoi{10.1088/0067-0049/209/1/3}

\bibitem[{{Kriek} \& {Conroy}(2013)}]{kriekconroy2013}
{Kriek}, M., \& {Conroy}, C. 2013, \apjl, 775, L16,
  \dodoi{10.1088/2041-8205/775/1/L16}

\bibitem[{{Kroupa}(2001)}]{kroupa2001}
{Kroupa}, P. 2001, \mnras, 322, 231, \dodoi{10.1046/j.1365-8711.2001.04022.x}

\bibitem[{{Leja} {et~al.}(2019){Leja}, {Carnall}, {Johnson}, {Conroy}, \&
  {Speagle}}]{leja2019}
{Leja}, J., {Carnall}, A.~C., {Johnson}, B.~D., {Conroy}, C., \& {Speagle},
  J.~S. 2019, \apj, 876, 3, \dodoi{10.3847/1538-4357/ab133c}

\bibitem[{{Lejeune} {et~al.}(1997){Lejeune}, {Cuisinier}, \&
  {Buser}}]{lejeune1997}
{Lejeune}, T., {Cuisinier}, F., \& {Buser}, R. 1997, \aaps, 125, 229,
  \dodoi{10.1051/aas:1997373}

\bibitem[{{Lejeune} {et~al.}(1998){Lejeune}, {Cuisinier}, \&
  {Buser}}]{lejeune1998}
---. 1998, \aaps, 130, 65, \dodoi{10.1051/aas:1998405}

\bibitem[{{Looser} {et~al.}(2023){Looser}, {D'Eugenio}, {Maiolino},
  {Tacchella}, {Curti}, {Arribas}, {Baker}, {Baum}, {Bonaventura}, {Boyett},
  {Bunker}, {Carniani}, {Charlot}, {Chevallard}, {Curtis-Lake}, {Danhaive},
  {Eisenstein}, {de Graaff}, {Hainline}, {Ji}, {Johnson}, {Kumari}, {Nelson},
  {Parlanti}, {Rix}, {Robertson}, {Rodr{\'\i}guez Del Pino}, {Sandles},
  {Scholtz}, {Smit}, {Stark}, {{\"U}bler}, {Williams}, {Willott}, \&
  {Witstok}}]{looser2023}
{Looser}, T.~J., {D'Eugenio}, F., {Maiolino}, R., {et~al.} 2023, arXiv
  e-prints, arXiv:2306.02470, \dodoi{10.48550/arXiv.2306.02470}

\bibitem[{{Madau}(1995)}]{madau1995}
{Madau}, P. 1995, \apj, 441, 18, \dodoi{10.1086/175332}

\bibitem[{{Maiolino} \& {Mannucci}(2019)}]{maiolino2019}
{Maiolino}, R., \& {Mannucci}, F. 2019, \aapr, 27, 3,
  \dodoi{10.1007/s00159-018-0112-2}

\bibitem[{{Maiolino} {et~al.}(2023){Maiolino}, {Scholtz}, {Witstok},
  {Carniani}, {D'Eugenio}, {de Graaff}, {Uebler}, {Tacchella}, {Curtis-Lake},
  {Arribas}, {Bunker}, {Charlot}, {Chevallard}, {Curti}, {Looser}, {Maseda},
  {Rawle}, {Rodriguez Del Pino}, {Willott}, {Egami}, {Eisenstein}, {Hainline},
  {Robertson}, {Williams}, {Willmer}, {Baker}, {Boyett}, {DeCoursey}, {Fabian},
  {Helton}, {Ji}, {Jones}, {Kumari}, {Laporte}, {Nelson}, {Perna}, {Sandles},
  {Shivaei}, \& {Sun}}]{maiolino2023}
{Maiolino}, R., {Scholtz}, J., {Witstok}, J., {et~al.} 2023, arXiv e-prints,
  arXiv:2305.12492, \dodoi{10.48550/arXiv.2305.12492}

\bibitem[{{Marques-Chaves} {et~al.}(2021){Marques-Chaves}, {Schaerer},
  {{\'A}lvarez-M{\'a}rquez}, {Colina}, {Dessauges-Zavadsky},
  {P{\'e}rez-Fournon}, {Saldana-Lopez}, \& {Verhamme}}]{marqueschaves2021}
{Marques-Chaves}, R., {Schaerer}, D., {{\'A}lvarez-M{\'a}rquez}, J., {et~al.}
  2021, \mnras, 507, 524, \dodoi{10.1093/mnras/stab2187}

\bibitem[{{Marques-Chaves} {et~al.}(2022){Marques-Chaves}, {Schaerer},
  {{\'A}lvarez-M{\'a}rquez}, {Verhamme}, {Ceverino}, {Chisholm}, {Colina},
  {Dessauges-Zavadsky}, {P{\'e}rez-Fournon}, {Saldana-Lopez}, {Upadhyaya}, \&
  {Vanzella}}]{marqueschaves2022}
---. 2022, \mnras, 517, 2972, \dodoi{10.1093/mnras/stac2893}

\bibitem[{{Marques-Chaves} {et~al.}(2024){Marques-Chaves}, {Schaerer},
  {Vanzella}, {Verhamme}, {Dessauges-Zavadsky}, {Chisholm}, {Leclercq},
  {Upadhyaya}, {Alvarez-Marquez}, {Colina}, {Garel}, \&
  {Messa}}]{marqueschaves2024}
{Marques-Chaves}, R., {Schaerer}, D., {Vanzella}, E., {et~al.} 2024, arXiv
  e-prints, arXiv:2407.18804, \dodoi{10.48550/arXiv.2407.18804}

\bibitem[{{Mason} \& {Gronke}(2020)}]{masongronke2020}
{Mason}, C.~A., \& {Gronke}, M. 2020, \mnras, 499, 1395,
  \dodoi{10.1093/mnras/staa2910}

\bibitem[{{Menon} {et~al.}(2024){Menon}, {Burkhart}, {Somerville}, {Thompson},
  \& {Sternberg}}]{menon2024}
{Menon}, S.~H., {Burkhart}, B., {Somerville}, R.~S., {Thompson}, T.~A., \&
  {Sternberg}, A. 2024, arXiv e-prints, arXiv:2408.14591,
  \dodoi{10.48550/arXiv.2408.14591}

\bibitem[{{Morales} {et~al.}(2024){Morales}, {Finkelstein}, {Leung}, {Bagley},
  {Cleri}, {Dave}, {Dickinson}, {Ferguson}, {Hathi}, {Jones}, {Koekemoer},
  {Papovich}, {P{\'e}rez-Gonz{\'a}lez}, {Pirzkal}, {Smith}, {Wilkins}, \&
  {Yung}}]{morales2024}
{Morales}, A.~M., {Finkelstein}, S.~L., {Leung}, G. C.~K., {et~al.} 2024,
  \apjl, 964, L24, \dodoi{10.3847/2041-8213/ad2de4}

\bibitem[{{Naidu} {et~al.}(2020){Naidu}, {Tacchella}, {Mason}, {Bose}, {Oesch},
  \& {Conroy}}]{naidu2020}
{Naidu}, R.~P., {Tacchella}, S., {Mason}, C.~A., {et~al.} 2020, \apj, 892, 109,
  \dodoi{10.3847/1538-4357/ab7cc9}

\bibitem[{{Naidu} {et~al.}(2022){Naidu}, {Oesch}, {van Dokkum}, {Nelson},
  {Suess}, {Brammer}, {Whitaker}, {Illingworth}, {Bouwens}, {Tacchella},
  {Matthee}, {Allen}, {Bezanson}, {Conroy}, {Labbe}, {Leja}, {Leonova},
  {Magee}, {Price}, {Setton}, {Strait}, {Stefanon}, {Toft}, {Weaver}, \&
  {Weibel}}]{naidu2022}
{Naidu}, R.~P., {Oesch}, P.~A., {van Dokkum}, P., {et~al.} 2022, \apjl, 940,
  L14, \dodoi{10.3847/2041-8213/ac9b22}

\bibitem[{{Nakajima} {et~al.}(2023){Nakajima}, {Ouchi}, {Isobe}, {Harikane},
  {Zhang}, {Ono}, {Umeda}, \& {Oguri}}]{nakajima2023}
{Nakajima}, K., {Ouchi}, M., {Isobe}, Y., {et~al.} 2023, \apjs, 269, 33,
  \dodoi{10.3847/1538-4365/acd556}

\bibitem[{{Nakajima} {et~al.}(2022){Nakajima}, {Ouchi}, {Xu}, {Rauch},
  {Harikane}, {Nishigaki}, {Isobe}, {Kusakabe}, {Nagao}, {Ono}, {Onodera},
  {Sugahara}, {Kim}, {Komiyama}, {Lee}, \& {Zahedy}}]{nakajima2022}
{Nakajima}, K., {Ouchi}, M., {Xu}, Y., {et~al.} 2022, \apjs, 262, 3,
  \dodoi{10.3847/1538-4365/ac7710}

\bibitem[{{Oesch} {et~al.}(2016){Oesch}, {Brammer}, {van Dokkum},
  {Illingworth}, {Bouwens}, {Labb{\'e}}, {Franx}, {Momcheva}, {Ashby}, {Fazio},
  {Gonzalez}, {Holden}, {Magee}, {Skelton}, {Smit}, {Spitler}, {Trenti}, \&
  {Willner}}]{oesch2016}
{Oesch}, P.~A., {Brammer}, G., {van Dokkum}, P.~G., {et~al.} 2016, \apj, 819,
  129, \dodoi{10.3847/0004-637X/819/2/129}

\bibitem[{{Oesch} {et~al.}(2023){Oesch}, {Brammer}, {Naidu}, {Bouwens},
  {Chisholm}, {Illingworth}, {Matthee}, {Nelson}, {Qin}, {Reddy}, {Shapley},
  {Shivaei}, {van Dokkum}, {Weibel}, {Whitaker}, {Wuyts}, {Covelo-Paz},
  {Endsley}, {Fudamoto}, {Giovinazzo}, {Herard-Demanche}, {Kerutt},
  {Kramarenko}, {Labbe}, {Leonova}, {Lin}, {Magee}, {Marchesini}, {Maseda},
  {Mason}, {Matharu}, {Meyer}, {Neufeld}, {Prieto Lyon}, {Schaerer}, {Sharma},
  {Shuntov}, {Smit}, {Stefanon}, {Wyithe}, \& {Xiao}}]{oesch2023}
{Oesch}, P.~A., {Brammer}, G., {Naidu}, R.~P., {et~al.} 2023, \mnras, 525,
  2864, \dodoi{10.1093/mnras/stad2411}

\bibitem[{{Oke}(1974)}]{oke1974}
{Oke}, J.~B. 1974, \apjs, 27, 21, \dodoi{10.1086/190287}

\bibitem[{{Oke} \& {Gunn}(1983)}]{oke1983}
{Oke}, J.~B., \& {Gunn}, J.~E. 1983, \apj, 266, 713, \dodoi{10.1086/160817}

\bibitem[{{Planck Collaboration} {et~al.}(2020){Planck Collaboration},
  {Aghanim}, {Akrami}, {Ashdown}, {Aumont}, {Baccigalupi}, {Ballardini},
  {Banday}, {Barreiro}, {Bartolo}, {Basak}, {Battye}, {Benabed}, {Bernard},
  {Bersanelli}, {Bielewicz}, {Bock}, {Bond}, {Borrill}, {Bouchet}, {Boulanger},
  {Bucher}, {Burigana}, {Butler}, {Calabrese}, {Cardoso}, {Carron},
  {Challinor}, {Chiang}, {Chluba}, {Colombo}, {Combet}, {Contreras}, {Crill},
  {Cuttaia}, {de Bernardis}, {de Zotti}, {Delabrouille}, {Delouis}, {Di
  Valentino}, {Diego}, {Dor{\'e}}, {Douspis}, {Ducout}, {Dupac}, {Dusini},
  {Efstathiou}, {Elsner}, {En{\ss}lin}, {Eriksen}, {Fantaye}, {Farhang},
  {Fergusson}, {Fernandez-Cobos}, {Finelli}, {Forastieri}, {Frailis},
  {Fraisse}, {Franceschi}, {Frolov}, {Galeotta}, {Galli}, {Ganga},
  {G{\'e}nova-Santos}, {Gerbino}, {Ghosh}, {Gonz{\'a}lez-Nuevo}, {G{\'o}rski},
  {Gratton}, {Gruppuso}, {Gudmundsson}, {Hamann}, {Handley}, {Hansen},
  {Herranz}, {Hildebrandt}, {Hivon}, {Huang}, {Jaffe}, {Jones}, {Karakci},
  {Keih{\"a}nen}, {Keskitalo}, {Kiiveri}, {Kim}, {Kisner}, {Knox},
  {Krachmalnicoff}, {Kunz}, {Kurki-Suonio}, {Lagache}, {Lamarre}, {Lasenby},
  {Lattanzi}, {Lawrence}, {Le Jeune}, {Lemos}, {Lesgourgues}, {Levrier},
  {Lewis}, {Liguori}, {Lilje}, {Lilley}, {Lindholm}, {L{\'o}pez-Caniego},
  {Lubin}, {Ma}, {Mac{\'\i}as-P{\'e}rez}, {Maggio}, {Maino}, {Mandolesi},
  {Mangilli}, {Marcos-Caballero}, {Maris}, {Martin}, {Martinelli},
  {Mart{\'\i}nez-Gonz{\'a}lez}, {Matarrese}, {Mauri}, {McEwen}, {Meinhold},
  {Melchiorri}, {Mennella}, {Migliaccio}, {Millea}, {Mitra},
  {Miville-Desch{\^e}nes}, {Molinari}, {Montier}, {Morgante}, {Moss}, {Natoli},
  {N{\o}rgaard-Nielsen}, {Pagano}, {Paoletti}, {Partridge}, {Patanchon},
  {Peiris}, {Perrotta}, {Pettorino}, {Piacentini}, {Polastri}, {Polenta},
  {Puget}, {Rachen}, {Reinecke}, {Remazeilles}, {Renzi}, {Rocha}, {Rosset},
  {Roudier}, {Rubi{\~n}o-Mart{\'\i}n}, {Ruiz-Granados}, {Salvati}, {Sandri},
  {Savelainen}, {Scott}, {Shellard}, {Sirignano}, {Sirri}, {Spencer},
  {Sunyaev}, {Suur-Uski}, {Tauber}, {Tavagnacco}, {Tenti}, {Toffolatti},
  {Tomasi}, {Trombetti}, {Valenziano}, {Valiviita}, {Van Tent}, {Vibert},
  {Vielva}, {Villa}, {Vittorio}, {Wandelt}, {Wehus}, {White}, {White},
  {Zacchei}, \& {Zonca}}]{planck2020}
{Planck Collaboration}, {Aghanim}, N., {Akrami}, Y., {et~al.} 2020, \aap, 641,
  A6, \dodoi{10.1051/0004-6361/201833910}

\bibitem[{{Plat} {et~al.}(2019){Plat}, {Charlot}, {Bruzual}, {Feltre},
  {Vidal-Garc{\'\i}a}, {Morisset}, {Chevallard}, \& {Todt}}]{plat2019}
{Plat}, A., {Charlot}, S., {Bruzual}, G., {et~al.} 2019, \mnras, 490, 978,
  \dodoi{10.1093/mnras/stz2616}

\bibitem[{{Rauscher} {et~al.}(2012){Rauscher}, {Arendt}, {Fixsen}, {Lander},
  {Lindler}, {Loose}, {Moseley}, {Wilson}, \& {Xenophontos}}]{rauscher2012}
{Rauscher}, B.~J., {Arendt}, R.~G., {Fixsen}, D.~J., {et~al.} 2012, in Society
  of Photo-Optical Instrumentation Engineers (SPIE) Conference Series, Vol.
  8453, High Energy, Optical, and Infrared Detectors for Astronomy V, ed. A.~D.
  {Holland} \& J.~W. {Beletic}, 84531F, \dodoi{10.1117/12.926089}

\bibitem[{{Rawle} {et~al.}(2022){Rawle}, {Giardino}, {Franz}, {Rapp}, {te
  Plate}, {Zincke}, {Abul-Huda}, {Alves de Oliveira}, {Bechtold}, {Beck},
  {Birkmann}, {B{\"o}ker}, {Ehrenwinkler}, {Ferruit}, {Garland}, {Jakobsen},
  {Karakla}, {Karl}, {Keyes}, {Koehler}, {Nimisha}, {L{\"u}tzgendorf},
  {Manjavacas}, {Marston}, {Moseley}, {Mosner}, {Muzerolle}, {Ogle},
  {Proffitt}, {Sabbi}, {Sirianni}, {Wahlgren}, {Wislowski}, {Wright}, {Wu}, \&
  {Zeidler}}]{rawle2022}
{Rawle}, T.~D., {Giardino}, G., {Franz}, D.~E., {et~al.} 2022, in Society of
  Photo-Optical Instrumentation Engineers (SPIE) Conference Series, Vol. 12180,
  Space Telescopes and Instrumentation 2022: Optical, Infrared, and Millimeter
  Wave, ed. L.~E. {Coyle}, S.~{Matsuura}, \& M.~D. {Perrin}, 121803R,
  \dodoi{10.1117/12.2629231}

\bibitem[{{Reddy} {et~al.}(2023){Reddy}, {Sanders}, {Shapley}, {Topping},
  {Kriek}, {Coil}, {Mobasher}, {Siana}, \& {Rezaee}}]{reddy2023}
{Reddy}, N.~A., {Sanders}, R.~L., {Shapley}, A.~E., {et~al.} 2023, \apj, 951,
  56, \dodoi{10.3847/1538-4357/acd0b1}

\bibitem[{{Rivera-Thorsen} {et~al.}(2019){Rivera-Thorsen}, {Dahle}, {Chisholm},
  {Florian}, {Gronke}, {Rigby}, {Gladders}, {Mahler}, {Sharon}, \&
  {Bayliss}}]{riverathorsen2019}
{Rivera-Thorsen}, T.~E., {Dahle}, H., {Chisholm}, J., {et~al.} 2019, Science,
  366, 738, \dodoi{10.1126/science.aaw0978}

\bibitem[{{Robertson} {et~al.}(2023){Robertson}, {Tacchella}, {Johnson},
  {Hainline}, {Whitler}, {Eisenstein}, {Endsley}, {Rieke}, {Stark}, {Alberts},
  {Dressler}, {Egami}, {Hausen}, {Rieke}, {Shivaei}, {Williams}, {Willmer},
  {Arribas}, {Bonaventura}, {Bunker}, {Cameron}, {Carniani}, {Charlot},
  {Chevallard}, {Curti}, {Curtis-Lake}, {D'Eugenio}, {Jakobsen}, {Looser},
  {L{\"u}tzgendorf}, {Maiolino}, {Maseda}, {Rawle}, {Rix}, {Smit}, {{\"U}bler},
  {Willott}, {Witstok}, {Baum}, {Bhatawdekar}, {Boyett}, {Chen}, {de Graaff},
  {Florian}, {Helton}, {Hviding}, {Ji}, {Kumari}, {Lyu}, {Nelson}, {Sandles},
  {Saxena}, {Suess}, {Sun}, {Topping}, \& {Wallace}}]{robertson2023}
{Robertson}, B.~E., {Tacchella}, S., {Johnson}, B.~D., {et~al.} 2023, Nature
  Astronomy, 7, 611, \dodoi{10.1038/s41550-023-01921-1}

\bibitem[{{S{\'a}nchez-Bl{\'a}zquez} {et~al.}(2006){S{\'a}nchez-Bl{\'a}zquez},
  {Peletier}, {Jim{\'e}nez-Vicente}, {Cardiel}, {Cenarro},
  {Falc{\'o}n-Barroso}, {Gorgas}, {Selam}, \& {Vazdekis}}]{sanchezblazquez2006}
{S{\'a}nchez-Bl{\'a}zquez}, P., {Peletier}, R.~F., {Jim{\'e}nez-Vicente}, J.,
  {et~al.} 2006, \mnras, 371, 703, \dodoi{10.1111/j.1365-2966.2006.10699.x}

\bibitem[{{Saxena} {et~al.}(2023){Saxena}, {Robertson}, {Bunker}, {Endsley},
  {Cameron}, {Charlot}, {Simmonds}, {Tacchella}, {Witstok}, {Willott},
  {Carniani}, {Curtis-Lake}, {Ferruit}, {Jakobsen}, {Arribas}, {Chevallard},
  {Curti}, {D'Eugenio}, {De Graaff}, {Jones}, {Looser}, {Maseda}, {Rawle},
  {Rix}, {Del Pino}, {Smit}, {{\"U}bler}, {Eisenstein}, {Hainline}, {Hausen},
  {Johnson}, {Rieke}, {Williams}, {Willmer}, {Baker}, {Bhatawdekar}, {Bowler},
  {Boyett}, {Chen}, {Egami}, {Ji}, {Kumari}, {Nelson}, {Perna}, {Sandles},
  {Scholtz}, \& {Shivaei}}]{saxena2023}
{Saxena}, A., {Robertson}, B.~E., {Bunker}, A.~J., {et~al.} 2023, \aap, 678,
  A68, \dodoi{10.1051/0004-6361/202346245}

\bibitem[{{Schaerer} {et~al.}(2022){Schaerer}, {Marques-Chaves}, {Barrufet},
  {Oesch}, {Izotov}, {Naidu}, {Guseva}, \& {Brammer}}]{schaerer2022}
{Schaerer}, D., {Marques-Chaves}, R., {Barrufet}, L., {et~al.} 2022, \aap, 665,
  L4, \dodoi{10.1051/0004-6361/202244556}

\bibitem[{{Senchyna} {et~al.}(2023){Senchyna}, {Plat}, {Stark}, \&
  {Rudie}}]{senchyna2023}
{Senchyna}, P., {Plat}, A., {Stark}, D.~P., \& {Rudie}, G.~C. 2023, arXiv
  e-prints, arXiv:2303.04179, \dodoi{10.48550/arXiv.2303.04179}

\bibitem[{{Shibuya} {et~al.}(2019){Shibuya}, {Ouchi}, {Harikane}, \&
  {Nakajima}}]{shibuya2019}
{Shibuya}, T., {Ouchi}, M., {Harikane}, Y., \& {Nakajima}, K. 2019, \apj, 871,
  164, \dodoi{10.3847/1538-4357/aaf64b}

\bibitem[{{Speagle} {et~al.}(2014){Speagle}, {Steinhardt}, {Capak}, \&
  {Silverman}}]{speagle2014}
{Speagle}, J.~S., {Steinhardt}, C.~L., {Capak}, P.~L., \& {Silverman}, J.~D.
  2014, \apjs, 214, 15, \dodoi{10.1088/0067-0049/214/2/15}

\bibitem[{{Stiavelli} {et~al.}(2023){Stiavelli}, {Morishita}, {Chiaberge},
  {Grillo}, {Leethochawalit}, {Rosati}, {Schuldt}, {Trenti}, \&
  {Treu}}]{stiavelli2023}
{Stiavelli}, M., {Morishita}, T., {Chiaberge}, M., {et~al.} 2023, \apjl, 957,
  L18, \dodoi{10.3847/2041-8213/ad0159}

\bibitem[{{Tacchella} {et~al.}(2022){Tacchella}, {Finkelstein}, {Bagley},
  {Dickinson}, {Ferguson}, {Giavalisco}, {Graziani}, {Grogin}, {Hathi},
  {Hutchison}, {Jung}, {Koekemoer}, {Larson}, {Papovich}, {Pirzkal},
  {Rojas-Ruiz}, {Song}, {Schneider}, {Somerville}, {Wilkins}, \&
  {Yung}}]{tacchella2022}
{Tacchella}, S., {Finkelstein}, S.~L., {Bagley}, M., {et~al.} 2022, \apj, 927,
  170, \dodoi{10.3847/1538-4357/ac4cad}

\bibitem[{{Tacchella} {et~al.}(2023{\natexlab{a}}){Tacchella}, {Johnson},
  {Robertson}, {Carniani}, {D'Eugenio}, {Kumari}, {Maiolino}, {Nelson},
  {Suess}, {{\"U}bler}, {Williams}, {Adebusola}, {Alberts}, {Arribas},
  {Bhatawdekar}, {Bonaventura}, {Bowler}, {Bunker}, {Cameron}, {Curti},
  {Egami}, {Eisenstein}, {Frye}, {Hainline}, {Helton}, {Ji}, {Looser}, {Lyu},
  {Perna}, {Rawle}, {Rieke}, {Rieke}, {Saxena}, {Sandles}, {Shivaei},
  {Simmonds}, {Sun}, {Willmer}, {Willott}, \& {Witstok}}]{tacchella2023}
{Tacchella}, S., {Johnson}, B.~D., {Robertson}, B.~E., {et~al.}
  2023{\natexlab{a}}, \mnras, 522, 6236, \dodoi{10.1093/mnras/stad1408}

\bibitem[{{Tacchella} {et~al.}(2023{\natexlab{b}}){Tacchella}, {Eisenstein},
  {Hainline}, {Johnson}, {Baker}, {Helton}, {Robertson}, {Suess}, {Chen},
  {Nelson}, {Pusk{\'a}s}, {Sun}, {Alberts}, {Egami}, {Hausen}, {Rieke},
  {Rieke}, {Shivaei}, {Williams}, {Willmer}, {Bunker}, {Cameron}, {Carniani},
  {Charlot}, {Curti}, {Curtis-Lake}, {Looser}, {Maiolino}, {Maseda}, {Rawle},
  {Rix}, {Smit}, {{\"U}bler}, {Willott}, {Witstok}, {Baum}, {Bhatawdekar},
  {Boyett}, {Danhaive}, {de Graaff}, {Endsley}, {Ji}, {Lyu}, {Sandles},
  {Saxena}, {Scholtz}, {Topping}, \& {Whitler}}]{tacchella2023b}
{Tacchella}, S., {Eisenstein}, D.~J., {Hainline}, K., {et~al.}
  2023{\natexlab{b}}, \apj, 952, 74, \dodoi{10.3847/1538-4357/acdbc6}

\bibitem[{{Topping} {et~al.}(2022){Topping}, {Stark}, {Endsley}, {Plat},
  {Whitler}, {Chen}, \& {Charlot}}]{topping2022}
{Topping}, M.~W., {Stark}, D.~P., {Endsley}, R., {et~al.} 2022, \apj, 941, 153,
  \dodoi{10.3847/1538-4357/aca522}

\bibitem[{{Topping} {et~al.}(2024){Topping}, {Stark}, {Endsley}, {Whitler},
  {Hainline}, {Johnson}, {Robertson}, {Tacchella}, {Chen}, {Alberts}, {Baker},
  {Bunker}, {Carniani}, {Charlot}, {Chevallard}, {Curtis-Lake}, {DeCoursey},
  {Egami}, {Eisenstein}, {Ji}, {Maiolino}, {Williams}, {Willmer}, {Willott}, \&
  {Witstok}}]{topping2024}
---. 2024, \mnras, 529, 4087, \dodoi{10.1093/mnras/stae800}

\bibitem[{{Umeda} {et~al.}(2023){Umeda}, {Ouchi}, {Nakajima}, {Harikane},
  {Ono}, {Xu}, {Isobe}, \& {Zhang}}]{umeda2023}
{Umeda}, H., {Ouchi}, M., {Nakajima}, K., {et~al.} 2023, arXiv e-prints,
  arXiv:2306.00487, \dodoi{10.48550/arXiv.2306.00487}

\bibitem[{{Vanzella} {et~al.}(2016){Vanzella}, {de Barros}, {Vasei}, {Alavi},
  {Giavalisco}, {Siana}, {Grazian}, {Hasinger}, {Suh}, {Cappelluti}, {Vito},
  {Amorin}, {Balestra}, {Brusa}, {Calura}, {Castellano}, {Comastri}, {Fontana},
  {Gilli}, {Mignoli}, {Pentericci}, {Vignali}, \& {Zamorani}}]{vanzella2016}
{Vanzella}, E., {de Barros}, S., {Vasei}, K., {et~al.} 2016, \apj, 825, 41,
  \dodoi{10.3847/0004-637X/825/1/41}

\bibitem[{{Vidal-Garc{\'\i}a} {et~al.}(2017){Vidal-Garc{\'\i}a}, {Charlot},
  {Bruzual}, \& {Hubeny}}]{Vidal2017}
{Vidal-Garc{\'\i}a}, A., {Charlot}, S., {Bruzual}, G., \& {Hubeny}, I. 2017,
  \mnras, 470, 3532, \dodoi{10.1093/mnras/stx1324}

\bibitem[{{Wang} {et~al.}(2023){Wang}, {Fujimoto}, {Labb{\'e}}, {Furtak},
  {Miller}, {Setton}, {Zitrin}, {Atek}, {Bezanson}, {Brammer}, {Leja}, {Oesch},
  {Price}, {Chemerynska}, {Cutler}, {Dayal}, {van Dokkum}, {Goulding},
  {Greene}, {Fudamoto}, {Khullar}, {Kokorev}, {Marchesini}, {Pan}, {Weaver},
  {Whitaker}, \& {Williams}}]{wang2023}
{Wang}, B., {Fujimoto}, S., {Labb{\'e}}, I., {et~al.} 2023, \apjl, 957, L34,
  \dodoi{10.3847/2041-8213/acfe07}

\bibitem[{{Watanabe} {et~al.}(2024){Watanabe}, {Ouchi}, {Nakajima}, {Isobe},
  {Tominaga}, {Suzuki}, {Ishigaki}, {Nomoto}, {Takahashi}, {Harikane},
  {Hatano}, {Kusakabe}, {Moriya}, {Nishigaki}, {Ono}, {Onodera}, \&
  {Sugahara}}]{watanabe2024}
{Watanabe}, K., {Ouchi}, M., {Nakajima}, K., {et~al.} 2024, \apj, 962, 50,
  \dodoi{10.3847/1538-4357/ad13ff}

\bibitem[{{Westera} {et~al.}(2002){Westera}, {Lejeune}, {Buser}, {Cuisinier},
  \& {Bruzual}}]{westera2002}
{Westera}, P., {Lejeune}, T., {Buser}, R., {Cuisinier}, F., \& {Bruzual}, G.
  2002, \aap, 381, 524, \dodoi{10.1051/0004-6361:20011493}

\bibitem[{{Whitaker} {et~al.}(2019){Whitaker}, {Ashas}, {Illingworth}, {Magee},
  {Leja}, {Oesch}, {van Dokkum}, {Mowla}, {Bouwens}, {Franx}, {Holden},
  {Labb{\'e}}, {Rafelski}, {Teplitz}, \& {Gonzalez}}]{whitaker2019}
{Whitaker}, K.~E., {Ashas}, M., {Illingworth}, G., {et~al.} 2019, \apjs, 244,
  16, \dodoi{10.3847/1538-4365/ab3853}

\bibitem[{{Williams} {et~al.}(2023){Williams}, {Tacchella}, {Maseda},
  {Robertson}, {Johnson}, {Willott}, {Eisenstein}, {Willmer}, {Ji}, {Hainline},
  {Helton}, {Alberts}, {Baum}, {Bhatawdekar}, {Boyett}, {Bunker}, {Carniani},
  {Charlot}, {Chevallard}, {Curtis-Lake}, {de Graaff}, {Egami}, {Franx},
  {Kumari}, {Maiolino}, {Nelson}, {Rieke}, {Sandles}, {Shivaei}, {Simmonds},
  {Smit}, {Suess}, {Sun}, {{\"U}bler}, \& {Witstok}}]{williams2023}
{Williams}, C.~C., {Tacchella}, S., {Maseda}, M.~V., {et~al.} 2023, \apjs, 268,
  64, \dodoi{10.3847/1538-4365/acf130}

\bibitem[{{Willott} {et~al.}(2023){Willott}, {Desprez}, {Asada}, {Sarrouh},
  {Abraham}, {Brada{\v{c}}}, {Brammer}, {Estrada-Carpenter}, {Iyer}, {Martis},
  {Matharu}, {Mowla}, {Muzzin}, {Noirot}, {Sawicki}, {Strait},
  {Rihtar{\v{s}}i{\v{c}}}, \& {Withers}}]{willott2023}
{Willott}, C.~J., {Desprez}, G., {Asada}, Y., {et~al.} 2023, arXiv e-prints,
  arXiv:2311.12234, \dodoi{10.48550/arXiv.2311.12234}

\bibitem[{{Witstok} {et~al.}(2023){Witstok}, {Smit}, {Saxena}, {Jones},
  {Helton}, {Sun}, {Maiolino}, {Kumari}, {Stark}, {Bunker}, {Arribas}, {Baker},
  {Bhatawdekar}, {Boyett}, {Cameron}, {Carniani}, {Charlot}, {Chevallard},
  {Curti}, {Curtis-Lake}, {Eisenstein}, {Endsley}, {Hainline}, {Ji}, {Johnson},
  {Looser}, {Nelson}, {Perna}, {Rix}, {Robertson}, {Sandles}, {Scholtz},
  {Simmonds}, {Tacchella}, {{\"U}bler}, {Williams}, {Willmer}, \&
  {Willott}}]{witstok2023}
{Witstok}, J., {Smit}, R., {Saxena}, A., {et~al.} 2023, arXiv e-prints,
  arXiv:2306.04627, \dodoi{10.48550/arXiv.2306.04627}

\bibitem[{{Wolfe} {et~al.}(2005){Wolfe}, {Gawiser}, \& {Prochaska}}]{wolfe2005}
{Wolfe}, A.~M., {Gawiser}, E., \& {Prochaska}, J.~X. 2005, \araa, 43, 861,
  \dodoi{10.1146/annurev.astro.42.053102.133950}

\bibitem[{{Zavala} {et~al.}(2024){Zavala}, {Castellano}, {Akins}, {Bakx},
  {Burgarella}, {Casey}, {Ch{\'a}vez Ortiz}, {Dickinson}, {Finkelstein},
  {Mitsuhashi}, {Nakajima}, {P{\'e}rez-Gonz{\'a}lez}, {Arrabal Haro}, {Buat},
  {Backhaus}, {Calabr{\`o}}, {Cleri}, {Fern{\'a}ndez-Arenas}, {Fontana},
  {Franco}, {Giavalisco}, {Grogin}, {Hathi}, {Hirschmann}, {Ikeda}, {Jung},
  {Kartaltepe}, {Koekemoer}, {Larson}, {McKinney}, {Papovich}, {Saito},
  {Santini}, {Terlevich}, {Terlevich}, {Treu}, \& {Yung}}]{zavala2024}
{Zavala}, J.~A., {Castellano}, M., {Akins}, H.~B., {et~al.} 2024, arXiv
  e-prints, arXiv:2403.10491, \dodoi{10.48550/arXiv.2403.10491}

\end{thebibliography}
\bibliographystyle{aasjournal}

\end{document}